\documentclass[twocolumn]{aastex62}
\received{June 4}
\revised{August 6}
\accepted{August 20}
\submitjournal{PASP}
\shorttitle{Extending Supernova Spectral Templates}
\shortauthors{Pierel et al.}

\newif{\ifchangetext}
\changetextfalse

\InputIfFileExists{\jobname-options}

\ifchangetext
  
  \newcommand{\changenote}[1]{\textcolor{blue}{ \bf #1}}x
\else
  
  \newcommand{\changenote}[1]{}
\fi

\hypersetup{
    colorlinks=True,
    citecolor=black,
    linkcolor=black,
    filecolor=magenta,      
    urlcolor=cyan,
}

\def\uprime{\ensuremath{{\textit{u}^\prime}}}
\def\rprime{\ensuremath{{\textit{r}^\prime}}}

\def\xone{\ensuremath{x_{1}}}

\newcolumntype{L}{>{$}l<{$}} %
\newcolumntype{C}{>{$}c<{$}} %
\newcolumntype{R}{>{$}r<{$}} %
\begin{document}

\title{Extending Supernova Spectral Templates for Next-Generation Space Telescope Observations}

\newcommand{\USC}{Department of Physics and Astronomy, University of South Carolina, 712 Main St., Columbia, SC 29208, USA}
\newcommand{\CfA}{Harvard-Smithsonian Center for Astrophysics, Cambridge, MA 02138, USA}
\newcommand{\Chicago}{Department of Physics, The University of Chicago, Chicago, IL 60637, USA}
\newcommand{\NYU}{Center for Cosmology and Particle Physics, New York University, New York, NY 10003, USA}
\newcommand{\UCSB}{Department of Physics, University of California, Santa Barbara, CA 93106-9530, USA}
\newcommand{\SantaBarbara}{\UCSB}
\newcommand{\UCSD}{Center for Astrophysics \& Space Sciences, University of California, San Diego, 9500 Gilman Drive, La Jolla, CA 92093, USA}
\newcommand{\IllinoisAstro}{Astronomy Department, University of Illinois at Urbana-Champaign, Urbana, IL 61801, USA }
\newcommand{\Cambridge}{Statistical Laboratory, DPMMS, University of Cambridge, Wilberforce Road, Cambridge, CB3 0WB, UK}
\newcommand{\KICPCambridge}{Institute of Astronomy and Kavli Institute for Cosmology, Madingley Road, Cambridge, CB3 0HA, UK}
\newcommand{\STScI}{Space Telescope Science Institute, 3700 San Martin Dr., Baltimore, MD 21218, USA}
\newcommand{\Berkeley}{Department of Astronomy, University of California, Berkeley, CA 94720-3411, USA}
\newcommand{\Miller}{Miller Senior Fellow, Miller Institute for Basic Research in Science, University of California, Berkeley, CA 94720, USA}
\newcommand{\KICPChicago}{The Kavli Institute for Cosmological Physics, Chicago, IL 60637, USA}
\newcommand{\Moore}{The Gordon and Betty Moore Foundation, 1661 Page Mill Road, Palo Alto, CA 94304}
\newcommand{\Rutgers}{Department of Physics and Astronomy, Rutgers, the State University of New Jersey, Piscataway, NJ 08854, USA}
\newcommand{\UCSC}{Department of Astronomy and Astrophysics, University of California, 1156 High St., Santa Cruz, CA 95064, USA}

\correspondingauthor{J.~D.~R.~Pierel}
\email{jr23@email.sc.edu}
\author{J.~D.~R.~Pierel}

\affil{Department of Physics and Astronomy, University of South Carolina, 712 Main St., Columbia, SC 29208, USA}
\author{S.~Rodney}

\affil{Department of Physics and Astronomy, University of South Carolina, 712 Main St., Columbia, SC 29208, USA}
\author{A.~Avelino}

\affil{Harvard-Smithsonian Center for Astrophysics, Cambridge, MA 02138, USA}
\author{F.~Bianco}

\affil{Center for Cosmology and Particle Physics, New York University, New York, NY 10003, USA}
\author{A.~V.~Filippenko}

\affil{Department of Astronomy, University of California, Berkeley, CA 94720-3411, USA}
\affil{Miller Senior Fellow, Miller Institute for Basic Research in Science, University of California, Berkeley, CA 94720, USA}
\author{R.~J.~Foley}

\affil{Department of Astronomy and Astrophysics, University of California, 1156 High St., Santa Cruz, CA 95064, USA}
\author{A.~Friedman}

\affil{Center for Astrophysics \& Space Sciences, University of California, San Diego, 9500 Gilman Drive, La Jolla, CA 92093, USA}
\author{M.~Hicken}

\affil{Harvard-Smithsonian Center for Astrophysics, Cambridge, MA 02138, USA}
\author{R.~Hounsell}

\affil{Department of Astronomy and Astrophysics, University of California, 1156 High St., Santa Cruz, CA 95064, USA}
\affil{Astronomy Department, University of Illinois at Urbana-Champaign, Urbana, IL 61801, USA }
\author{S.~W.~Jha}

\affil{Department of Physics and Astronomy, Rutgers, the State University of New Jersey, Piscataway, NJ 08854, USA}
\author{R.~Kessler}

\affil{The Kavli Institute for Cosmological Physics, Chicago, IL 60637, USA}
\author{R.~P.~Kirshner}

\affil{Harvard-Smithsonian Center for Astrophysics, Cambridge, MA 02138, USA}
\affil{The Gordon and Betty Moore Foundation, 1661 Page Mill Road, Palo Alto, CA 94304}
\author{K.~Mandel}

\affil{Statistical Laboratory, DPMMS, University of Cambridge, Wilberforce Road, Cambridge, CB3 0WB, UK}
\affil{Institute of Astronomy and Kavli Institute for Cosmology, Madingley Road, Cambridge, CB3 0HA, UK}
\author{G.~Narayan}

\affil{Space Telescope Science Institute, 3700 San Martin Dr., Baltimore, MD 21218, USA}
\author{D.~Scolnic}

\affil{The Kavli Institute for Cosmological Physics, Chicago, IL 60637, USA}
\author{L.~Strolger}

\affil{Space Telescope Science Institute, 3700 San Martin Dr., Baltimore, MD 21218, USA}

\begin{center}

\end{center}
\begin{abstract}
Empirical  models of supernova (SN)  spectral energy  distributions
(SEDs) are widely used for SN survey simulations and photometric classifications. The existing library of SED models has excellent optical templates but limited, poorly constrained coverage of ultraviolet (UV) and infrared (IR) wavelengths. However, both regimes are critical for
the design and operation of future SN surveys, particularly at IR wavelengths that will be accessible with the James Webb Space Telescope
(JWST) and the Wide-Field Infrared Survey Telescope (WFIRST).  We create a public repository of improved empirical SED templates using a sampling of
Type Ia and core-collapse  (CC)  photometric  light curves to extend the Type Ia parameterized SALT2 model and a set of SN Ib, SN Ic, and SN II
SED templates into the UV and near-IR. We apply this new repository of  extrapolated SN SED models to examine how future surveys  can
discriminate  between CC  and  Type  Ia SNe at UV and IR wavelengths, and present an open-source software package written  in Python, \textit{SNSEDextend}, that enables a user
to generate their  own extrapolated SEDs.

\end{abstract}

\

\section{Introduction}
\label{sub:intro}
Photometric templates and models of supernova (SN) spectral energy distributions (SEDs) are critical tools for gleaning physical properties of supernovae (SNe) from observations, determining how those properties evolve over time, and performing SN classifications.  Many SN analysis tools, such as 
the widely-used SuperNova ANAlysis \citep[SNANA][]{Kessler:2009a} and SNCosmo \citep{Barbary:2014} software packages, utilize a common set of empirically-derived SEDs that represent a variety of core-collapse (CC) and Type Ia SNe. 
Most existing template SEDs, however, are only constrained by data at optical wavelengths \citep{Blondin:2007}. 
Many of the software packages for photometric SN classification rely on these template SEDs \citep{Kessler:2010,Sako:2011}.  Extending the SED templates into near-infrared (NIR) wavelengths is necessary for those classification tools to be applicable for the next
generation of telescopes---such as the James Web Space Telescope (JWST), the Wide-Field Infrared Survey Telescope (WFIRST), and the Large Synoptic Survey Telescope (LSST)---which will provide a
plethora of new SN observations that span wavelengths from the optical
to the far-IR \citep{Dahlen:1999,Mesinger:2006,Ivezic:2008,Spergel:2015}.  

A preliminary extension of the SN SED library into NIR bands \citep{Pierel:2018a}\footnote{\href{dx.doi.org/10.5281/zenodo.1250492}{DOI:10.5281/zenodo.1250492}} has already been used for the analysis of SN discoveries in the Cosmic Assembly Near-infrared Deep Extragalactic Legacy Survey \citep[CANDELS,][]{Rodney:2014} and the Cluster Lensing And Supernova survey with Hubble \citep[CLASH,][]{Graur:2014}.   The simplistic modified SN SED templates employed for that work have also been used to explore various survey strategies for the WFIRST SN program  \citep{Hounsell:2017}.

In this work we provide a more rigorous extension of ultraviolet (UV) and NIR coverage for current SEDs.  First, we describe a new
open-source software tool, \textit{SNSEDextend}, that is capable of extrapolating SN SEDs to match photometric observations.  The data and methodology for extending CC SN SEDs are presented in Section~\ref{sec:ccsn}.  Extrapolation of the Type Ia model SALT2 \citep{Guy:2010} is described in Section~\ref{sec:typeIa}.  We then provide a new repository of SEDs extrapolated to 
cover the wavelength range $\sim 1700$--25,000~\mbox{\normalfont\AA}, and in Section~\ref{sec:results} we apply these SEDs to explore photometric SN classifications in IR bands. 

As the number of UV and IR
observations of SNe increases, the accuracy of the extrapolations will continue to improve and the \textit{SNSEDextend}
package will be available to supplement the repository with updated and new SED templates.  Meanwhile, the
intention is that these extrapolated SEDs will be used by the wider SN
research community for simulations and photometric classifications.  We note, however, that our extrapolations of the SALT2 Type Ia SN model to UV and NIR wavelengths are not intended to make SALT2 capable of light-curve fitting in those wavelength regimes for cosmological distance measurements.  That would require retraining of the model, which is beyond the scope of this work.

\
\section{Core-Collapse Supernovae}
\label{sec:ccsn}
Classifications of CC~SNe can be broadly grouped into three types, each with their own set of subclasses. Type Ib and Ic SNe are characterized with their early-time spectra first by a lack of hydrogen (Type I), and then by the absence of strong Si II and the presence of He I (Ib), as well as the absence of both strong Si II and He I (Ic) \citep[see, e.g.,][]{Filippenko:1997}. The classification of Type II is broadly inclusive of SNe containing hydrogen in their spectra, and is then further split based on optical spectral and light-curve properties into II-P, II-L, IIb, and IIn \citep[e.g.,][]{Filippenko:1997}. The existing CC~SN SED library comprises 11 SNe~Ib, 8 SNe~Ic, 28 SNe~II-P, 3 SNe~IIn, 1 SN~II-L, and 0 SN~IIb templates created from observations of 48 objects. Owing to the sparsity of SED templates and existing optical+NIR light curves for SNe~IIn, II-L, and IIb, we have excluded them from this analysis. The \textit{SNSEDextend}\ code can be used in the future to perform these necessary extrapolations when more data are available. Some SNe included in our analysis have a classification of Type II, but no further subclassification. As these SNe are not clearly SNe~IIb, II-L, or IIn, and recent CC~SN frequency studies have concluded that 83--95\% of SNe~II can be classified as Type II-P \citep[e.g.,][]
{Smartt:2009}, we have grouped these SNe together with the SNe~II-P and have removed any that the SN~II-P optical templates are unable to fit.

\begin{figure*}[t]
\centering
\includegraphics[width=.7\textwidth]{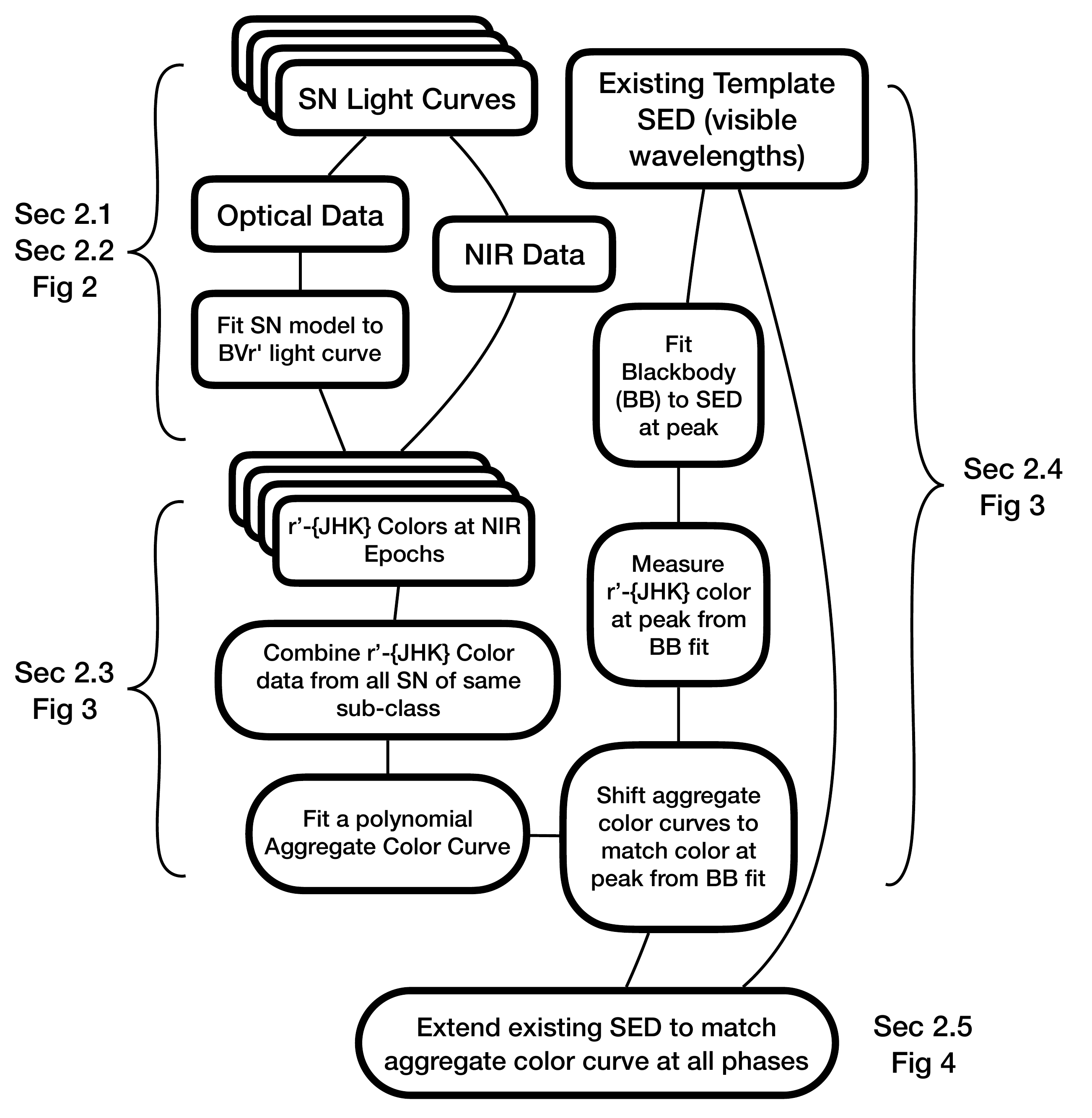}
\caption{\label{fig:flow} A flowchart summarizing the process for extending an existing CC~SN template SED out to NIR wavelengths.  The identical process can be applied for CC~SN SED extrapolation to UV wavelengths.}
\end{figure*}

The CC~SN SED template library was created using light curves with excellent optical data, but only 7 of the 44 objects were observed in the NIR \citep{Kessler:2010,Sako:2011}. Our goal is to improve these templates by extending their coverage to NIR wavelengths using a set of SNe observed in the optical and NIR. The steps for these CC~SN template SED extrapolations are summarized in Figure~\ref{fig:flow}.   
Our process begins with a collection of SN light curves that include both optical and NIR photometric data, detailed in Section \ref{sub:cc_data}. 
These light curves are then grouped by SN subclass (Type Ib, Ic, II+IIP). For each SN in a given subclass we use the existing templates to fit the well-sampled optical light-curve data (Section \ref{sub:fitting}), which enables interpolation over the light curve in the optical bands. The interpolation is necessary as we combine these fits with the discrete observed UV and NIR photometry to derive color-evolution curves for each CC~SN subclass (Section \ref{sub:curves}). 
These optical-NIR colors are then used to constrain the extrapolations for each existing SED template (Sections \ref{sub:bb} and \ref{sub:extrap}).

\subsection{Core-Collapse Supernova Data}
\label{sub:cc_data}

The existing library of CC~SN SEDs that we are extending, developed for a photometric classification challenge in 2010 \citep{Kessler:2010} and used by \citet{Sako:2011}, consists of templates for 11 SNe~Ib, 8 SNe~Ic, and 28 SNe~II/II-P. These templates were formed from smoothed spectral time series for SNe of a given subtype, which were then warped to match photometric observations of different SNe of the same subtype \citep{Kessler:2010}. None of the SN light curves used to guide the warping of these CC~SN SED templates had well-sampled NIR data, so any template extending redward beyond optical wavelengths is poorly constrained.  These CC~SN templates have reasonable UV constraints, reaching down to 2000~\AA.  We therefore have not modified or extrapolated them on the blue side, but the \textit{SNSEDextend}\ package has the capability to do UV extrapolations, following the algorithm outlined in Figure~\ref{fig:flow} for the IR side.

To constrain our SED template extrapolations, we require that the color evolution of each extrapolated template SED matches the best available observed optical and NIR photometry for SNe of the same subtype.   This approach requires a set of CC~SNe with both well-sampled optical light curves and some NIR photometry.  For this purpose we adopt a collection of  photometric data from low-redshift 
CC~SNe, taken from \citet{Bianco:2014} and
\citet{Hicken:2017}, both collected at the Fred L. Whipple Observatory
(FLWO) (Table \ref{Atab:cc_data}). The  \citet{Hicken:2017} data include SNe~II and II-P (Figure \ref{Afig:hickencurves}), while the \citet{Bianco:2014} data are from stripped-envelope SNe of Types Ib
and Ic (Figure \ref{Afig:biancocurves}). Any SNe in these samples that do not have data in at least one of the $U$, $J$, $H$, or $K$ bands are discarded.  This assemblage includes 9 SNe~Ib, 1 SN~IIb, 7 SNe~Ic, 8 SN~II, and 2 SNe~II-P,
for a total of 27 objects, and the SNe~II and II-P
are combined into a single group for this process. As stated above, extrapolation into the UV is not necessary for this work, but light curves with UV data are included so that the \textit{SNSEDextend}\ package is prepared to perform future UV extrapolations. 

\begin{figure}[ht!]
\centering
\includegraphics[width=.5\textwidth]{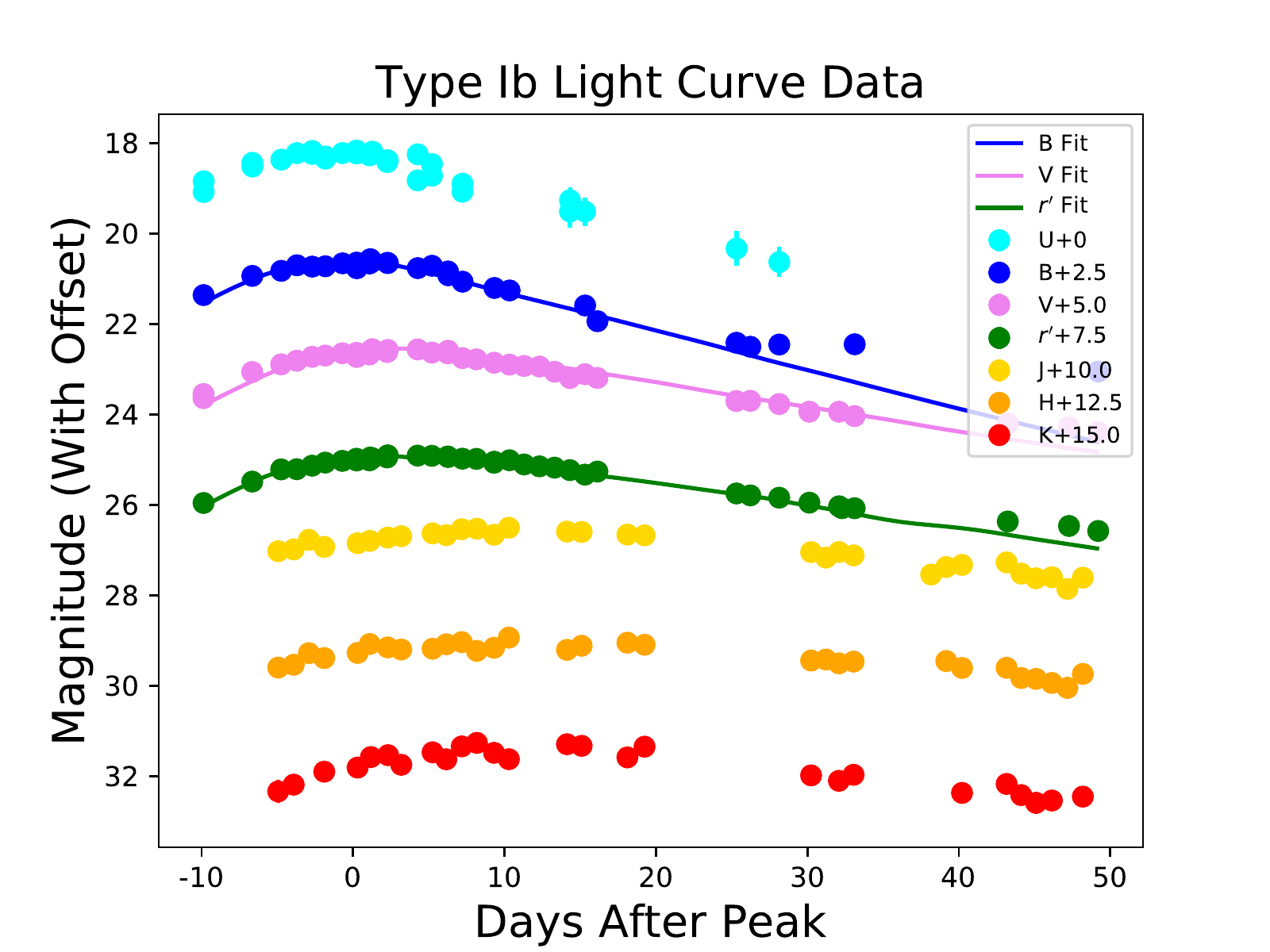}
\caption{\label{fig:cc_lc} 
The light curve of SN 2005hg, a well-sampled SN~Ib used in this work.  Filled circles show observed photometry in UV, optical, and NIR bands.  Solid lines illustrate fits to the optical bandpasses, derived with SNCosmo. The $U$, $J$, $H$, and $K$ bands are not fit. Magnitude offsets have been added to the data in order to easily distinguish between each band, and to clearly show the bands being fit in Section \ref{sub:fitting}. The offsets used are listed in the legend.}
\end{figure}

A particularly well-sampled SN~Ib light curve, that of SN 2005hg \citep{Bianco:2014}, is shown in Figure \ref{fig:cc_lc}.  The observed data are overlaid with light-curve fits in optical bands, which are described in Section \ref{sub:fitting}.  Similar light-curve plots for all the CC~SNe used in this work are provided in the Appendix.

\

\subsection{Light-Curve Fitting}
\label{sub:fitting}

In the \textit{SNSEDextend}\ package we use the SNCosmo\footnote{\href{https://sncosmo.readthedocs.io/en/v1.6.x/}{SNCosmo Version 1.6}} Python toolkit \citep{Barbary:2014} for light-curve fitting, with the important change that we have added the FLWO PAIRITEL $J$, $H$, and $K_s$ bands to the SNCosmo bandpass registry. A best-fit optical light-curve model is found for each SN in our photometric dataset by fitting one of the existing spectrophotometric SED models to the observed Bessell $B$, $V$, and SDSS $\rprime$ photometric data.  More details of the fitting process are given in the Appendix.  These optical models are used to obtain $\rprime-J$, $\rprime-H$, and $\rprime-K$ colors for use in extrapolation. 

\renewcommand{\labelitemii}{-}

\subsection{Color Table Generation}
\label{sub:curves}
Measuring, for example, an $\rprime-J$ color over time is called an $\rprime-J$ ``color curve.'' We will use $\rprime-\{JHK\}$ to refer to the ensemble of $\rprime-J$, $\rprime-H$, and $\rprime-K$ color curves. The purpose of fitting the optical SN light curves in Section \ref{sub:fitting} is to generate color curves for each SN type, which are used to define SED extrapolations into the NIR (Section \ref{sub:extrap}). In order to minimize uncertainties that arise from comparing different colors (e.g., $\rprime-J$ and $i^\prime-J$), we have chosen the most prevalent ``red" wavelength band in our dataset (SDSS $\rprime$) as the optical anchor for all of our NIR color measurements.

The sparsity of the NIR data for each SN type necessitates calculating colors for each SN, and merging SNe of like classification to obtain a well-sampled color curve. The \textit{SNSEDextend}\ package does this by creating a merged ``color table'' for each SN type, which simply defines the discrete set of colors measured for all SNe of that type (Table \ref{Atab:color_table}). 
These data giving observed color over time are fit with a polynomial ``color curve.''   Details of the polynomial fitting are given in the Appendix. 
An example of deriving a continuous color curve for a single SN subtype is shown in Figure \ref{fig:shifted}.  
Note that the SN~Ib color curve is flat for phases beyond the temporal extent of our dataset, as we have no color constraints in those regions. 
Previous work suggests that the shape of a SN color curve near peak brightness should not be used to predict the color at phases far from peak \citep{Krisciunas:2009}.

\

\subsection{Host-Galaxy Extinction and Intrinsic Color Variation}
\label{sub:varation}
In this work we are grouping SNe of like subtype together to perform extrapolations, which means that intrinsic and extrinsic variability of NIR colors within a SN subtype must be accounted for before defining the extrapolations. Recent work suggests that the majority of SN~II color diversity is intrinsic and not due to host-galaxy extinction \citep{jaeger:2018}, while similar analyses with stripped-envelope SNe seem to assume the exact opposite \citep[e.g.][]{Taddia:2018}. Although further investigation is warranted, it is clear that SNe of all subtypes will suffer from some measure of both intrinsic \citep[e.g.,][]{Filippenko:1997} and extrinsic \citep[e.g.,][]{Kelly:2012} variability.

To account for extrinsic variation in the SN colors, a correction for host-galaxy extinction is made during the light-curve fitting process. We adopt the dust law defined by \citet{Cardelli:1989} with $R_V=3.1$, and fit for the host-galaxy $E(B-V)$ using SNCosmo (see Section \ref{sub:fitting} and the Appendix). No color-variation parameter exists in the models capable of accounting for the intrinsic color variation of SNe from the same subclass. Instead, we use the diversity of colors present in the extrapolated SED templates to represent the inherent color variation, for which we account in Section \ref{sub:bb}.

\

\subsection{Fixing the Peak Color with Blackbody Fits}
\label{sub:bb}
After creating an aggregate color curve for each SN subtype in Section \ref{sub:curves}, the intrinsic variation of NIR colors within each SN subtype is addressed.  To do this, we use a blackbody fit to each optical template SED at peak.   Although a blackbody spectrum is not necessarily an accurate model for SN SEDs at early and late times \citep{Baron:2004,Shussman:2016}, it will generally provide a valid approximation close to the time of peak luminosity \citep{Hershkowitz:1986}. 

To test the assumption of a blackbody approximation at peak brightness, we collected each of the SNe with optical+NIR data that were originally used to create the base template SED repository. As discussed in Section \ref{sec:ccsn}, only 7 of these 44 SNe were observed in the NIR, and one of the 7 is discarded owing to insufficient NIR coverage. This leaves 3 SNe~Ib and 3 SNe~Ic, none of which have sufficient data in the $K_s$ band, to be used in the comparison.

\begin{figure}
\centering
\includegraphics[width=.5\textwidth]{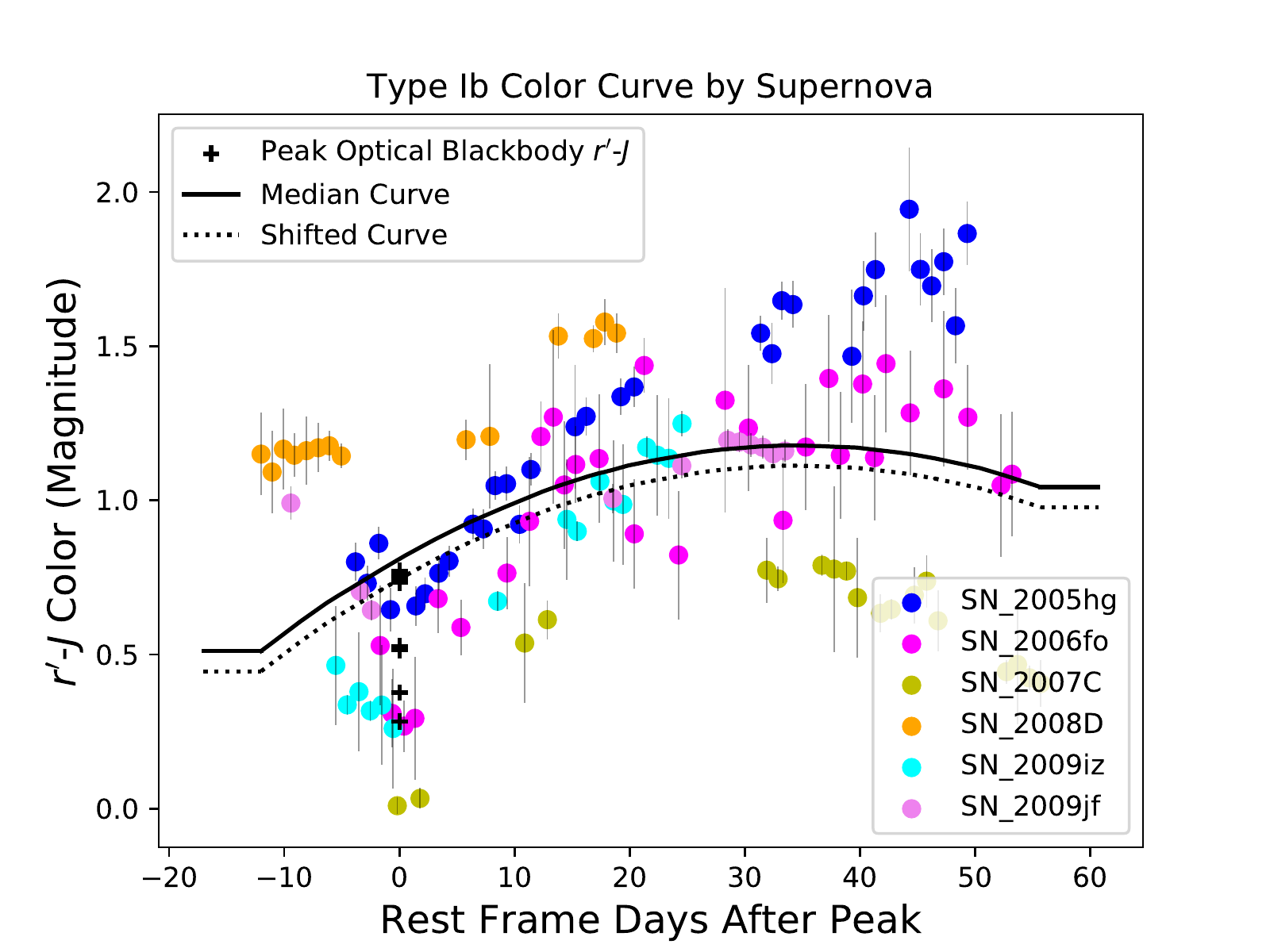}
\caption{\label{fig:shifted} 
Example of defining an optical-NIR color curve for SNe~Ib.  Colored circles show the $\rprime-J$ color from six SNe~Ib with well-sampled optical and NIR light curves. For each color data point, the \rprime magnitude is interpolated from the light-curve fit to optical bands and the $J$ magnitude is a directly observed value.   The solid line shows a polynomial fit to the aggregate data from all six SNe with NIR light-curve data in this subclass.  Black crosses show the $\rprime-J$ color at peak brightness for the 11 existing SN~Ib SED templates, derived by fitting a blackbody to the optical wavelength region of each SED in the template library.  The dashed line shows an example of the aggregate color curve after shifting to match the peak $\rprime-J$ color of one of those 11 SED templates.  This dashed curve is then used to constrain the NIR extrapolation of that SED template at all phases.  }
\end{figure}

The SED template corresponding to each SN at peak brightness is fit with a blackbody spectrum in the optical wavelength range ($\sim4200--7500~\mbox{\normalfont\AA}$). The templates are not fit with a blackbody at UV wavelengths owing to the well-documented effects of line blanketing \citep[e.g.,][]{Marion:2014}. The best-fit blackbody spectrum defines the fluxes through the $J$ and $H$ bandpasses ($\sim 10,500$--18,500~\mbox{\normalfont\AA}), which are then used to define $\rprime-{JH}$ colors. By comparing the observed $\rprime-{JH}$ color to the $\rprime-{JH}$ color predicted by the blackbody fit, we conclude that the scatter introduced by the blackbody fitting (Table \ref{tab:bb_comp}) is $\sim2$--$3\sigma$ less than the intrinsic scatter for the whole sample in each subtype ($\sim0.5$; e.g., Figure \ref{fig:shifted}).

\begin{table*}[t]
\centering
\caption{Scatter introduced by the blackbody fitting.$^a$}
\begin{tabular}{crr}
\toprule
\multicolumn{1}{l}{SN Subtype} & \multicolumn{1}{c}{|($\rprime-J)_{\rm obs}-(\rprime-J)_{\rm BB}$|} & \multicolumn{1}{c}{|$(\rprime-H)_{\rm obs}-(\rprime-H)_{\rm BB}$|} \\\hline
Ib & $0.16\pm0.09$ & $0.20\pm0.13$ \\
Ic & $0.14\pm0.11$ & $0.11\pm0.11$
\\\hline
\end{tabular}
\label{tab:bb_comp}
\\
{$^a$Scatter measured by comparing the observed and blackbody-predicted NIR colors for 3 SNe~Ib and 3 SNe~Ic at peak brightness. These uncertainties are much less than the intrinsic scatter observable in each subclass ($\sim0.5$; e.g., Figure \ref{fig:shifted}), indicating that fitting a blackbody spectrum to a SN SED provides a reasonable approximation of peak-brightness NIR colors for SNe~Ib and Ic.}
\end{table*}

\begin{figure}
\centering
\includegraphics[width=.5\textwidth]{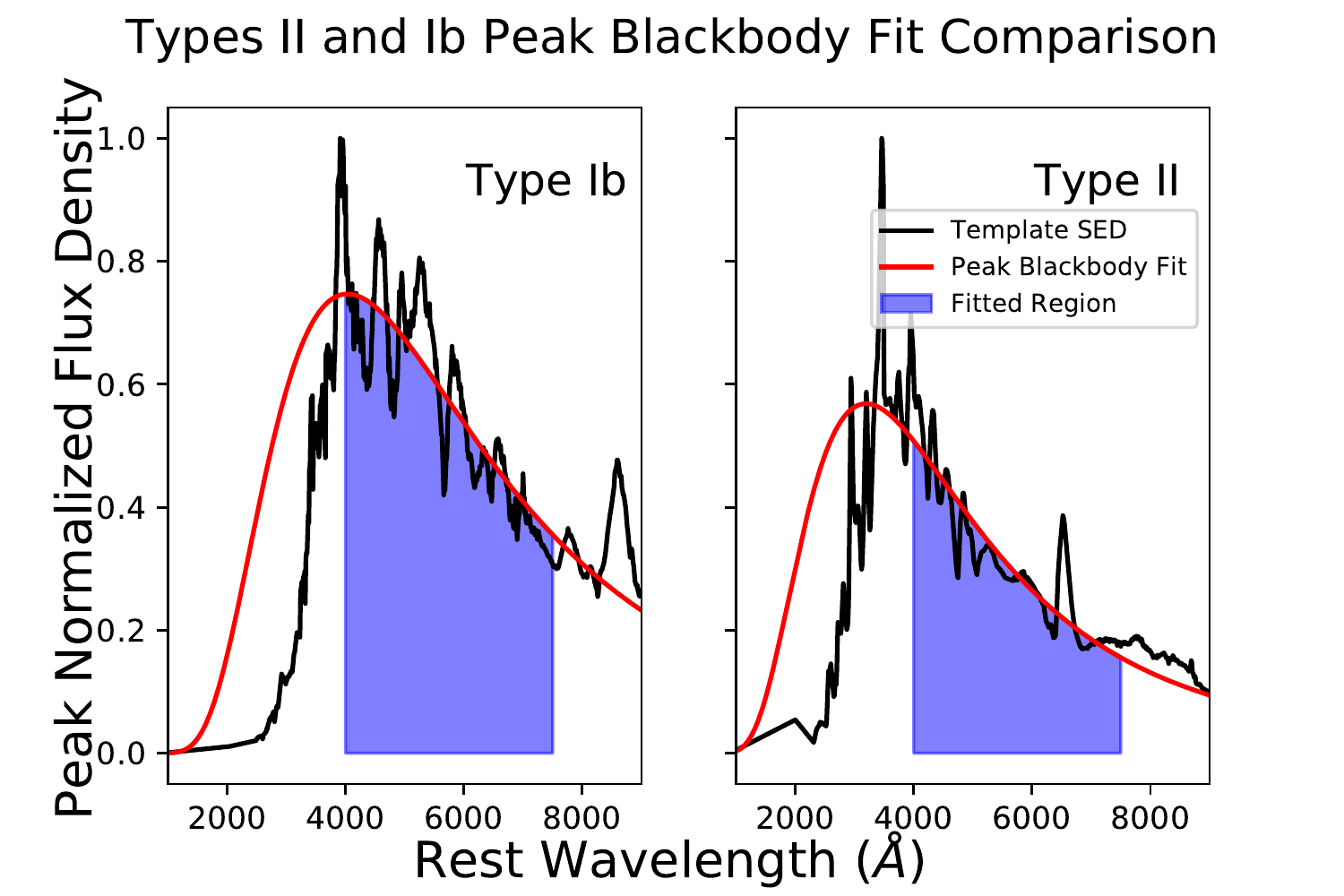}
\caption{\label{fig:bb_fits} An example of a blackbody spectrum fit to the optical wavelengths of a SN~Ib (left) and SN~II-P (right) template SED at peak. Although none of the Type II-P template SNe have NIR data available to compare the observed and blackbody-predicted colors directly (as is done in Table \ref{tab:bb_comp}), the fact that a blackbody can be as successfully fit to a SN~II SED as to a SN~Ib SED is both encouraging and the only currently obtainable evidence that such an approximation is reasonable.}
\end{figure}

Although none of the template SNe~II/II-P were observed in the NIR, we must somehow anchor the color curves to take intrinsic variation into consideration. It would clearly be valuable to check the assumption that the blackbody approximation is also valid for SNe~II/II-P with the same test done for SNe~Ib and Ic. However, without templates containing NIR data, we are only able to confirm that the blackbody fitting method is as successful for SNe~II/II-P as for SNe~Ib and Ic (Figure \ref{fig:bb_fits}); the SN~II/II-P intrinsic variation methodology should be updated as needed when more data become available.

As shown in Figure \ref{fig:shifted}, the aggregate color curve for each SN subtype, derived in Section \ref{sub:curves}, is shifted vertically with the shape remaining intact, such that it intersects with the corresponding peak blackbody color.  This means that the change in color over time is the same for all the SN SED extrapolations in a given SN subclass, but each extrapolation is anchored to a different set of $\rprime-{JHK}$ colors at peak brightness, derived from the blackbody fits.

\

\subsection{SED Extrapolation}
\label{sub:extrap}
\begin{figure}
\centering
\includegraphics[width=.5\textwidth]{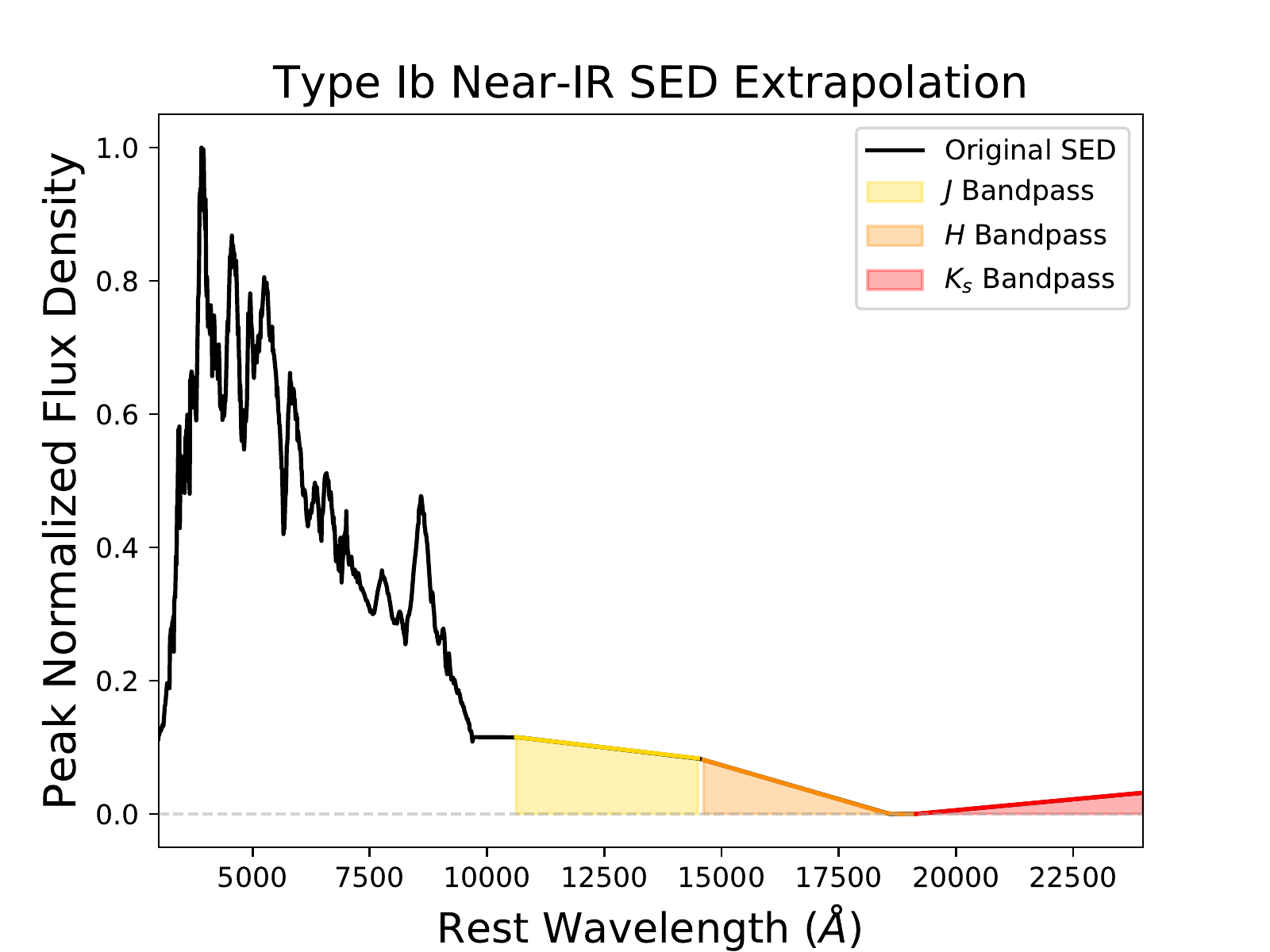}
\caption{\label{fig:sedExtrap} Example of a SN~Ib SED at peak brightness, with linear extrapolations into the UV and IR. The flux through each bandpass, represented by shaded regions, is constrained to match the color curves generated by the \textit{SNSEDextend} \ package for the corresponding SN subtype (see Section \ref{sub:curves} and Figure~\ref{fig:shifted}).}
\end{figure}

For each CC~SN Type Ib, Ic, and II/II-P, the SED templates described in Section \ref{sub:cc_data} are extrapolated to match the color curves defined in Sections \ref{sub:curves} and \ref{sub:bb}. 
From the shifted aggregate color curve (Figure~\ref{fig:shifted}), we extract a requirement for the $\rprime-{JHK}$ color for every epoch defined in the template SED timeseries.  
We apply a piecewise linear extrapolation to the SED  such that the extrapolated SED's colors match these measured colors, as shown in Figure \ref{fig:sedExtrap}.   Although our color data extend only to the $K$ band, we continue the extrapolation to even longer wavelengths by arbitrarily setting the flux to be zero at 55,000~$\mbox{\normalfont\AA}$ in all phases.  Similarly, we extrapolate on the UV side to reach zero at 1200~$\mbox{\normalfont\AA}$. 
The final result is a
set of extrapolated SEDs that cover the full wavelength range of
1200--55,000~\mbox{\normalfont\AA}, where the updated SEDs' colors in the NIR correspond to those measured in Sections \ref{sub:curves} and \ref{sub:bb}. 
These extrapolated SEDs are available for download from an online repository\footnote{\href{dx.doi.org/10.5281/zenodo.1250492}{DOI:10.5281/zenodo.1250492}}, and are included in the latest versions of the SNANA and SNCosmo packages.

\
\subsection{SED Spectroscopic Features}
\label{sub:spec}
Once the linear extrapolations from Section \ref{sub:extrap} are applied, one may wish to add absorption and emission features to the extrapolated regions of the SED. 
This is implemented in \textit{SNSEDextend}\ by drawing from a repository of flattened SN spectra, such as those in the Supernova Identification software package \citep[SNID;][]{Blondin:2007}. 
A flattened SNID spectrum is normalized to match the template SED, and then added to the linear \textit{SNSEDextend} extrapolations. The broad-band colors defined in Sections \ref{sub:curves}--\ref{sub:extrap} are maintained, as the mean flux values in the flattened SNID spectra can be normalized to zero for any wavelength bin \citep{Blondin:2007}. A spectral feature with high equivalent width, such as H$\alpha$, would affect broad-band colors despite the normalization. However, there are no features with similarly high equivalent width in NIR CC~SN spectra where this method is to be applied. \citep[e.g.,][]{Dessart:2012,Dessart:2013} .  

For the repository of extrapolated SN SEDs described here, all of the input SED templates were already well constrained at UV wavelengths, and the SNID template library has essentially no information about spectral features in the NIR. Therefore, this capability is not used in this work, but is available for future applications of the \textit{SNSEDextend}\ package. 

\

\subsection{K-Corrections}
\label{sub:kcorr}
The ``K-correction'' allows for the comparison of flux measurements between objects at different redshifts \citep{Humason:1956,Oke:1968}.  With accurate K-corrections at NIR bands, one could use a larger dataset of high-redshift ($z\gtrsim0.1$) CC~SN light curves to define the color curves that constrain our SED extrapolations.  However, deriving an accurate K-correction requires knowledge of the SED at the wavelength range of interest, which is the goal of the extrapolation.  To avoid this circularity, we use exclusively low-redshift SNe ($z<0.04$) to guide our SED extrapolations, and assume that the K-corrections are negligible.

After implementing the extrapolations described above, we check the magnitude of the K-correction that each extrapolated CC~SN SED would predict.   We find that none of these post-hoc K-corrections would be greater than $\sim0.04$ mag, which is much less than the intrinsic scatter in the colors of the CC~SN population.  If the \textit{SNSEDextend}\ package is applied in the future using CC~SNe of higher redshift, then the initial NIR extrapolations presented here could be used to define K-corrections for those high-$z$ light curves.

\

\

\section{Type Ia Supernovae}
\label{sec:typeIa}

As our baseline spectrophotometric model for SNe~Ia, we adopt SALT2 \citep{Guy:2007}, a parametric model for SN light-curve fitting.  This model gives the SN~Ia flux density $F$ as a function of phase $\phi$ and wavelength $\lambda$ according to \citep{Guy:2007}
\begin{equation}
\label{eq:salt2}
F(\phi,\lambda)=x_0\big[M_0(\phi,\lambda)+x_1M_1(\phi,\lambda)\big]\exp\big[c\times CL(\lambda)\big] .
\end{equation}
The core components of the SALT2 model are $M_0$, $M_1$, and $CL$, which are fixed components, common for every SN~Ia. The free parameters are $x_0$, $x_1,$ and $c$, which are fit to match the luminosity, light-curve shape, and color of each individual SN~Ia. 

Our goal is to extend the wavelength range over which the SALT2 model is defined, growing from the current 2000--9200~\mbox{\normalfont\AA} \ to 1700--25,000~\mbox{\normalfont\AA}.  
This extension improves the utility of the SALT2 model for SN survey simulations and photometric classifications. 
As the original SALT2 training had limited constraints below $\sim3500$~\mbox{\normalfont\AA} \ and above $\sim8000$~\mbox{\normalfont\AA} \ \citep{Guy:2007}, we also seek to improve the accuracy of the blue and red ends of the SALT2 model, in the 2000--3500~\mbox{\normalfont\AA} \ and 8000--9200~\mbox{\normalfont\AA} \ wavelength ranges. 
The original SALT2 model is known to produce negative fluxes, particularly at early UV and late NIR epochs \citep[e.g.,][]{Mosher:2014}. 
We do not correct this issue, as the SALT2 model between 3500 and 8000~\mbox{\normalfont\AA} \ and the overall framework are left untouched so as to not disturb its current applications. 
However, we have ensured that the UV and NIR extrapolations do not produce negative fluxes at any epoch.

\begin{figure*}[t]
\centering \includegraphics[width=0.8\textwidth]{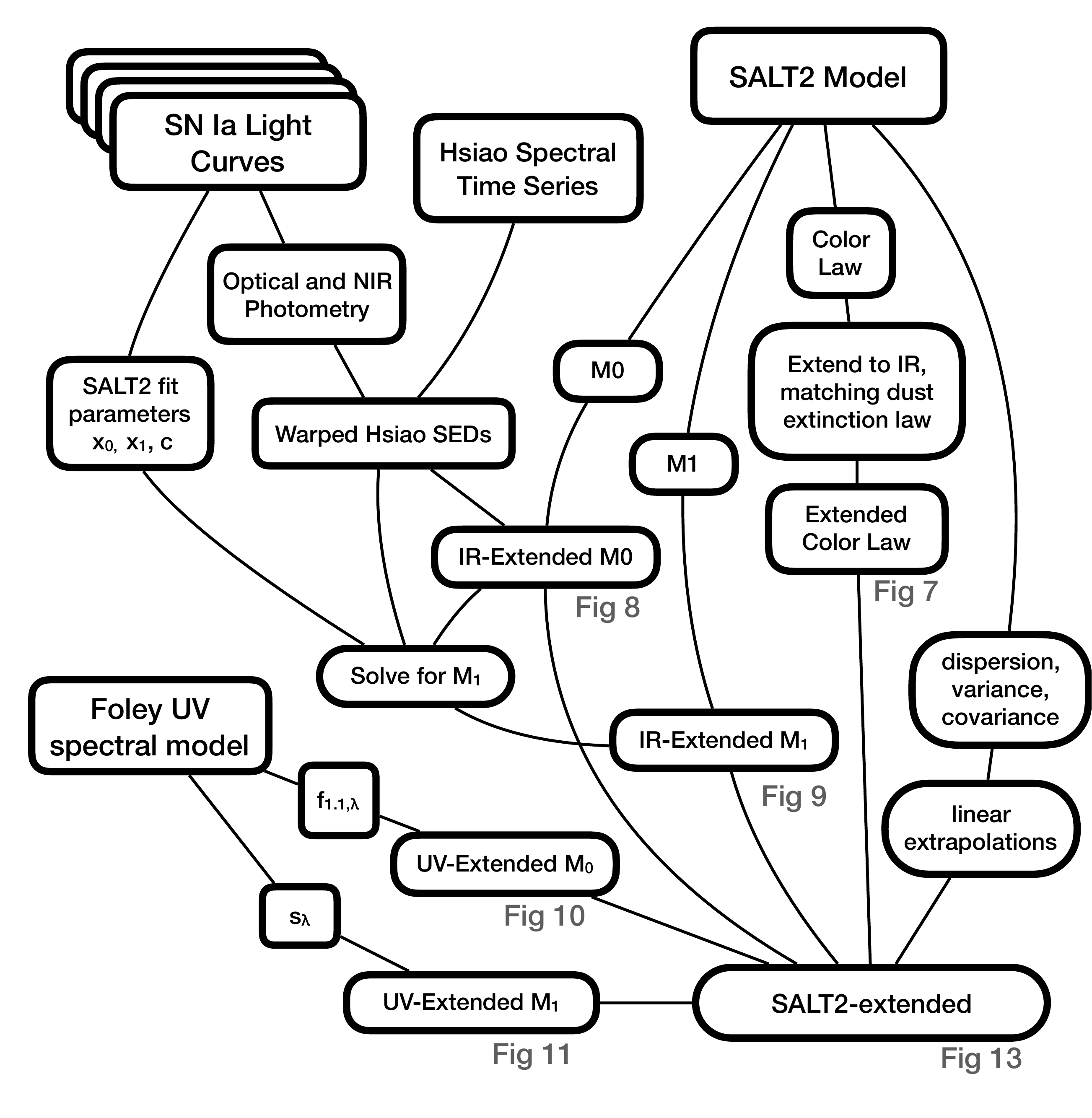}
\caption{\label{fig:flowchartIa} Flowchart summary of the  process for extending the SALT2 model to UV and NIR wavelengths.}
\end{figure*}

Figure \ref{fig:flowchartIa} summarizes the steps for the extrapolation of the SALT2 model to NIR and UV wavelengths. In Section \ref{sub:ia_nir} we extend the SALT2 model to NIR wavelengths, first addressing the color law, and then extending the $M_0$ and $M_1$ components using constraints derived from a sample of low-redshift SNe~Ia.   
In Section \ref{sub:ia_uv}, we use the parameterized SN~Ia light-curve model of \citet{Foley:2016} to guide the extrapolation of SALT2 to UV wavelengths.  Our extension of the dispersion, variance, and covariance terms of SALT2 is described in Section \ref{sub:ia_variance}.

\subsection{Extending SALT2 to Near-Infrared Wavelengths}
\label{sub:ia_nir}

\subsubsection{The Color Law}
\label{sub:ir_colorlaw}
The SALT2 color law incorporates any
wavelength-dependent color variations that are independent
of epoch, and is defined as a polynomial in wavelength with no time dependence \citep{Guy:2007}. For SALT2 the color law was defined over the wavelength range 2800--7000~\mbox{\normalfont\AA}, and it extends with a linear extrapolation to any wavelength outside of that range \citep{Guy:2007,Guy:2010}. In practice those linear extension regions are much more limited, since the SALT2 model is only defined over the range 2000--9200~\mbox{\normalfont\AA}.  We have left the SALT2 color law effectively unchanged on the UV side, by maintaining approximately the same slope at the 2800~\mbox{\normalfont\AA}\ left edge of the polynomial.  To allow the SALT2 model to apply beyond 9200~\mbox{\normalfont\AA}, we have updated the coefficients of the color law polynomial, with a physically motivated constraint.

\begin{figure}
\centering
\includegraphics[ width=0.26\textwidth]{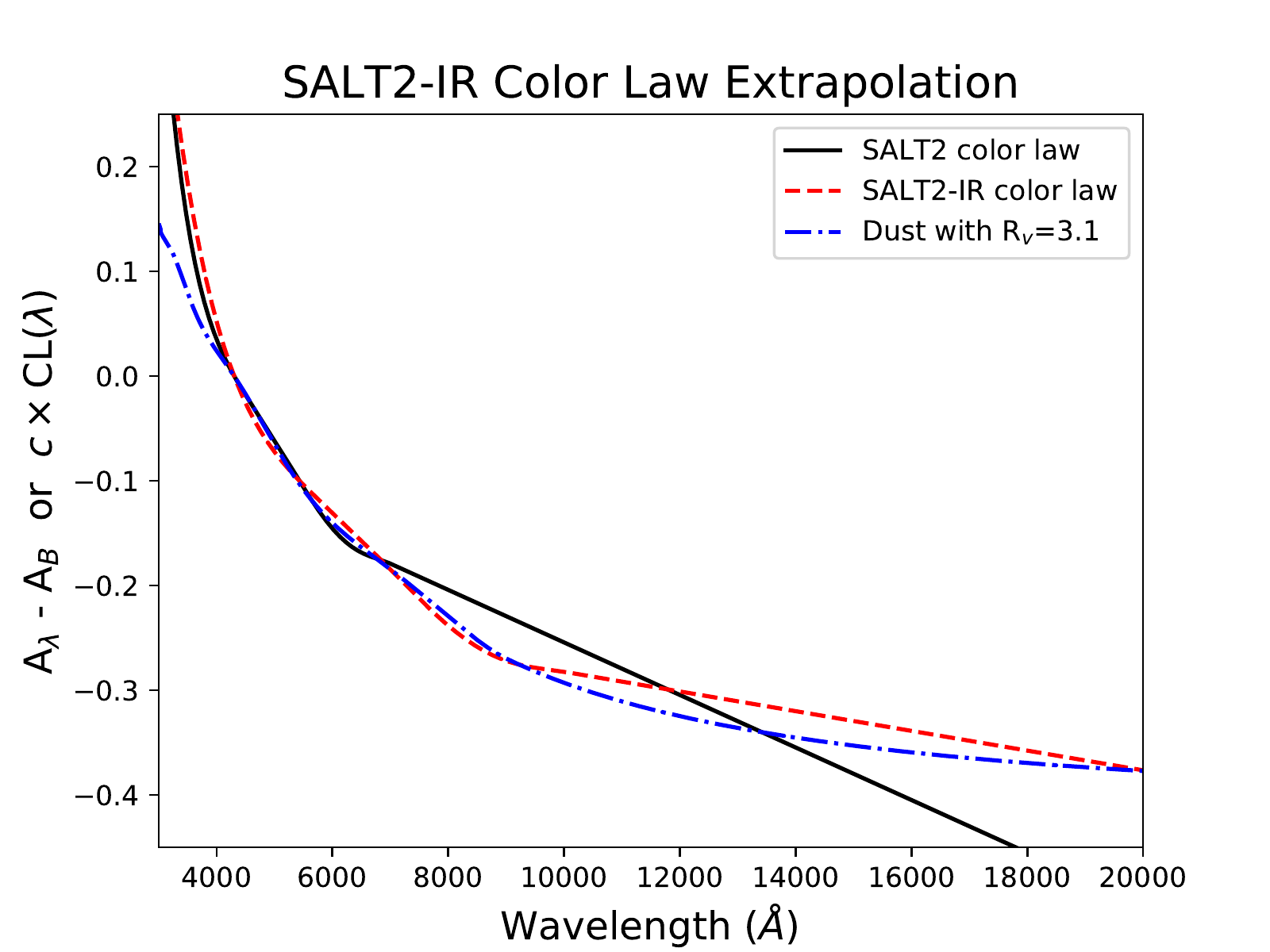}
\caption{\label{fig:SALT2_colorlaw} Extrapolation of the SALT2 color law to IR wavelengths. The solid black line and dashed red line show the SALT2 color law multiplied by the SALT2 color parameter $c$ (set to $c=0.1$ for this plot), for the input SALT2 model and the modified SALT2-IR model, respectively.  The blue dash-dot line shows the color excess $A_\lambda-A_B$ as a function of wavelength $\lambda$ for a dust extinction law with $R_V=3.1$, following \citet{Cardelli:1989} and \citet{odonnell:1994}. The SALT2-IR color law is designed to follow the original SALT2 color law closely until it reaches 7000~\mbox{\normalfont\AA}, then approximate the $A_\lambda-A_B$ curve at NIR wavelengths (Section \ref{sub:ir_colorlaw}). The SALT2-IR color law does not precisely equal the $A_\lambda-A_B$ curve at NIR wavelengths as it is defined as a third-order polynomial, which is necessary to maintain the framework of the original SALT2 model}.
\end{figure}

 Our constraint is to make the assumption that the correlation between color and peak luminosity at IR wavelengths is dominated by dust extinction, not by color variation that is intrinsic to the SN.  This is similar to the assumptions underpinning the ``color excess model" developed by \citet{Phillips:1999} and employed by \citet{Burns:2011}.  To encode this assumption in the extrapolated SALT2 model, we want the revised color law to be close to the existing SALT2 color law up to 7000~\mbox{\normalfont\AA}, and then at redder wavelengths the model should respond to changes in the color parameter $c$ as if that color change is caused by dust extinction. To enforce this constraint, we set the color law polynomial coefficients such that for a SN with a moderate color parameter $c = 0.1$ we will get an extinction factor $c\times CL(\lambda)$ that behaves approximately like the Milky Way dust extinction curve from \cite{Cardelli:1989} and \cite{odonnell:1994}. Figure \ref{fig:SALT2_colorlaw} shows the final SALT2-IR color law, given by the red curve, and its relationship to the original SALT2 color law (black) and the Milky Way dust extinction curve (blue).

Investigations of the use of NIR light curves for SNe~Ia as standardizable candles have adopted other approaches for how to model the color-luminosity relationship \citep[e.g.,][]{Mandel:2009,Mandel:2011,Burns:2011,Kattner:2012,Dhawan:2018}. 
These studies have not found a strong correlation of NIR colors with peak luminosity that would invalidate our assumption.  However, the available datasets are limited, and further work is certainly warranted to explore whether the structure of the SALT2 model should be modified to handle NIR colors in a fundamentally different way.

\subsubsection{NIR Type Ia Supernova Data}
\label{sub:ia_data}

To constrain the extrapolation of the SALT2 $M_0$ and $M_1$ components to IR wavelengths, we use optical and NIR photometric data from a low-$z$ SN~Ia sample to construct a set of synthetic SEDs based on the \citet{Hsiao:2007} spectral model.  We begin with a set of SN light curves collected by the CfA and CSP surveys, as well as some other SNe~Ia reported in the literature (Table \ref{Atab:ia_selections}; CfA:  \citealt{Woodvasey:2008}; \citealt{Hicken:2009b}; \citealt{Hicken:2012}; \citealt{Friedman:2015}; \citealt{Marion:2016}. 
CSP: \citealt{Contreras:2010}; \citealt{Stritzinger:2010}; \citealt{Stritzinger:2011}; \citealt{Krisciunas:2017}. 
Other Groups: \citealt{Krisciunas:2004};  \citealt{Valentini:2003}; \citealt{Krisciunas:2003}; \citealt{Pignata:2008}; \citealt{Stanishev:2007}; \citealt{Leloudas:2009}; \citealt{Krisciunas:2007}).

This sample consists of 86{} spectroscopically confirmed SNe~Ia in the redshift range  $0.0028 < z_{\rm helio} <  0.0390$ obtained during the years 1998--2011,  and that passed the quality requirements (cuts) $0.8 < \Delta m_{15} < 1.6$ mag, $E(B-V)_{\rm host} < 0.4$ mag, $E(B-V)_{\rm Milky~Way} < 1$ mag, and only \textit{normal} SNe~Ia as identified by SNID.
We further restrict the sample to exclude SNe with extreme values of their light-curve shape and color parameters. An ideal but unattainable sample would only contain SNe with fitting parameters $x_1=c=0$. To approximate this, we limit our sample to those SNe with SALT2 parameters of $-1<x_1<1$ and $-0.1<c<0.3$.  These cuts allow a close approximation to $x_1=c=0$, while maintaining a statistically significant sample of 45 low-$z$ SNe for this analysis from the original 86{}.

For each SN in this sample, we generate a  synthetic SN~Ia SED time series that spans the entire wavelength range of interest, 1700--25,000~\mbox{\normalfont\AA}, and 
is defined at every phase in which we have  multiband photometric data. 
Each synthetic SED starts with the Hsiao spectrophotometric model \citep{Hsiao:2007}, which is multiplied by a smooth function created with a ``tension spline'' \citep{Renka:1987} to match the actual observer-frame colors of the SN in our light-curve sample (with appropriate K-corrections applied).  
These warped SEDs are ``dereddened"--- meaning that they are corrected for reddening due to dust extinction in the host galaxy and Milky Way dust along the line of sight.  
The $E(B-V)$ host-galaxy extinction is derived from the optical and NIR light-curve fits using the \textit{SNooPy} code \citep{Burns:2011}, and the Milky Way extinction is determined from the dust maps of \citet{Schlafly:2011} via the NASA/IPAC Extragalactic Database (NED)\footnote{\href{http://ned.ipac.caltech.edu/}{http://ned.ipac.caltech.edu/}} . 

\subsubsection{NIR Extrapolation of the SALT2 M\textsubscript{0} and M\textsubscript{1} Components}
\label{sub:ir_m0}
\begin{figure}
\centering
\includegraphics[width=.5\textwidth]{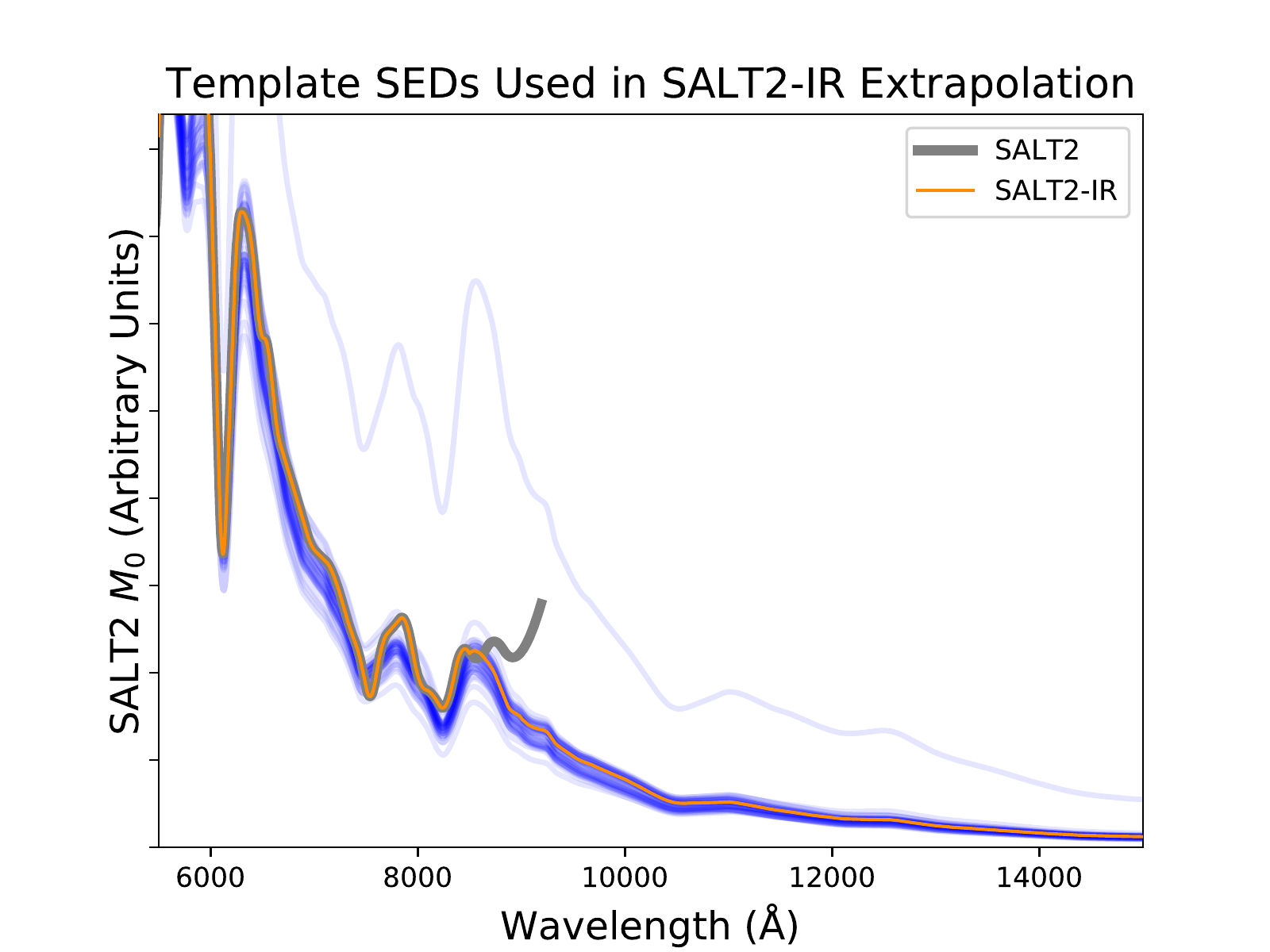}
\includegraphics[width=.5\textwidth]{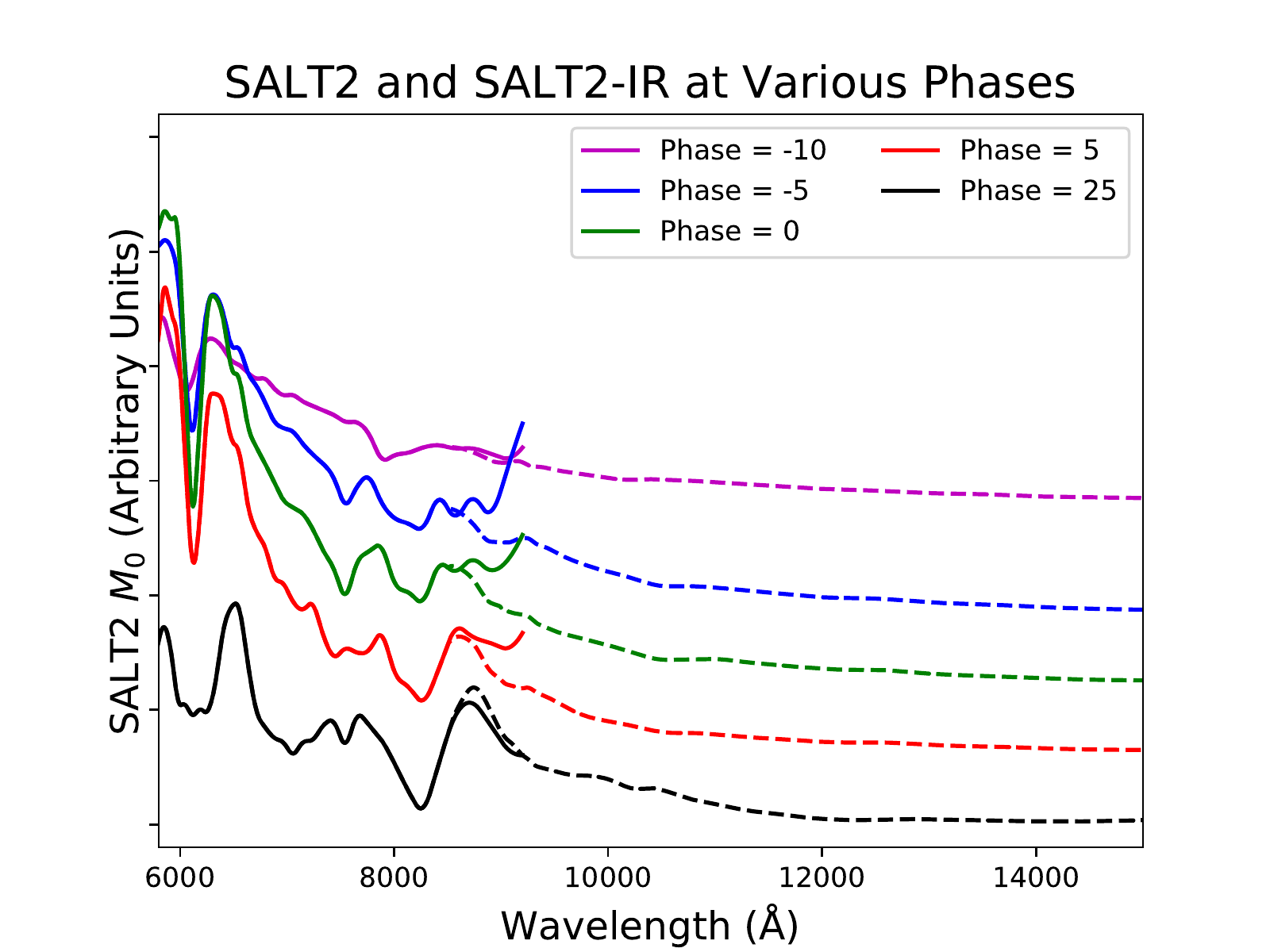}
\caption{\label{fig:SALT2_template0_peak} Extrapolation of the SALT2 $M_0$ model component to IR wavelengths. In the top panel the extrapolation is shown at phase = 0 (peak $B$-band brightness), with thin blue curves showing the input SN~Ia synthetic spectra, thick gray line showing the existing SALT2 $M_0$ template, and the thin orange line showing the extrapolation of the SALT2-IR model.  The bottom panel shows the same extrapolation at five separate phases, with the existing SALT2 model as solid lines, and the SALT2-IR revision as dashed lines.}
\end{figure}

The $M_0$ component of SALT2 is a reference SN~Ia SED at each epoch, with the free parameters $x_1=0$, $c=0$. The NIR extrapolation for any given phase is defined as the median flux at each wavelength from the full sample of all the input SN~Ia SEDs (Figure \ref{fig:SALT2_template0_peak}). This new definition of $M_0$ is used in the parameterization of the SALT2 model beyond its present extent of 9200~\mbox{\normalfont\AA}.

To extend the SALT2 $M_1$ component, the newly-extended $M_0$ component is adopted for wavelengths beyond the SALT2 range of 9200~\mbox{\normalfont\AA}. For each input SN Ia SED, the best-fit values for the $x_0$, $x_1$, and $c$ SALT2 fitting parameters, measured by the Pantheon analysis \citep{Scolnic:2017}, are used to solve the SALT2 model (Equation \ref{eq:salt2}) for $M_1$. Thus, for the $i^{th}$ input SN SED with known fitting parameters ($x_{0,i}$, $x_{1,i}$, $c_i$) and flux density $F_i(\phi,\lambda)$, the $M_1$ component is given by 
\begin{equation}
M_{1,i}(\phi,\lambda)=\frac{1}{x_{1,i}}\Big[\frac{F_i(\phi,\lambda)}{x_{0,i}\,\exp({c_i\times CL(\lambda)})}-M_{0,i}(\phi,\lambda)\Big].
\end{equation}
where $M_0^{(i)}(\phi,\lambda)$ is the parameter measured above and shown in Figure \ref{fig:SALT2_template0_peak}. Because the $M_0$ component is a function of phase and wavelength, we now have a set of $M_1$ values over a two-dimensional (2D) grid in phase and wavelength space, for each input SN~Ia in our light-curve sample. The spacing of the grid in the phase dimension is defined by the spectral sampling of the input low-$z$ SNe~Ia, and the spacing in wavelength space is 10~\mbox{\normalfont\AA}. A median of the $M_1$ grid across all 45 input objects is extracted, and a Savitzky-Golay smoothing function is applied with a 100~\mbox{\normalfont\AA} \ smoothing window in wavelength and a 5-day window in time.

\begin{figure}
\centering
\includegraphics[width=.5\textwidth]{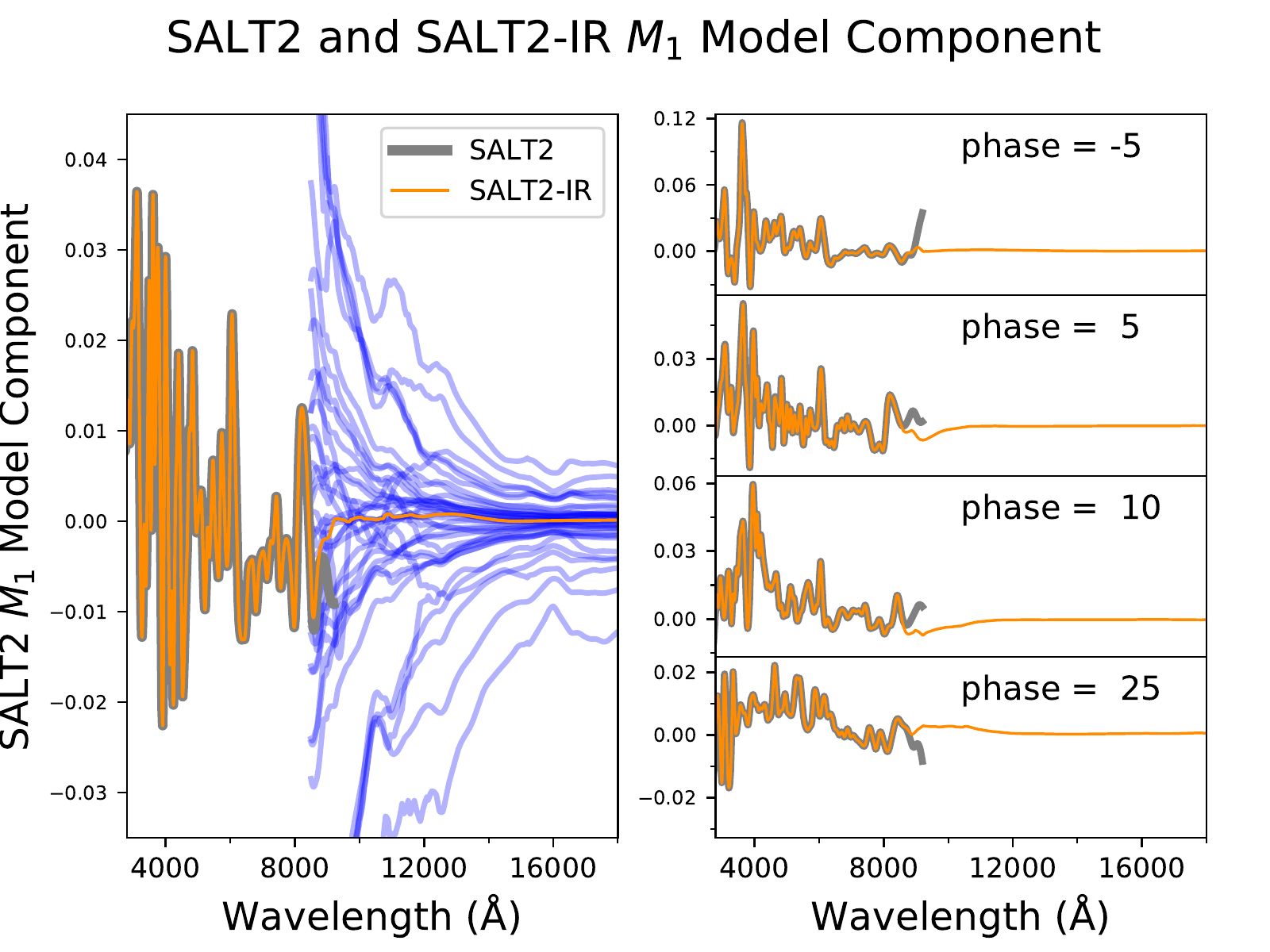}
\caption{\label{fig:SALT2_template1} Extrapolation of the SALT2 $M_1$ model component to IR wavelengths. In the left panel, which is at peak, the solid gray line shows the $M_1$ component of the original SALT2 model. The thin blue curves show the $M_1$ model components for each of the input SN Ia synthetic spectra. The orange line, which overlaps the original SALT2 model over optical wavelengths, represents the SALT2-IR $M_1$ model component for NIR wavelengths. The right panel is the same as the left panel, without the blue curves showing the input data, for various phases.}
\end{figure}

To make a smooth join with the existing template, a weighted average of the median curve with the existing SALT2 $M_1$ component is used, with the fractional weight given to the new median curve smoothly varying from 0 at 8500~\mbox{\normalfont\AA} \ to 1 at 9200~\mbox{\normalfont\AA} \ (Figure \ref{fig:SALT2_template1}).

\subsection{Extending SALT2 to Ultraviolet Wavelengths}
\label{sub:ia_uv}
For the extrapolation of SALT2 into the UV, we use the spectral model defined by \cite[][hereafter F16]{Foley:2016},  

\begin{equation}
\label{eq:foley}
	f(0,\lambda)=f_{1.1}(0,\lambda)+s(0,\lambda)\,(\Delta m_{15}(B)-1.1),
\end{equation}

\noindent where $f_{1.1}(0,\lambda)$ represents the spectrum of a nominal $\Delta m_{15}=1.1$ mag  SN~Ia at peak phase, and $s(0,\lambda)$ is the deviation from that spectrum for a hypothetical $\Delta m_{15}=2.1$ mag SN~Ia at peak. These parameters were measured by F16, and a table can be found in their electronic edition.\footnote{\href{https://academic.oup.com/mnras/article-abstract/461/2/1308/2608545}{https://academic.oup.com/mnras/article-abstract/461/2/1308/2608545}}. In order to create a parameterized extrapolation for the SALT2 model (Equation \ref{eq:salt2}), we recast the F16 model using the base SALT2 parameters $M_0$, $M_1$, $x_0$, and $x_1$. The F16 model is defined over a wavelength range of 1700--5695~\mbox{\normalfont\AA}\ at peak only, while the SALT2 model extends down to 2000~\mbox{\normalfont\AA}\ at phases of $-20$ to +50 days from peak.  We do not modify the SALT2 color law on the UV side, because we do not have sufficient data about the UV color variation of SNe~Ia to guide an informed extrapolation.  Instead, we maintain the blueward linear extrapolation of the original SALT2 color law, as described in Section~\ref{sub:ir_colorlaw}.

\subsubsection{UV Extrapolation of the SALT2 M\textsubscript{0} and M\textsubscript{1} Components}
\label{sub:uv_m0}
\begin{figure}
\centering
\includegraphics[ width=.26\textwidth]{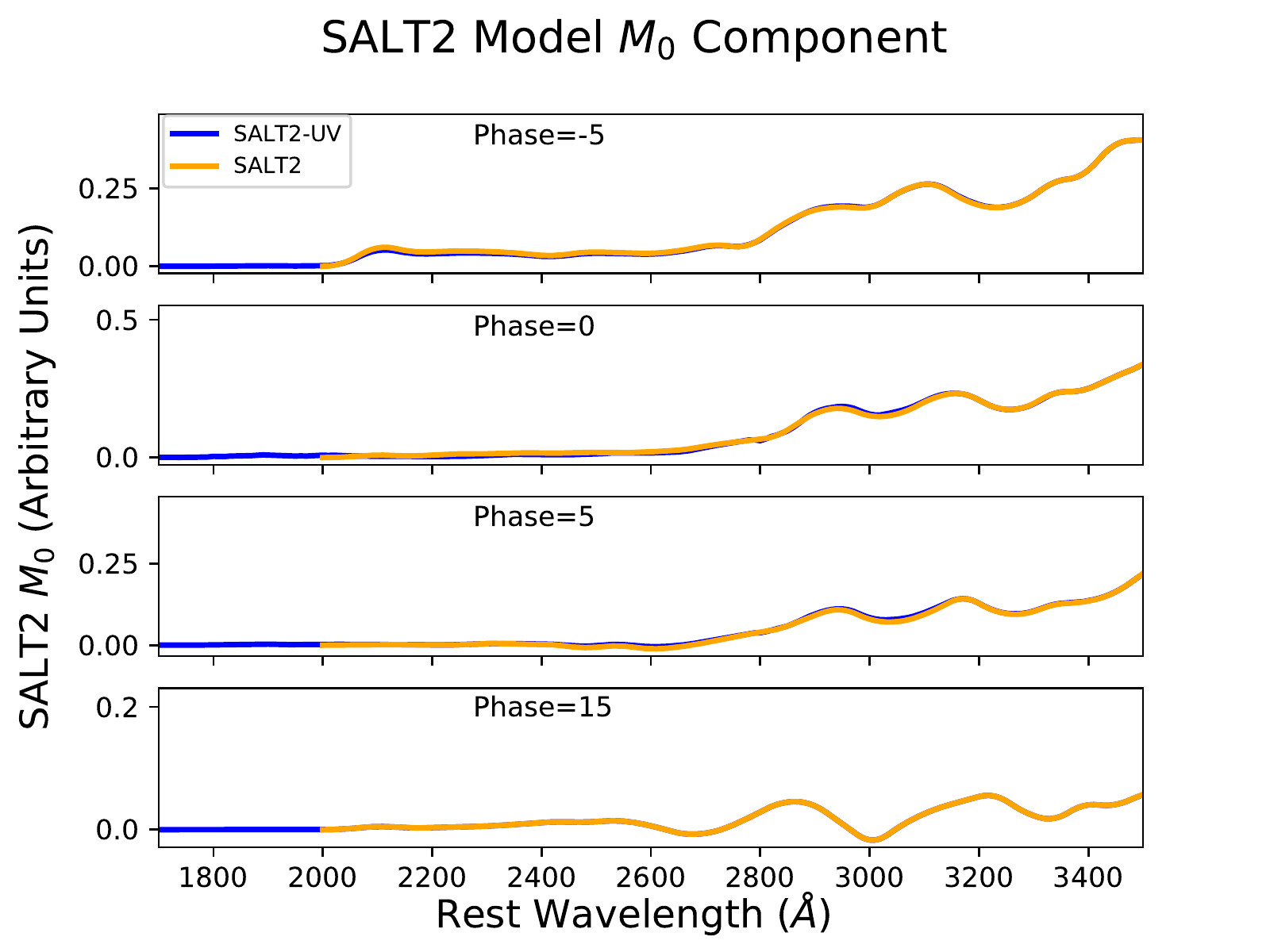}
\caption{\label{fig:uv_m0} Extrapolation of the SALT2 $M_0$ model component to UV wavelengths at various phases from peak using Equation \ref{eq:foley} from F16, and the weighting function defined by Equation \ref{eq:weight}. The blue line is the updated $M_0$ component of the SALT2-UV model, and the orange line is the original $M_0$ component of the SALT2 model. Note that for phases lower than $-5$ or greater than +10, as well as wavelengths greater than 3500~\mbox{\normalfont\AA}, the model reverts back to the original SALT2 $M_0$ component. }
\end{figure}

We first merge the baseline spectral components of the F16 and SALT2 models using a weighted average of the $f_{1.1}$ parameter in F16 and the $M_0$ component of SALT2. This 2D weighted average is given by
\begin{equation}
\label{eq:weight}
M_0(\phi,\lambda)=(1-w_\phi w_\lambda)M_0^{(SALT2)}(\phi,\lambda)+w_\phi w_\lambda f_{1.1}^{(F16)}(0,\lambda).
\end{equation}
Here $w_\phi$ is the weighting function for the phase $\phi$, and $w_\lambda$ is the weighting function for the wavelength $\lambda$. The function $w_\phi$ varies smoothly from 0 (all original SALT2 $M_0$) at $\phi=-5$, to 1 (all F16-defined $M_0$) at peak ($\phi=0$), back to 0 at $\phi=10$. The function $w_\lambda$ is 1 in the range 1700--2000~\mbox{\normalfont\AA}\ where the original SALT2 is not defined, then varies smoothly from 1 at 2000~\mbox{\normalfont\AA} \ to 0 at 3500~\mbox{\normalfont\AA}, which marks the region where SALT2 becomes less reliable. The result of using this weighted average to perform the $M_0$ extrapolation is shown in Figure \ref{fig:uv_m0}.

\begin{figure}
\centering
\includegraphics[ width=.26\textwidth]{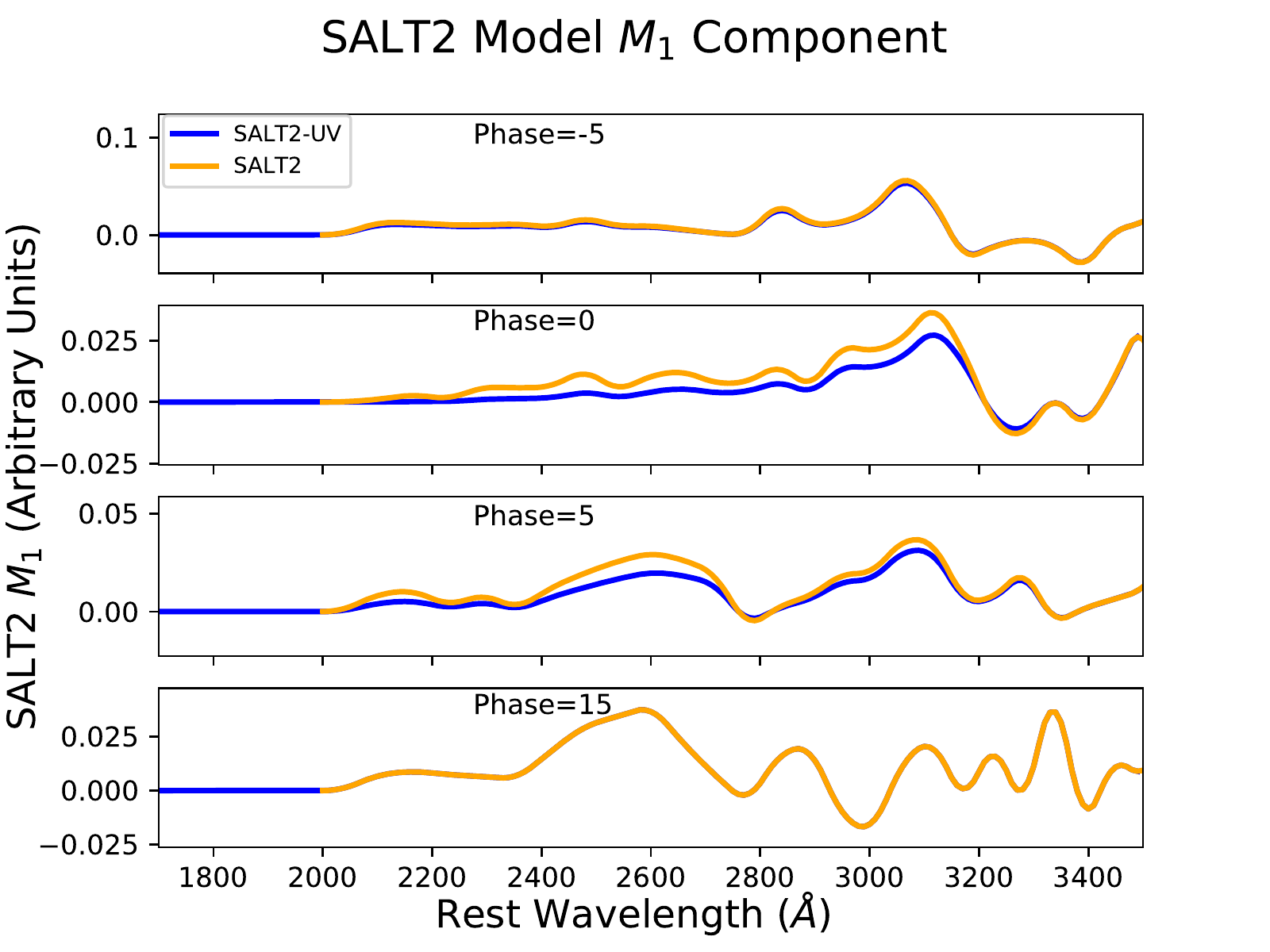}
\caption{\label{fig:uv_m1} Extrapolation of the SALT2 $M_1$ model component to UV wavelengths at various phases from peak using Equation \ref{eq:m1}, and the weighting function defined by Equation \ref{eq:weight}. The blue line is the updated $M_1$ component of the SALT2-UV model, and the orange line is the original $M_1$ component of the SALT2 model. Note that for phases lower than $-5$ or greater than +10, as well as wavelengths greater than 3500~\mbox{\normalfont\AA}, the model reverts back to the original SALT2 $M_1$ component.}
\end{figure}

To modify the SALT2 $M_1$ component at UV wavelengths, we first fix the SALT2 color parameter $c$ at zero, then require that the SALT2 model flux $F$ at the time of peak brightness (Equation \ref{eq:salt2}) must be equal to the F16 flux density $f_{\lambda}$ (Equation \ref{eq:foley}), for all wavelengths $\lambda$.   Solving for the $M_1$ parameter as a scaled version of $s_\lambda$ gives 
\begin{equation}
\label{eq:m1}
	M_1(0,\lambda)=\frac{s(0,\lambda)\,(\Delta m_{15}-1.1)}{x_1}.
\end{equation}
The same weighting functions over wavelength and phase described above for $M_0$ are used here for $M_1$ as well (Figure \ref{fig:uv_m1}). These new SALT2 parameters can then be combined using Equation \ref{eq:salt2} to define the resulting flux density as a function of wavelength and phase (see Section \ref{sec:results}).

\

\subsection{UV and NIR Extrapolation of the SALT2 Dispersion, Variance, and Covariance}
\label{sub:ia_variance}

Finally, to complete the extrapolation of the SALT2 model into NIR and UV wavelengths, we extend the dispersion, variance, and covariance components of the SALT2 model. To make the model suitable for precision cosmology applications, one would need to execute a full retraining of the SALT2 model with a large sample of UV and NIR SN Ia light curves. That is beyond the scope of this work, since we are only aiming to modify the model for the purposes of simulation and classification. Therefore, we adopt simple linear extrapolations of the dispersion, variance, and covariance tables instead. 

The variance and covariance for the $M_0$ and $M_1$ components are encoded in SALT2 as 2D arrays, over wavelength and phase. We treat each phase independently, and extrapolate to both UV and IR wavelengths by holding flat at a constant value, matching the value of the existing SALT2 model at 3500\,\AA\ and 8500\,\AA\ (Figure \ref{Afig:SALT2_variance}).  Our extrapolations apply to the variance and covariance components used when applying the SALT2 model to broadband light curves \citep{Guy:2010}.  The SALT2 model also allows for a separate set of 2D arrays that define the spectral variance and covariance, which can be used for simultaneously fitting an observed SN spectrum alongside the light-curve fitting \citep{Guy:2007}.  We have not modified the spectral variance and covariance components, because we do not intend for this extension of the SALT2 model to be used for spectral fitting.

The dispersion of the SALT2 model is characterized by a light-curve dispersion scaling array --- a 2D grid in phase and wavelength --- and also a color dispersion array.  The latter accounts for the uncertainty in the model due to limited color information, and also encodes the expected intrinsic scatter in SNe~Ia that is color dependent. 
On the NIR side we apply a linear extrapolation that has both of these dispersion components decreasing toward increasing wavelengths, to reflect the understanding that SNe~Ia are better standard candles at NIR wavelengths \citep[e.g.,][]{Dhawan:2017}.   On the UV side, we we use a flat-line extrapolation --- holding the value at 3500~\AA\ constant over all UV wavelengths down to 1700~\mbox{\normalfont\AA}.   These extrapolations are shown in Figure \ref{Afig:SALT2_dispersion}.

When the SALT2 model is used to generate a simulated population of SNe~Ia it is necessary to add additional scatter to the simulated light curves, reflecting the intrinsic scatter in SN~Ia luminosities of $\sigma_{\rm int} \approx 0.1$ mag \citep{Chotard:2011,Scolnic:2014a,Kessler:2017,Zheng:2018}.  Simulations using our extrapolated SALT2 model should reflect the understanding that the intrinsic scatter in SN~Ia luminosities is lower at NIR wavelengths than at visible bands.
As a simple way to implement this, we suggest adopting an intrinsic scatter model that is fixed to a specific value at optical wavelengths (e.g., $\sigma_{\rm int,opt}=0.1$), and decreases linearly to a fixed value at 25,000~\mbox{\normalfont\AA}\ 
(e.g., $\sigma_{\rm int,ir}=0.02$).\footnote{This capability has been implemented in releases of the {\tt SNANA} software for v10.61b or later, and is activated by setting multiple values for the {\tt SIGCOH} variable in the SALT2.INFO file.}

\

\

\section{Results and Discussion}
\label{sec:results}

\begin{figure}
\centering
\includegraphics[width=.5\textwidth]{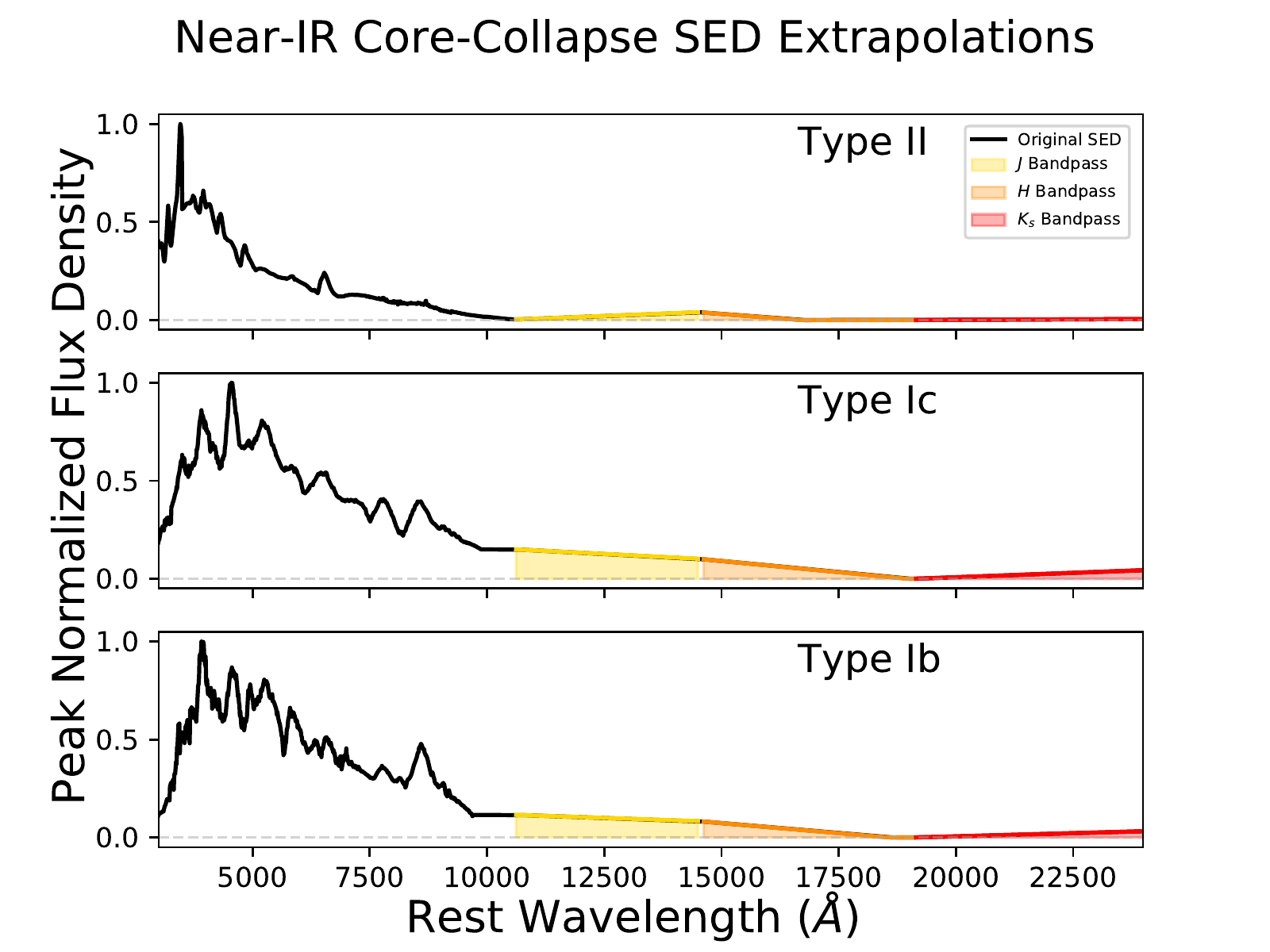}
\caption{\label{fig:allSEDs} A representative set of the results produced by the \textit{SNSEDextend} package, for CC~SNe at peak brightness. In each panel the original SED flux density (black) has been normalized, and the shaded regions represent the flux through each bandpass that has been set by the \textit{SNSEDextend}\ package to match the broad-band colors measured in Sections \ref{sub:curves} and \ref{sub:bb}. }
\end{figure}

\begin{figure}
\centering
\includegraphics[width=.5\textwidth]{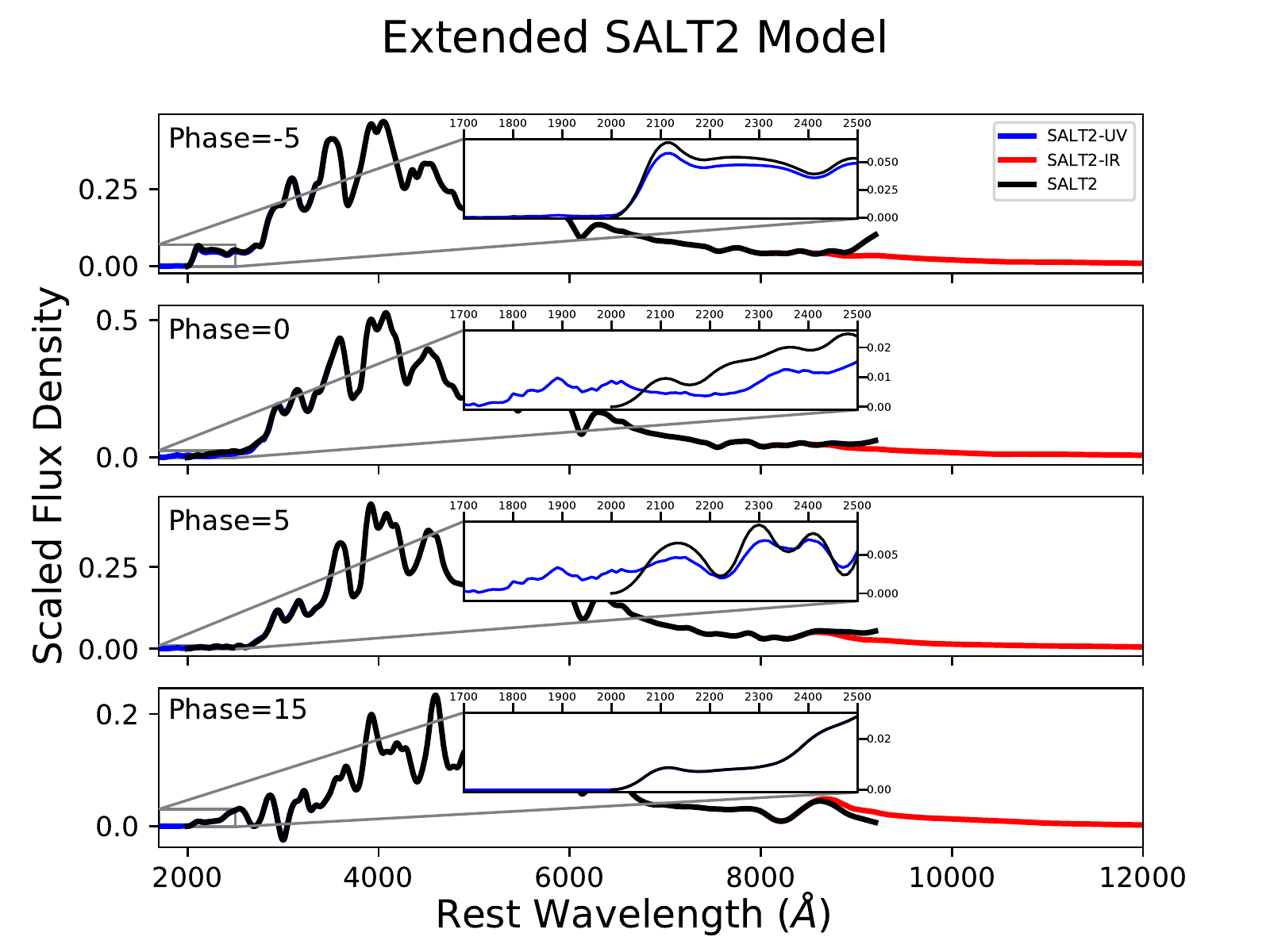}
\caption{\label{fig:salt2-extended} The fully extrapolated SALT2 model, which combines the SALT2-IR model (red) derived in Section \ref{sub:ia_nir} and the SALT2-UV model (blue) derived in Section \ref{sub:ia_uv}. The original SALT2 model (black) was previously defined over the wavelength range 2000--9200~\mbox{\normalfont\AA} \ (though had increased error below 3500~\mbox{\normalfont\AA} \ and above 8000~\mbox{\normalfont\AA} \ [\cite{Guy:2007}]), and now extends over the wavelength range 1700--25,000~\mbox{\normalfont\AA}.}
\end{figure}

The primary output of this work is the open-source \textit{SNSEDextend}\ software package,
 which is written in Python \citep{Pierel:2018a}\footnote{Code: \href{https://github.com/jpierel14/snsed}{github.com/jpierel14/snsed}; Documentation: \href{http://snsedextend.readthedocs.io}{snsedextend.readthedocs.io} }. The package makes extensive use of the SNCosmo Python package for fitting, calculations, its template library, and more. \textit{SNSEDextend} \ is fully documented, and gives users the capability to generate color curves as in Section \ref{sub:curves} with their own data,
and use their own SN templates to produce SED extrapolations into the UV
and IR.  Using the new open-source \textit{SNSEDextend} \ package, we have produced an initial
set of 47 SEDs for SNe~II, Ib, and Ic, plus an extrapolated version of the SALT2 model for SNe~Ia (Figures \ref{fig:allSEDs}, \ref{fig:salt2-extended}). The fully documented and
complete repository of these SEDs can be found at an online repository\footnote{\href{dx.doi.org/10.5281/zenodo.1250492}{DOI:10.5281/zenodo.1250492}}.

\

Time-domain science is a key driver in the design of the next generation of observatories, including LSST, JWST, and WFIRST.  These telescopes will provide the community with thousands of new SN observations, many of which will include rest-frame UV and NIR photometry. \citep[e.g.,][]{Mesinger:2006,Oguri:2010,najita:2016,Hounsell:2017}. The WFIRST mission, for example, could observe as many as 8,000 SNe at $z<0.8$ with filters covering a rest-frame wavelength range of $\sim7000$--20,000~\mbox{\normalfont\AA}\ \citep{Hounsell:2017}. 
The new repository of empirically derived SED extrapolations presented here is immediately useful for simulations of these future SN surveys, which are used to optimize survey strategies. 

The huge number of SN discoveries coming in the next decade will necessitate photometric classifications \citep[e.g.,][]{Kessler:2010,Sako:2011,Campbell:2013,Jones:2018}, as it will not be possible to perform spectroscopic follow-up observations for each object.  The extension of the SN SED template library into rest-frame UV and NIR wavelengths may be especially valuable for photometrically isolating SNe~Ia because of their distinguishing spectrophotometric features in those regions.  At UV wavelengths, SNe~Ia exhibit a distinct flux deficit relative to CC~SNe \citep{Riess:2004a,Milne:2013}. In NIR bands, SNe~Ia are considered to be excellent standard candles \citep{Dhawan:2017} and exhibit a distinctive secondary maximum in their broadband light curves \citep[e.g.,][]{Leibundgut:1988,Ford:1993}.  

\begin{figure}
\includegraphics[width=.5\textwidth]{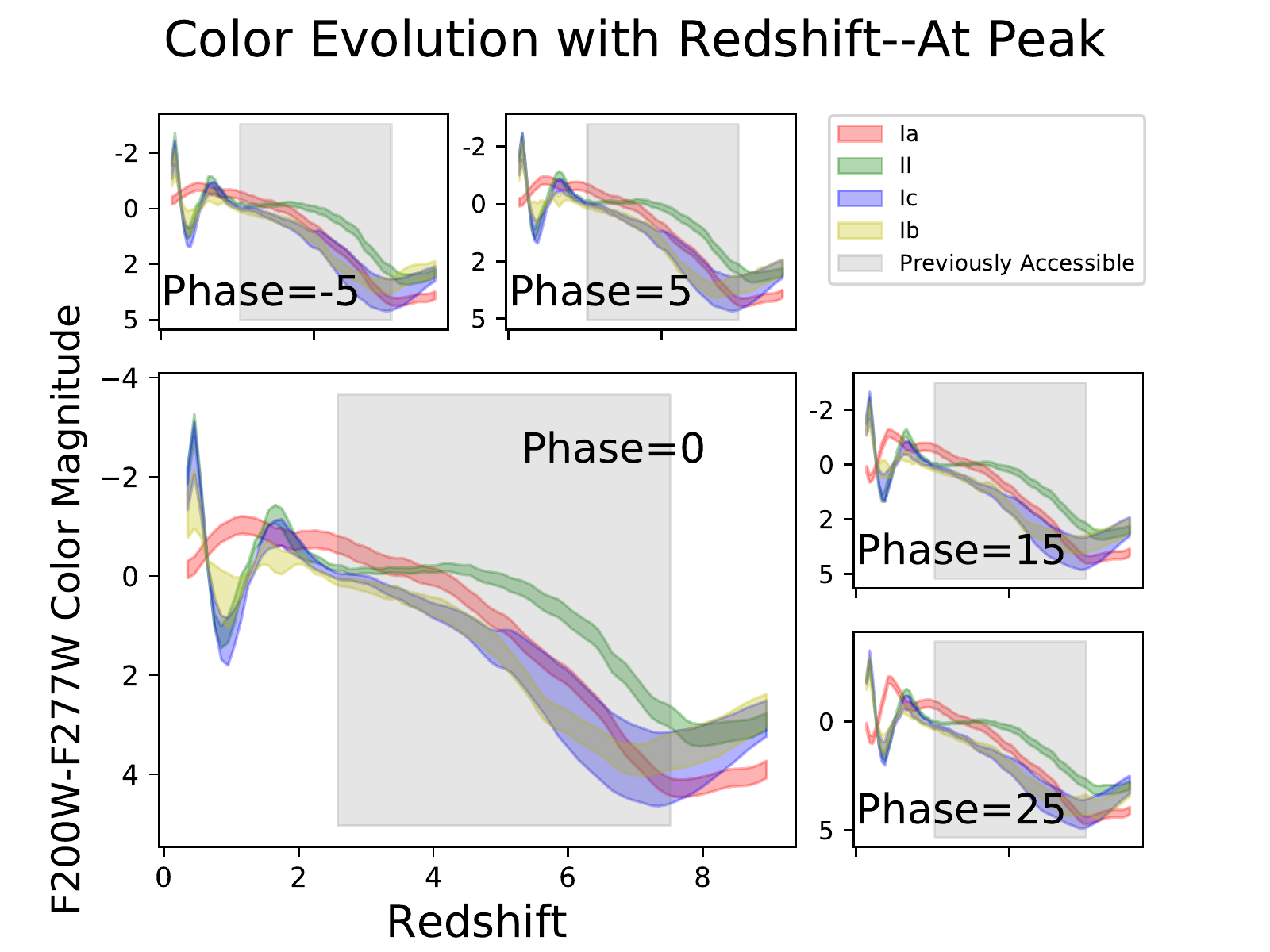}\caption{\label{fig:jwst} The evolution of CC~SN and SN~Ia colors defined by the JWST F200W and F277W filters (effective wavelengths of $\sim 20,000$~\mbox{\normalfont\AA}\ and $\sim27,700$~\mbox{\normalfont\AA}, respectively), at multiple epochs. Regions where there is separation between the SN~Ia color and all CC~SN colors correspond to redshifts at which distinguishing between SNe~Ia and CC~SNe at that redshift should be possible with the new extrapolated SEDs. The shaded region shows the redshift range over which the nonextrapolated SEDs would have provided color information, and the nonshaded region shows what has been made accessible by the extrapolated SEDs.}
\end{figure}

\begin{figure}
\includegraphics[width=.5\textwidth]{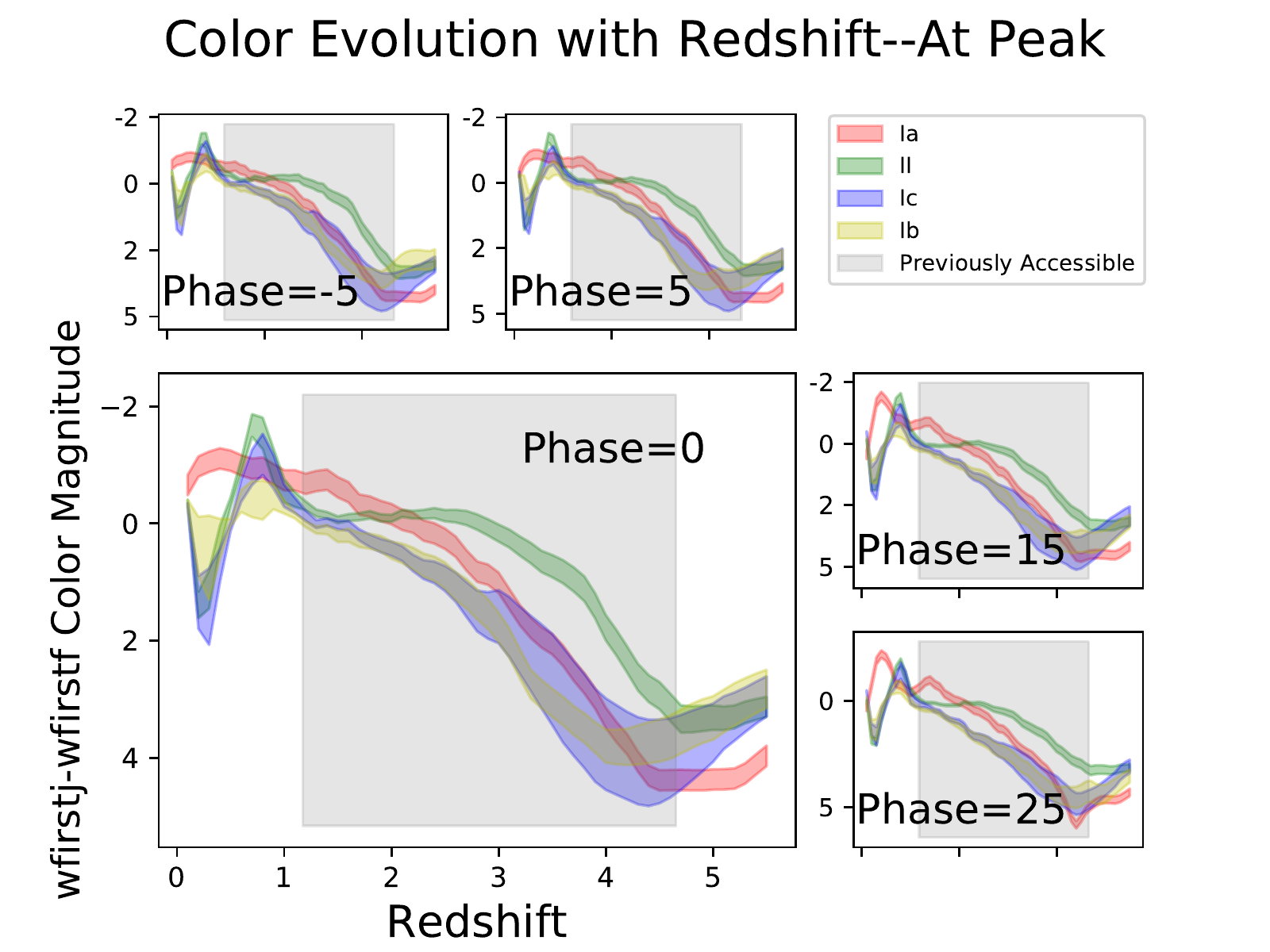}
\caption{\label{fig:wfirst} The evolution of CC~SN and SN~Ia colors defined by the WFIRST $J$ and $F$ filters (effective wavelengths of $\sim 12,900$~\mbox{\normalfont\AA}\ and $\sim18,500$~\mbox{\normalfont\AA}, respectively), at multiple epochs. Regions where there is separation between the SN~Ia color and all CC~SN colors correspond to redshifts at which distinguishing between SNe~Ia and CC~SNe at that redshift should be possible with the new extrapolated SEDs. Filter transmission functions were taken from \cite{Hounsell:2017}, which in turn references the WFIRST Cycle 6 instrument parameter release. The shaded region shows the redshift range over which the nonextrapolated SEDs would have provided color information, and the nonshaded region shows what has been made accessible by the extrapolated SEDs. }
\end{figure}

\begin{figure}
\includegraphics[width=.5\textwidth]{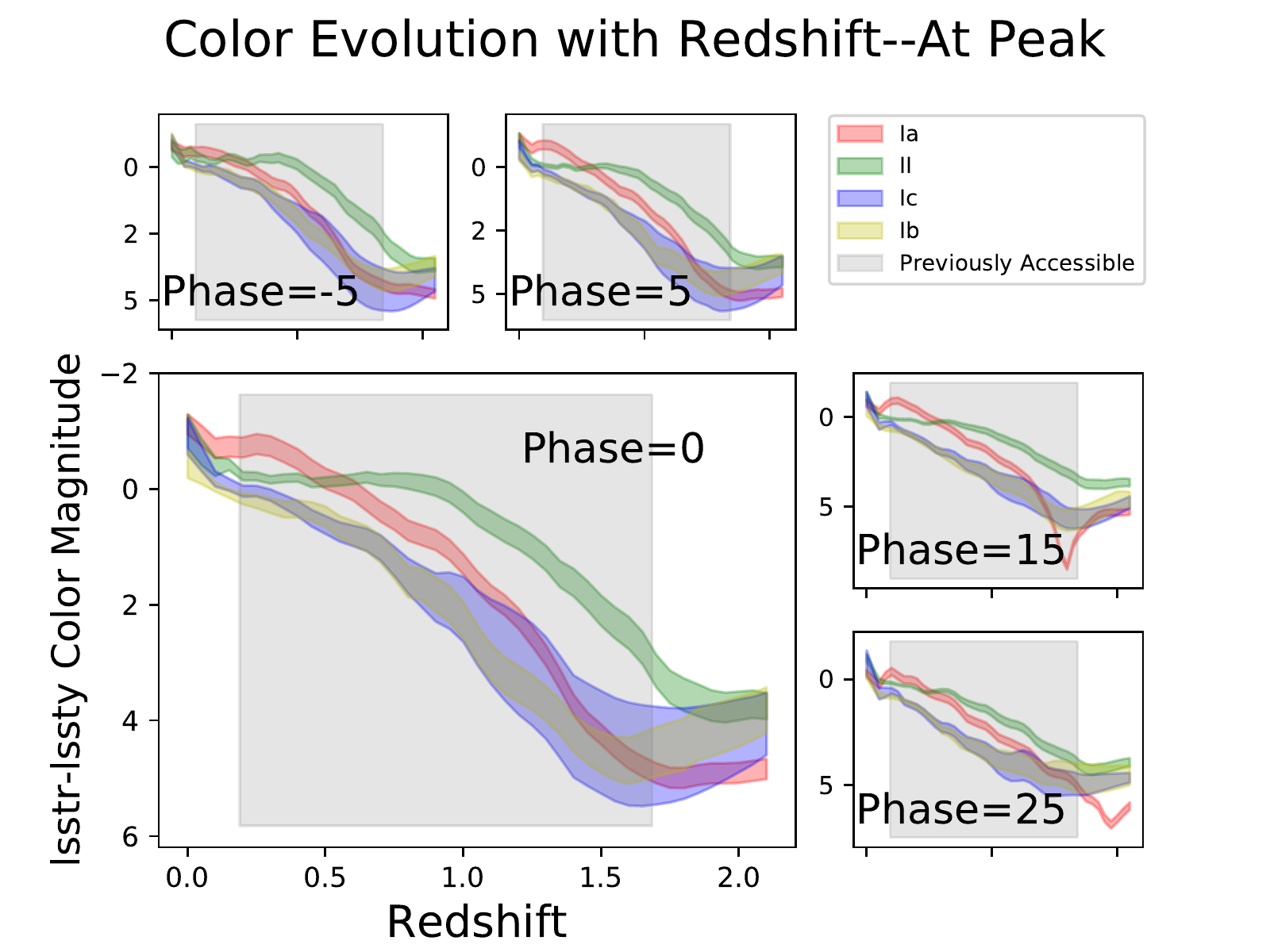}
\caption{\label{fig:lsst} The evolution of CC~SN and SN~Ia colors defined by the LSST $R$ and $Y$ filters (effective wavelengths of $\sim 6200~\mbox{\normalfont\AA}$ and $\sim9700~\mbox{\normalfont\AA}$, respectively), at multiple epochs. Regions where there is separation between the SN~Ia color and all CC~SN colors correspond to redshifts at which distinguishing between SNe~Ia and CC~SNe at that redshift should be possible with the new extrapolated SEDs. The shaded region shows the redshift range over which the nonextrapolated SEDs would have provided color information, and the nonshaded region shows what has been made accessible by the extrapolated SEDs.}
 \end{figure}

A full examination of the accuracy and efficiency of the improvement to photometric classifications is beyond the scope of this work. We have, however, made a preliminary exploration of how the extrapolated SED library will affect color-based classifications with JWST, WFIRST, and LSST (Figures \ref{fig:jwst}, \ref{fig:wfirst}, \ref{fig:lsst}). Using UV-NIR colors at peak brightness (an approach similar to the single-epoch classifications of \citet{Poznanski:2007}), the next generation of telescopes will now be able to distinguish a SN~Ia from a CC~SN with 95\% confidence over 47\% (JWST), 47\% (WFIRST), and 44\% (LSST) of the redshift ranges shown in Figures \ref{fig:jwst}, \ref{fig:wfirst}, and \ref{fig:lsst}, respectively. These rates are a significant improvement upon what was previously possible over the same redshift ranges: 21\% (JWST), 25\% (WFIRST), and 37\% (LSST).

To continue improving the extrapolated SED templates, the most valuable additions will be from well-sampled NIR light curves of CC~SNe and UV light curves of SNe~Ia.  Aided by the publicly available \textit{SNSEDextend}\ software package, new photometric time-series data can be easily adopted to update the SED templates.  These improvements can propagate into more informative simulations for the optimization of future SN surveys, and more accurate photometric classification tools for the analysis of the thousands of SNe those surveys will deliver.


\bigskip

\

\acknowledgments

J.E.D.P. was supported for this work by a NASA grant through the South Carolina NASA EPSCoR Research Grant Program.
This work is supported in part by NASA grant NNG17PX03C. 
The UCSC team is supported in part by NASA grant NNG17PX03C, NSF
grant AST-1518052, the Gordon \& Betty Moore Foundation, the Heising-Simons Foundation, and by fellowships from the Alfred P.\
Sloan Foundation and the David and Lucile Packard Foundation to R.J.F. This supernova research at Rutgers University is 
supported by NASA grant NNG17PX03C and US Department of Energy award DE-SC0011636. Support for A.V.F.'s work has been provided by the TABASGO Foundation, the Christopher R. Redlich Fund, and the Miller Institute for Basic Research in Science (U.C. Berkeley). A.S.F. acknowledges support from NSF grant PHYS-1541160. Research on supernova cosmology at Harvard is supported by NSF AST-1516854, NASA gran NNX15AJSSG and the Gordon and Betty Moore Foundation. A.A. acknowledges Fundacion Mexico en Harvard, CONACyT, and Hubble Space Telescope Awards HST GO-14216 and HST GO-13046 supporting the HST RAISIN program.

\bibliographystyle{Overflow/aas}

\begin{thebibliography}{}
\expandafter\ifx\csname natexlab\endcsname\relax\def\natexlab#1{#1}\fi
\providecommand{\url}[1]{\href{#1}{#1}}
\providecommand{\dodoi}[1]{doi:~\href{http://doi.org/#1}{\nolinkurl{#1}}}
\providecommand{\doeprint}[1]{\href{http://ascl.net/#1}{\nolinkurl{http://ascl.net/#1}}}
\providecommand{\doarXiv}[1]{\href{https://arxiv.org/abs/#1}{\nolinkurl{https://arxiv.org/abs/#1}}}

\bibitem[{Barbary(2014)}]{Barbary:2014}
Barbary, K. 2014, \dodoi{{10.5281/zenodo.11938}}

\bibitem[{{Baron} {et~al.}(2004){Baron}, {Nugent}, {Branch}, \&
  {Hauschildt}}]{Baron:2004}
{Baron}, E., {Nugent}, P.~E., {Branch}, D., \& {Hauschildt}, P.~H. 2004, \apjl,
  616, L91, \dodoi{10.1086/426506}

\bibitem[{Bianco {et~al.}(2014)Bianco, Modjaz, Hicken, Friedman, Kirshner,
  Bloom, Challis, Marion, Wood-Vasey, \& Rest}]{Bianco:2014}
Bianco, F.~B., Modjaz, M., Hicken, M., {et~al.} 2014, Astrophys. Journal,
  Suppl. Ser., 213, \dodoi{10.1088/0067-0049/213/2/19}

\bibitem[{Blondin \& Tonry(2007)}]{Blondin:2007}
Blondin, S., \& Tonry, J. 2007, $\backslash$apj, 666, 1024,
  \dodoi{10.1086/520494}

\bibitem[{{Burns} {et~al.}(2011){Burns}, {Stritzinger}, {Phillips}, {Kattner},
  {Persson}, {Madore}, {Freedman}, {Boldt}, {Campillay}, {Contreras},
  {Folatelli}, {Gonzalez}, {Krzeminski}, {Morrell}, {Salgado}, \&
  {Suntzeff}}]{Burns:2011}
{Burns}, C.~R., {Stritzinger}, M., {Phillips}, M.~M., {et~al.} 2011, The
  Astronomical Journal, 141, 19, \dodoi{10.1088/0004-6256/141/1/19}

\bibitem[{{Campbell} {et~al.}(2013){Campbell}, {D'Andrea}, {Nichol}, {Sako},
  {Smith}, {Lampeitl}, {Olmstead}, {Bassett}, {Biswas}, {Brown}, {Cinabro},
  {Dawson}, {Dilday}, {Foley}, {Frieman}, {Garnavich}, {Hlozek}, {Jha},
  {Kuhlmann}, {Kunz}, {Marriner}, {Miquel}, {Richmond}, {Riess}, {Schneider},
  {Sollerman}, {Taylor}, \& {Zhao}}]{Campbell:2013}
{Campbell}, H., {D'Andrea}, C.~B., {Nichol}, R.~C., {et~al.} 2013, \apj, 763,
  88, \dodoi{10.1088/0004-637X/763/2/88}

\bibitem[{Cardelli {et~al.}(1989)Cardelli, Clayton, \& Mathis}]{Cardelli:1989}
Cardelli, J.~A., Clayton, G.~C., \& Mathis, J.~S. 1989, \apj, 345, 245

\bibitem[{{Chotard} {et~al.}(2011){Chotard}, {Gangler}, {Aldering},
  {Antilogus}, {Aragon}, {Bailey}, {Baltay}, {Bongard}, {Buton}, {Canto},
  {Childress}, {Copin}, {Fakhouri}, {Hsiao}, {Kerschhaggl}, {Kowalski},
  {Loken}, {Nugent}, {Paech}, {Pain}, {Pecontal}, {Pereira}, {Perlmutter},
  {Rabinowitz}, {Runge}, {Scalzo}, {Smadja}, {Tao}, {Thomas}, {Weaver}, {Wu},
  \& {Nearby Supernova Factory}}]{Chotard:2011}
{Chotard}, N., {Gangler}, E., {Aldering}, G., {et~al.} 2011, \aap, 529, L4,
  \dodoi{10.1051/0004-6361/201116723}

\bibitem[{{Contreras} {et~al.}(2010){Contreras}, {Hamuy}, {Phillips},
  {Folatelli}, {Suntzeff}, {Persson}, {Stritzinger}, {Boldt}, {Gonz{\'a}lez},
  {Krzeminski}, {Morrell}, {Roth}, {Salgado}, {Jos{\'e} Maureira}, {Burns},
  {Freedman}, {Madore}, {Murphy}, {Wyatt}, {Li}, \&
  {Filippenko}}]{Contreras:2010}
{Contreras}, C., {Hamuy}, M., {Phillips}, M.~M., {et~al.} 2010, \aj, 139, 519,
  \dodoi{10.1088/0004-6256/139/2/519}

\bibitem[{{Dahl{\'e}n} \& {Fransson}(1999)}]{Dahlen:1999}
{Dahl{\'e}n}, T., \& {Fransson}, C. 1999, \aap, 350, 349

\bibitem[{de~Jaeger {et~al.}(2018)de~Jaeger, Anderson, Galbany,
  Gonz{\'{a}}lez-Gait{\'{a}}n, Hamuy, Phillips, Stritzinger, Contreras,
  Folatelli, Guti{\'{e}}rrez, Hsiao, Morrell, Suntzeff, Dessart, \&
  Filippenko}]{jaeger:2018}
de~Jaeger, T., Anderson, J.~P., Galbany, L., {et~al.} 2018, Mon. Not. R.
  Astron. Soc., 476, 4592, \dodoi{10.1093/mnras/sty508}

\bibitem[{Dessart {et~al.}(2012)Dessart, Hillier, Li, \&
  Woosley}]{Dessart:2012}
Dessart, L., Hillier, D.~J., Li, C., \& Woosley, S. 2012, Mon. Not. R. Astron.
  Soc., 424, 2139, \dodoi{10.1111/j.1365-2966.2012.21374.x}

\bibitem[{Dessart {et~al.}(2013)Dessart, Hillier, Waldman, \&
  Livne}]{Dessart:2013}
Dessart, L., Hillier, D.~J., Waldman, R., \& Livne, E. 2013, Mon. Not. R.
  Astron. Soc., 433, 1745, \dodoi{10.1093/mnras/stt861}

\bibitem[{Dhawan {et~al.}(2017)Dhawan, Jha, \& Leibundgut}]{Dhawan:2017}
Dhawan, S., Jha, S.~W., \& Leibundgut, B. 2017,
  \dodoi{10.1051/0004-6361/201731501}

\bibitem[{{Dhawan} {et~al.}(2018){Dhawan}, {Jha}, \&
  {Leibundgut}}]{Dhawan:2018}
{Dhawan}, S., {Jha}, S.~W., \& {Leibundgut}, B. 2018, \aap, 609, A72,
  \dodoi{10.1051/0004-6361/201731501}

\bibitem[{{Filippenko}(1997)}]{Filippenko:1997}
{Filippenko}, A.~V. 1997, \araa, 35, 309,
  \dodoi{10.1146/annurev.astro.35.1.309}

\bibitem[{Foley {et~al.}(2016)Foley, Pan, Brown, Filippenko, Fox, Hillebrandt,
  Kirshner, Marion, Milne, Parrent, Pignata, \& Stritzinger}]{Foley:2016}
Foley, R.~J., Pan, Y.~C., Brown, P., {et~al.} 2016, Mon. Not. R. Astron. Soc.,
  461, 1308, \dodoi{10.1093/mnras/stw1440}

\bibitem[{{Ford} {et~al.}(1993){Ford}, {Herbst}, {Richmond}, {Baker},
  {Filippenko}, {Treffers}, {Paik}, \& {Benson}}]{Ford:1993}
{Ford}, C.~H., {Herbst}, W., {Richmond}, M.~W., {et~al.} 1993, \aj, 106, 1101,
  \dodoi{10.1086/116708}

\bibitem[{{Friedman} {et~al.}(2015){Friedman}, {Wood-Vasey}, {Marion},
  {Challis}, {Mandel}, {Bloom}, {Modjaz}, {Narayan}, {Hicken}, {Foley},
  {Klein}, {Starr}, {Morgan}, {Rest}, {Blake}, {Miller}, {Falco}, {Wyatt},
  {Mink}, {Skrutskie}, \& {Kirshner}}]{Friedman:2015}
{Friedman}, A.~S., {Wood-Vasey}, W.~M., {Marion}, G.~H., {et~al.} 2015, \apjs,
  220, 9, \dodoi{10.1088/0067-0049/220/1/9}

\bibitem[{{Graur} {et~al.}(2014){Graur}, {Rodney}, {Maoz}, {Riess}, {Jha},
  {Postman}, {Dahlen}, {Holoien}, {McCully}, {Patel}, {Strolger},
  {Ben{\'{\i}}tez}, {Coe}, {Jouvel}, {Medezinski}, {Molino}, {Nonino},
  {Bradley}, {Koekemoer}, {Balestra}, {Cenko}, {Clubb}, {Dickinson},
  {Filippenko}, {Frederiksen}, {Garnavich}, {Hjorth}, {Jones}, {Leibundgut},
  {Matheson}, {Mobasher}, {Rosati}, {Silverman}, {U}, {Jedruszczuk}, {Li},
  {Lin}, {Mirmelstein}, {Neustadt}, {Ovadia}, \& {Rogers}}]{Graur:2014}
{Graur}, O., {Rodney}, S.~A., {Maoz}, D., {et~al.} 2014, \apj, 783, 28,
  \dodoi{10.1088/0004-637X/783/1/28}

\bibitem[{Guy {et~al.}(2007)Guy, Astier, Baumont, Hardin, Pain, Regnault, Basa,
  Carlberg, Conley, Fabbro, Fouchez, Hook, Howell, Perrett, Pritchet, Rich,
  Sullivan, Antilogus, Aubourg, Bazin, Bronder, Filiol, Palanque-Delabrouille,
  Ripoche, \& Ruhlmann-Kleider}]{Guy:2007}
Guy, J., Astier, P., Baumont, S., {et~al.} 2007, \aap, 466, 11

\bibitem[{{Guy} {et~al.}(2010){Guy}, {Sullivan}, {Conley}, {Regnault},
  {Astier}, {Balland}, {Basa}, {Carlberg}, {Fouchez}, {Hardin}, {Hook},
  {Howell}, {Pain}, {Palanque-Delabrouille}, {Perrett}, {Pritchet}, {Rich},
  {Ruhlmann-Kleider}, {Balam}, {Baumont}, {Ellis}, {Fabbro}, {Fakhouri},
  {Fourmanoit}, {Gonz{\'a}lez-Gait{\'a}n}, {Graham}, {Hsiao}, {Kronborg},
  {Lidman}, {Mourao}, {Perlmutter}, {Ripoche}, {Suzuki}, \&
  {Walker}}]{Guy:2010}
{Guy}, J., {Sullivan}, M., {Conley}, A., {et~al.} 2010, \aap, 523, A7,
  \dodoi{10.1051/0004-6361/201014468}

\bibitem[{Hershkowitz {et~al.}(1986)Hershkowitz, Linder, \&
  Wagoner}]{Hershkowitz:1986}
Hershkowitz, S., Linder, E., \& Wagoner, R.~V. 1986, Astrophys. J., 220, 220

\bibitem[{{Hicken} {et~al.}(2009){Hicken}, {Challis}, {Jha}, {Kirshner},
  {Matheson}, {Modjaz}, {Rest}, {Michael Wood-Vasey}, {Bakos}, {Barton},
  {Berlind}, {Bragg}, {Brice{\~n}o}, {Brown}, {Caldwell}, {Calkins}, {Cho},
  {Ciupik}, {Contreras}, {Dendy}, {Dosaj}, {Durham}, {Eriksen}, {Esquerdo},
  {Everett}, {Falco}, {Fernandez}, {Gaba}, {Garnavich}, {Graves}, {Green},
  {Groner}, {Hergenrother}, {Holman}, {Hradecky}, {Huchra}, {Hutchison},
  {Jerius}, {Jordan}, {Kilgard}, {Krauss}, {Luhman}, {Macri}, {Marrone},
  {McDowell}, {McIntosh}, {McNamara}, {Megeath}, {Mochejska}, {Munoz},
  {Muzerolle}, {Naranjo}, {Narayan}, {Pahre}, {Peters}, {Peterson}, {Rines},
  {Ripman}, {Roussanova}, {Schild}, {Sicilia-Aguilar}, {Sokoloski}, {Smalley},
  {Smith}, {Spahr}, {Stanek}, {Barmby}, {Blondin}, {Stubbs}, {Szentgyorgyi},
  {Torres}, {Vaz}, {Vikhlinin}, {Wang}, {Westover}, {Woods}, \&
  {Zhao}}]{Hicken:2009b}
{Hicken}, M., {Challis}, P., {Jha}, S., {et~al.} 2009, \apj, 700, 331,
  \dodoi{10.1088/0004-637X/700/1/331}

\bibitem[{{Hicken} {et~al.}(2012){Hicken}, {Challis}, {Kirshner}, {Rest},
  {Cramer}, {Wood-Vasey}, {Bakos}, {Berlind}, {Brown}, {Caldwell}, {Calkins},
  {Currie}, {de Kleer}, {Esquerdo}, {Everett}, {Falco}, {Fernandez},
  {Friedman}, {Groner}, {Hartman}, {Holman}, {Hutchins}, {Keys}, {Kipping},
  {Latham}, {Marion}, {Narayan}, {Pahre}, {Pal}, {Peters}, {Perumpilly},
  {Ripman}, {Sipocz}, {Szentgyorgyi}, {Tang}, {Torres}, {Vaz}, {Wolk}, \&
  {Zezas}}]{Hicken:2012}
{Hicken}, M., {Challis}, P., {Kirshner}, R.~P., {et~al.} 2012, \apjs, 200, 12,
  \dodoi{10.1088/0067-0049/200/2/12}

\bibitem[{Hicken {et~al.}(2017)Hicken, Friedman, Challis, Berlind, Blondin,
  Calkins, Esquerdo, Matheson, Modjaz, Rest, \& Kirshner}]{Hicken:2017}
Hicken, M., Friedman, A.~S., Challis, P., {et~al.} 2017, Astrophys. J. Suppl.
  Ser., 233, 6, \dodoi{10.3847/1538-4365/aa8ef4}

\bibitem[{Hounsell {et~al.}(2017)Hounsell, Scolnic, Foley, Kessler, Miranda,
  Avelino, Bohlin, Filippenko, Frieman, Jha, Kelly, Kirshner, Mandel, Rest,
  Riess, Rodney, \& Strolger}]{Hounsell:2017}
Hounsell, R., Scolnic, D., Foley, R.~J., {et~al.} 2017, 29, 1.
\newblock \doarXiv{1702.01747}

\bibitem[{Hsiao {et~al.}(2007)Hsiao, Conley, Howell, Sullivan, Pritchet,
  Carlberg, Nugent, \& Phillips}]{Hsiao:2007}
Hsiao, E.~Y., Conley, A., Howell, D.~A., {et~al.} 2007, \apj, 663, 1187

\bibitem[{{Humason} {et~al.}(1956){Humason}, {Mayall}, \&
  {Sandage}}]{Humason:1956}
{Humason}, M.~L., {Mayall}, N.~U., \& {Sandage}, A.~R. 1956, \aj, 61, 97,
  \dodoi{10.1086/107297}

\bibitem[{Ivezic {et~al.}(2008)Ivezic, Tyson, Abel, Acosta, Allsman, Alsayyad,
  Anderson, Andrew, Axelrod, Angeli, Ansari, Antilogus, Arndt, Astier, Barr,
  Barrau, Bartlett, Bauman, Baumont, Becker, Becla, Bellavia, Blanc, Blandford,
  Bloom, Bogart, Borne, Bosch, Boutigny, Brandt, Brown, Bullock, Burchat,
  Burke, Cagnoli, Chandrasekharan, Chesley, Cheu, Chiang, Claver, Connolly,
  Cooray, Covey, Cribbs, Cui, Cutri, Daubard, Daues, Delgado, Doherty, Dubois,
  Durech, Eracleous, Ferguson, Freemon, Gangler, Gawiser, Geary, Gee, Geha,
  Gibson, Glanzman, Goodenow, Gressler, Gris, Guyonnet, Hascall, Haupt,
  Hernandez, Hogan, Huang, Huffer, Innes, Jacoby, Jain, Jee, Johns, Jones,
  Kahn, Kallivayalil, Kalmbach, Kantor, Kasliwal, Kessler, Kirkby, Kuczewski,
  Kulkarni, Kotov, Krabbendam, Krughoff, Guillou, Levine, Liang, Lintott,
  Lupton, Mahabal, Marshall, Marshall, May, Mckercher, Migliore, Miller, Mills,
  Moniez, Neill, Nomerotski, Nordby, Oliver, Olivier, Olsen, Ortiz, Owen, Pain,
  Peterson, Petry, Pierfederici, Pike, Pinto, Plante, Plate, Price, Prouza,
  Radeka, Rasmussen, Regnault, Ridgway, Ritz, Rosing, Roucelle, Russo, Saha,
  Sassolas, Schalk, Schindler, Schneider, Sebag, Sembroski, Seppala, Shipsey,
  Silvestri, Smith, Strauss, Stubbs, Sweeney, Szalay, Takacs, Thaler, Berk,
  Vetter, Virieux, Xin, Walkowicz, Walter, Wang, Warner, Willman, Wittman,
  Wolff, Yoachim, \& Zhan}]{Ivezic:2008}
Ivezic, Z., Tyson, J.~A., Abel, B., {et~al.} 2008, ArXiv e-prints, 1.
\newblock \doarXiv{arXiv:0805.2366}

\bibitem[{{Jones} {et~al.}(2018){Jones}, {Scolnic}, {Riess}, {Rest},
  {Kirshner}, {Berger}, {Kessler}, {Pan}, {Foley}, {Chornock}, {Ortega},
  {Challis}, {Burgett}, {Chambers}, {Draper}, {Flewelling}, {Huber}, {Kaiser},
  {Kudritzki}, {Metcalfe}, {Tonry}, {Wainscoat}, {Waters}, {Gall}, {Kotak},
  {McCrum}, {Smartt}, \& {Smith}}]{Jones:2018}
{Jones}, D.~O., {Scolnic}, D.~M., {Riess}, A.~G., {et~al.} 2018, \apj, 857, 51,
  \dodoi{10.3847/1538-4357/aab6b1}

\bibitem[{{Kattner} {et~al.}(2012){Kattner}, {Leonard}, {Burns}, {Phillips},
  {Folatelli}, {Morrell}, {Stritzinger}, {Hamuy}, {Freedman}, {Persson},
  {Roth}, \& {Suntzeff}}]{Kattner:2012}
{Kattner}, S., {Leonard}, D.~C., {Burns}, C.~R., {et~al.} 2012, Publications of
  the Astronomical Society of the Pacific, 124, 114, \dodoi{10.1086/664734}

\bibitem[{Kelly \& Kirshner(2012)}]{Kelly:2012}
Kelly, P.~L., \& Kirshner, R.~P. 2012, Astrophys. J., 759, 107,
  \dodoi{10.1088/0004-637X/759/2/107}

\bibitem[{{Kessler} \& {Scolnic}(2017)}]{Kessler:2017}
{Kessler}, R., \& {Scolnic}, D. 2017, \apj, 836, 56,
  \dodoi{10.3847/1538-4357/836/1/56}

\bibitem[{{Kessler} {et~al.}(2009){Kessler}, {Bernstein}, {Cinabro}, {Dilday},
  {Frieman}, {Jha}, {Kuhlmann}, {Miknaitis}, {Sako}, {Taylor}, \&
  {Vanderplas}}]{Kessler:2009a}
{Kessler}, R., {Bernstein}, J.~P., {Cinabro}, D., {et~al.} 2009, \pasp, 121,
  1028, \dodoi{10.1086/605984}

\bibitem[{{Kessler} {et~al.}(2010){Kessler}, {Bassett}, {Belov}, {Bhatnagar},
  {Campbell}, {Conley}, {Frieman}, {Glazov}, {Gonz{\'a}lez-Gait{\'a}n},
  {Hlozek}, {Jha}, {Kuhlmann}, {Kunz}, {Lampeitl}, {Mahabal}, {Newling},
  {Nichol}, {Parkinson}, {Philip}, {Poznanski}, {Richards}, {Rodney}, {Sako},
  {Schneider}, {Smith}, {Stritzinger}, \& {Varughese}}]{Kessler:2010}
{Kessler}, R., {Bassett}, B., {Belov}, P., {et~al.} 2010, \pasp, 122, 1415,
  \dodoi{10.1086/657607}

\bibitem[{{Krisciunas} {et~al.}(2003){Krisciunas}, {Suntzeff}, {Candia},
  {Arenas}, {Espinoza}, {Gonzalez}, {Gonzalez}, {H{\"o}flich}, {Landolt},
  {Phillips}, \& {Pizarro}}]{Krisciunas:2003}
{Krisciunas}, K., {Suntzeff}, N.~B., {Candia}, P., {et~al.} 2003, \aj, 125,
  166, \dodoi{10.1086/345571}

\bibitem[{{Krisciunas} {et~al.}(2004){Krisciunas}, {Suntzeff}, {Phillips},
  {Candia}, {Prieto}, {Antezana}, {Chassagne}, {Chen}, {Dickinson},
  {Eisenhardt}, {Espinoza}, {Garnavich}, {Gonz{\'a}lez}, {Harrison}, {Hamuy},
  {Ivanov}, {Krzemi{\'n}ski}, {Kulesa}, {McCarthy}, {Moro-Mart{\'{\i}}n},
  {Muena}, {Noriega-Crespo}, {Persson}, {Pinto}, {Roth}, {Rubenstein},
  {Stanford}, {Stringfellow}, {Zapata}, {Porter}, \&
  {Wischnjewsky}}]{Krisciunas:2004}
{Krisciunas}, K., {Suntzeff}, N.~B., {Phillips}, M.~M., {et~al.} 2004, \aj,
  128, 3034, \dodoi{10.1086/425629}

\bibitem[{{Krisciunas} {et~al.}(2007){Krisciunas}, {Garnavich}, {Stanishev},
  {Suntzeff}, {Prieto}, {Espinoza}, {Gonzalez}, {Salvo}, {Elias de la Rosa},
  {Smartt}, {Maund}, \& {Kudritzki}}]{Krisciunas:2007}
{Krisciunas}, K., {Garnavich}, P.~M., {Stanishev}, V., {et~al.} 2007, \aj, 133,
  58, \dodoi{10.1086/509126}

\bibitem[{Krisciunas {et~al.}(2009)Krisciunas, Hamuy, Suntzeff, Espinoza,
  Gonzalez, Gonzalez, Gonzalez, Koviak, Krzeminski, Morrell, Phillips, Roth, \&
  Thomas-Osip}]{Krisciunas:2009}
Krisciunas, K., Hamuy, M., Suntzeff, N.~B., {et~al.} 2009, Astron. J., 137, 34,
  \dodoi{10.1088/0004-6256/137/1/34}

\bibitem[{{Krisciunas} {et~al.}(2017){Krisciunas}, {Contreras}, {Burns},
  {Phillips}, {Stritzinger}, {Morrell}, {Hamuy}, {Anais}, {Boldt}, {Busta},
  {Campillay}, {Castellon}, {Folatelli}, {Freedman}, {Gonzalez}, {Hsiao},
  {Krzeminski}, {Persson}, {Roth}, {Salgado}, {Seron}, {Suntzeff}, {Torres},
  {Filippenko}, {Li}, {Madore}, {DePoy}, {Marshall}, {Rheault}, \&
  {Villanueva}}]{Krisciunas:2017}
{Krisciunas}, K., {Contreras}, C., {Burns}, C.~R., {et~al.} 2017, ArXiv
  e-prints.
\newblock \doarXiv{1709.05146}

\bibitem[{{Leibundgut}(1988)}]{Leibundgut:1988}
{Leibundgut}, B. 1988, PhD thesis, -

\bibitem[{{Leloudas} {et~al.}(2009){Leloudas}, {Stritzinger}, {Sollerman},
  {Burns}, {Kozma}, {Krisciunas}, {Maund}, {Milne}, {Filippenko}, {Fransson},
  {Ganeshalingam}, {Hamuy}, {Li}, {Phillips}, {Schmidt}, {Skottfelt},
  {Taubenberger}, {Boldt}, {Fynbo}, {Gonzalez}, {Salvo}, \&
  {Thomas-Osip}}]{Leloudas:2009}
{Leloudas}, G., {Stritzinger}, M.~D., {Sollerman}, J., {et~al.} 2009, \aap,
  505, 265, \dodoi{10.1051/0004-6361/200912364}

\bibitem[{{Mandel} {et~al.}(2011){Mandel}, {Narayan}, \&
  {Kirshner}}]{Mandel:2011}
{Mandel}, K.~S., {Narayan}, G., \& {Kirshner}, R.~P. 2011, \apj, 731, 120,
  \dodoi{10.1088/0004-637X/731/2/120}

\bibitem[{{Mandel} {et~al.}(2009){Mandel}, {Wood-Vasey}, {Friedman}, \&
  {Kirshner}}]{Mandel:2009}
{Mandel}, K.~S., {Wood-Vasey}, W.~M., {Friedman}, A.~S., \& {Kirshner}, R.~P.
  2009, \apj, 704, 629, \dodoi{10.1088/0004-637X/704/1/629}

\bibitem[{Marion {et~al.}(2014)Marion, Vinko, Kirshner, Foley, Berlind,
  Bieryla, Bloom, Calkins, Challis, Chevalier, Chornock, Culliton, Curtis,
  Esquerdo, Everett, Falco, France, Fransson, Friedman, Garnavich, Leibundgut,
  Meyer, Smith, Soderberg, Sollerman, Starr, Szklenar, Takats, \&
  Wheeler}]{Marion:2014}
Marion, G.~H., Vinko, J., Kirshner, R.~P., {et~al.} 2014, Astrophys. J., 781,
  \dodoi{10.1088/0004-637X/781/2/69}

\bibitem[{{Marion} {et~al.}(2016){Marion}, {Brown}, {Vink{\'o}}, {Silverman},
  {Sand}, {Challis}, {Kirshner}, {Wheeler}, {Berlind}, {Brown}, {Calkins},
  {Camacho}, {Dhungana}, {Foley}, {Friedman}, {Graham}, {Howell}, {Hsiao},
  {Irwin}, {Jha}, {Kehoe}, {Macri}, {Maeda}, {Mandel}, {McCully}, {Pandya},
  {Rines}, {Wilhelmy}, \& {Zheng}}]{Marion:2016}
{Marion}, G.~H., {Brown}, P.~J., {Vink{\'o}}, J., {et~al.} 2016, \apj, 820, 92,
  \dodoi{10.3847/0004-637X/820/2/92}

\bibitem[{Mesinger \& Johnson(2006)}]{Mesinger:2006}
Mesinger, A., \& Johnson, B.~D. 2006, 80

\bibitem[{{Milne} {et~al.}(2013){Milne}, {Brown}, {Roming}, {Bufano}, \&
  {Gehrels}}]{Milne:2013}
{Milne}, P.~A., {Brown}, P.~J., {Roming}, P. W.~A., {Bufano}, F., \& {Gehrels},
  N. 2013, \apj, 779, 23, \dodoi{10.1088/0004-637X/779/1/23}

\bibitem[{{Mosher} {et~al.}(2014){Mosher}, {Guy}, {Kessler}, {Astier},
  {Marriner}, {Betoule}, {Sako}, {El-Hage}, {Biswas}, {Pain}, {Kuhlmann},
  {Regnault}, {Frieman}, \& {Schneider}}]{Mosher:2014}
{Mosher}, J., {Guy}, J., {Kessler}, R., {et~al.} 2014, \apj, 793, 16,
  \dodoi{10.1088/0004-637X/793/1/16}

\bibitem[{Najita {et~al.}(2016)Najita, Willman, Finkbeiner, Foley, Hawley,
  Newman, Rudnick, Simon, Trilling, Street, Bolton, Angus, Bell, Buzasi,
  Ciardi, Davenport, Dawson, Dickinson, Drlica-Wagner, Elias, Erb, Feaga, Fong,
  Gawiser, Giampapa, Guhathakurta, Hoffman, Hsieh, Jennings, Johnston, Kashyap,
  Li, Linder, Mandelbaum, Marshall, Matheson, Meibom, Miller, O'Meara, Reddy,
  Ridgway, Rockosi, Sand, Schafer, Schmidt, Sesar, Sheppard, Thomas, Tollerud,
  Trump, \& von~der Linden}]{najita:2016}
Najita, J., Willman, B., Finkbeiner, D.~P., {et~al.} 2016, 174.
\newblock \doarXiv{1610.01661}

\bibitem[{O'Donnell(1994)}]{odonnell:1994}
O'Donnell, J.~E. 1994, Astrophys. J., 422, 158, \dodoi{10.1086/173713}

\bibitem[{Oguri \& Marshall(2010)}]{Oguri:2010}
Oguri, M., \& Marshall, P.~J. 2010, 2593, 2579,
  \dodoi{10.1111/j.1365-2966.2010.16639.x}

\bibitem[{{Oke} \& {Sandage}(1968)}]{Oke:1968}
{Oke}, J.~B., \& {Sandage}, A. 1968, \apj, 154, 21, \dodoi{10.1086/149737}

\bibitem[{Phillips {et~al.}(1999)Phillips, Lira, Suntzeff, Schommer, Hamuy, \&
  Maza}]{Phillips:1999}
Phillips, M.~M., Lira, P., Suntzeff, N.~B., {et~al.} 1999, \aj, 118, 1766

\bibitem[{{Pierel} {et~al.}(2018){Pierel}, {Rodney}, {Avelino}, {Bianco},
  {Foley}, {Friedman}, {Hicken}, {Hounsell}, {Jha}, {Kessler}, {Kirshner},
  {Mandel}, {Narayan}, {Filippenko}, {Scolnic}, \& {Strolger}}]{Pierel:2018a}
{Pierel}, J.\, D.~R., {Rodney}, S.\, A., {Avelino}, A., {et~al.} 2018,
  {SNSEDextend}, Astrophysics Source Code Library.
\newblock \doeprint{1805.017}

\bibitem[{{Pignata} {et~al.}(2008){Pignata}, {Benetti}, {Mazzali}, {Kotak},
  {Patat}, {Meikle}, {Stehle}, {Leibundgut}, {Suntzeff}, {Buson}, {Cappellaro},
  {Clocchiatti}, {Hamuy}, {Maza}, {Mendez}, {Ruiz-Lapuente}, {Salvo},
  {Schmidt}, {Turatto}, \& {Hillebrandt}}]{Pignata:2008}
{Pignata}, G., {Benetti}, S., {Mazzali}, P.~A., {et~al.} 2008, \mnras, 388,
  971, \dodoi{10.1111/j.1365-2966.2008.13434.x}

\bibitem[{{Poznanski} {et~al.}(2007){Poznanski}, {Maoz}, {Yasuda}, {Foley},
  {Doi}, {Filippenko}, {Fukugita}, {Gal-Yam}, {Jannuzi}, {Morokuma}, {Oda},
  {Schweiker}, {Sharon}, {Silverman}, \& {Totani}}]{Poznanski:2007}
{Poznanski}, D., {Maoz}, D., {Yasuda}, N., {et~al.} 2007, \mnras, 382, 1169,
  \dodoi{10.1111/j.1365-2966.2007.12424.x}

\bibitem[{Renka(1987)}]{Renka:1987}
Renka, R.~J. 1987, Society for Industrial and Applied Mathematics, 8, 012

\bibitem[{Riess {et~al.}(2004)Riess, Strolger, Tonry, Tsvetanov, Casertano,
  Ferguson, Mobasher, Challis, Panagia, Filippenko, Li, Chornock, Kirshner,
  Leibundgut, Dickinson, Koekemoer, Grogin, \& Giavalisco}]{Riess:2004a}
Riess, A.~G., Strolger, L.-G., Tonry, J., {et~al.} 2004, \apj, 600, L163

\bibitem[{{Rodney} {et~al.}(2014){Rodney}, {Riess}, {Strolger}, {Dahlen},
  {Graur}, {Casertano}, {Dickinson}, {Ferguson}, {Garnavich}, {Hayden}, {Jha},
  {Jones}, {Kirshner}, {Koekemoer}, {McCully}, {Mobasher}, {Patel}, {Weiner},
  {Cenko}, {Clubb}, {Cooper}, {Filippenko}, {Frederiksen}, {Hjorth},
  {Leibundgut}, {Matheson}, {Nayyeri}, {Penner}, {Trump}, {Silverman}, {U},
  {Azalee Bostroem}, {Challis}, {Rajan}, {Wolff}, {Faber}, {Grogin}, \&
  {Kocevski}}]{Rodney:2014}
{Rodney}, S.~A., {Riess}, A.~G., {Strolger}, L.-G., {et~al.} 2014, \aj, 148,
  13, \dodoi{10.1088/0004-6256/148/1/13}

\bibitem[{{Sako} {et~al.}(2011){Sako}, {Bassett}, {Connolly}, {Dilday},
  {Cambell}, {Frieman}, {Gladney}, {Kessler}, {Lampeitl}, {Marriner}, {Miquel},
  {Nichol}, {Schneider}, {Smith}, \& {Sollerman}}]{Sako:2011}
{Sako}, M., {Bassett}, B., {Connolly}, B., {et~al.} 2011, \apj, 738, 162,
  \dodoi{10.1088/0004-637X/738/2/162}

\bibitem[{Salvatier {et~al.}(2015)Salvatier, Wiecki, \&
  Fonnesbeck}]{Salvatier:2015}
Salvatier, J., Wiecki, T., \& Fonnesbeck, C. 2015, 1,
  \dodoi{10.7717/peerj-cs.55}

\bibitem[{Schlafly \& Finkbeiner(2011)}]{Schlafly:2011}
Schlafly, E.~F., \& Finkbeiner, D.~P. 2011, Astrophys. J., 737,
  \dodoi{10.1088/0004-637X/737/2/103}

\bibitem[{{Scolnic} {et~al.}(2014){Scolnic}, {Riess}, {Foley}, {Rest},
  {Rodney}, {Brout}, \& {Jones}}]{Scolnic:2014a}
{Scolnic}, D.~M., {Riess}, A.~G., {Foley}, R.~J., {et~al.} 2014, \apj, 780, 37,
  \dodoi{10.1088/0004-637X/780/1/37}

\bibitem[{Scolnic {et~al.}(2017)Scolnic, Jones, Rest, Pan, Chornock, Foley,
  Huber, Kessler, Narayan, Riess, Rodney, Berger, Challis, Drout, Finkbeiner,
  Lunnan, Kirshner, Sanders, Schlafly, Smartt, Stubbs, Tonry, Wood-Vasey,
  Foley, Hand, Johnson, Burgett, Chambers, Draper, Hodapp, Kaiser, Kudritzki,
  Magnier, Metcalfe, Bresolin, Gall, Kotak, McCrum, \& Smith}]{Scolnic:2017}
Scolnic, D.~M., Jones, D.~O., Rest, A., {et~al.} 2017, \dodoi{10.17909/T95Q4X}

\bibitem[{Shussman {et~al.}(2016)Shussman, Waldman, \& Nakar}]{Shussman:2016}
Shussman, T., Waldman, R., \& Nakar, E. 2016, 21, 1.
\newblock \doarXiv{1610.05323}

\bibitem[{{Smartt} {et~al.}(2009){Smartt}, {Eldridge}, {Crockett}, \&
  {Maund}}]{Smartt:2009}
{Smartt}, S.~J., {Eldridge}, J.~J., {Crockett}, R.~M., \& {Maund}, J.~R. 2009,
  \mnras, 395, 1409, \dodoi{10.1111/j.1365-2966.2009.14506.x}

\bibitem[{Spergel {et~al.}(2015)Spergel, Gehrels, Baltay, Bennett,
  Breckinridge, Donahue, Dressler, Gaudi, Greene, Guyon, Hirata, Kalirai,
  Kasdin, Macintosh, Moos, Perlmutter, Postman, Rauscher, Rhodes, Wang,
  Weinberg, Benford, Hudson, Jeong, Mellier, Traub, Yamada, Capak, Colbert,
  Masters, Penny, Savransky, Stern, Zimmerman, Barry, Bartusek, Carpenter,
  Cheng, Content, Dekens, Demers, Grady, Jackson, Kuan, Kruk, Melton, Nemati,
  Parvin, Poberezhskiy, Peddie, Ruffa, Wallace, Whipple, Wollack, \&
  Zhao}]{Spergel:2015}
Spergel, D., Gehrels, N., Baltay, C., {et~al.} 2015.
\newblock \doarXiv{1503.03757}

\bibitem[{{Stanishev} {et~al.}(2007){Stanishev}, {Goobar}, {Benetti}, {Kotak},
  {Pignata}, {Navasardyan}, {Mazzali}, {Amanullah}, {Garavini}, {Nobili},
  {Qiu}, {Elias-Rosa}, {Ruiz-Lapuente}, {Mendez}, {Meikle}, {Patat},
  {Pastorello}, {Altavilla}, {Gustafsson}, {Harutyunyan}, {Iijima},
  {Jakobsson}, {Kichizhieva}, {Lundqvist}, {Mattila}, {Melinder}, {Pavlenko},
  {Pavlyuk}, {Sollerman}, {Tsvetkov}, {Turatto}, \&
  {Hillebrandt}}]{Stanishev:2007}
{Stanishev}, V., {Goobar}, A., {Benetti}, S., {et~al.} 2007, \aap, 469, 645,
  \dodoi{10.1051/0004-6361:20066020}

\bibitem[{{Stritzinger} {et~al.}(2010){Stritzinger}, {Filippenko}, {Folatelli},
  {Foley}, {Hamuy}, {Li}, {Mazzali}, {Phillips}, \&
  {Pignata}}]{Stritzinger:2010}
{Stritzinger}, M., {Filippenko}, A., {Folatelli}, G., {et~al.} 2010,
  {Multi-wavelength spectroscopic study of young Type Ia supernovae}, NOAO
  Proposal

\bibitem[{{Stritzinger} {et~al.}(2011){Stritzinger}, {Phillips}, {Boldt},
  {Burns}, {Campillay}, {Contreras}, {Gonzalez}, {Folatelli}, {Morrell},
  {Krzeminski}, {Roth}, {Salgado}, {DePoy}, {Hamuy}, {Freedman}, {Madore},
  {Marshall}, {Persson}, {Rheault}, {Suntzeff}, {Villanueva}, {Li}, \&
  {Filippenko}}]{Stritzinger:2011}
{Stritzinger}, M.~D., {Phillips}, M.~M., {Boldt}, L.~N., {et~al.} 2011, \aj,
  142, 156, \dodoi{10.1088/0004-6256/142/5/156}

\bibitem[{Taddia {et~al.}(2018)Taddia, Stritzinger, Bersten, Baron, Burns,
  Contreras, Holmbo, Hsiao, Morrell, Phillips, Sollerman, \&
  Suntzeff}]{Taddia:2018}
Taddia, F., Stritzinger, M.~D., Bersten, M., {et~al.} 2018, Astron. Astrophys.,
  609

\bibitem[{{Valentini} {et~al.}(2003){Valentini}, {Di Carlo}, {Massi}, {Dolci},
  {Arkharov}, {Larionov}, {Pastorello}, {Di Paola}, {Benetti}, {Cappellaro},
  {Turatto}, {Pedichini}, {D'Alessio}, {Caratti o Garatti}, {Li Causi},
  {Speziali}, {Danziger}, \& {Tornamb{\'e}}}]{Valentini:2003}
{Valentini}, G., {Di Carlo}, E., {Massi}, F., {et~al.} 2003, \apj, 595, 779,
  \dodoi{10.1086/377448}

\bibitem[{{Wood-Vasey} {et~al.}(2008){Wood-Vasey}, {Friedman}, {Bloom},
  {Hicken}, {Modjaz}, {Kirshner}, {Starr}, {Blake}, {Falco}, {Szentgyorgyi},
  {Challis}, {Blondin}, {Mandel}, \& {Rest}}]{Woodvasey:2008}
{Wood-Vasey}, W.~M., {Friedman}, A.~S., {Bloom}, J.~S., {et~al.} 2008, \apj,
  689, 377, \dodoi{10.1086/592374}

\bibitem[{{Zheng} {et~al.}(2018){Zheng}, {Kelly}, \& {Filippenko}}]{Zheng:2018}
{Zheng}, W., {Kelly}, P.~L., \& {Filippenko}, A.~V. 2018, \apj, 858, 104,
  \dodoi{10.3847/1538-4357/aabaeb}

\end{thebibliography}

\newpage

\appendix
\setcounter{table}{0}
\renewcommand{\thetable}{A\arabic{table}}

\setcounter{figure}{0}
\renewcommand{\thefigure}{A\arabic{figure}}

\begin{table*}
\centering
\caption{Summary of the supernovae used in this analysis.} 
\resizebox{\textwidth}{!}{%
\begin{tabular}{@{}LCCRCRC@{}}
\toprule
{\rm SNID} & {\rm Subtype} & {\rm Colors} & {\rm MW}~E(B-V) & z & {\rm Obs.~MJD~Peak} & {\rm Reference} \\ \hline
\rm SN 2002bx & \rm II & U-B & 0.0106 & 0.007539 & 52368.27\pm0.25 & \rm H17 \\
\rm SN 2004aw & \rm Ic & U-B & 0.0180 & 0.015911 & 53089.79\pm0.08 & \rm B14 \\
\rm SN 2004gq & \rm Ib & U-B & 0.0627 & 0.006401 & 53357.59\pm0.21 & B14 \\
\rm SN 2005hg & \rm Ib & U-B,r'-J,r'-H,r'-K & 0.0901 & 0.003389 & 53667.10\pm0.05 & \rm B14 \\
\rm SN 2005kl & \rm Ic & r'-J,r'-H,r'-K & 0.0219 & 0.026761 & 53703.28\pm0.10 & \rm B14 \\
\rm SN 2005mf & \rm Ic & U-B,r'-J,r'-H,r'-K & 0.0153 & 0.026761 & 53734.02\pm0.57 & \rm B14 \\
\rm SN 2006aj & \rm Ic & U-B,r'-J,r'-H,r'-K & 0.1267 & 0.033529 & 53792.94\pm0.14 & \rm B14 \\
\rm SN 2006ca & \rm II & U-B & 0.1990 & 0.008903 & 53866.30\pm0.16 & \rm H17 \\
\rm SN 2006cd & \rm IIP & U-B & 0.0407 & 0.037116 & 53852.51\pm0.56 & \rm H17 \\
\rm SN 2006F & \rm Ib & U-B & 0.1635 & 0.013999 & 53749.72\pm0.52 & \rm B14 \\
\rm SN 2006it & \rm IIP & U-B & 0.0850 & 0.015511 & 54015.13\pm0.05 & \rm H17 \\
\rm SN 2006fo & \rm Ib & r'-J,r'-H,r'-K & 0.0250 & 0.020728 & 53991.88\pm0.18 & \rm B14 \\
\rm SN 2006T & \rm IIb & U-B & 0.0647 & 0.007992 & 53765.11\pm0.05 & \rm B14 \\
\rm SN 2007C & \rm Ib & r'-J,r'-H,r'-K & 0.0363 & 0.005894 & 54114.10\pm0.20 & \rm B14 \\
\rm SN 2007D & \rm Ic & r'-J,r'-H,r'-K & 0.2881 & 0.023163 & 54119.68\pm0.61 & \rm B14 \\
\rm SN 2007gr & \rm Ic & U-B,r'-J,r'-H,r'-K & 0.0535 & 0.001727 & 54339.81\pm0.06 & \rm B14 \\
\rm SN 2007I & \rm Ic & r'-J,r'-H,r'-K & 0.0250 & 0.021638 & 54118.55\pm0.74 & \rm B14 \\
\rm SN 2008aj & \rm II & U-B & 0.0128 & 0.024963 & 54484.47\pm0.01 & \rm H17 \\
\rm SN 2008aq & \rm Ib & U-B & 0.0383 & 0.007969 & 54530.87\pm0.10 & \rm B14 \\
\rm SN 2008bj & \rm II & U-B & 0.0233 & 0.018965 & 54553.23\pm0.36 & \rm H17 \\
\rm SN 2008bn & \rm II & U-B & 0.0154 & 0.024220 & 54555.05\pm0.47 & \rm H17 \\
\rm SN 2008D & \rm Ib & r'-J,r'-H,r'-K & 0.0194 & 0.007004 & 54474.28\pm0.03 & \rm B14 \\
\rm SN 2008ip & \rm II & r'-J,r'-H,r'-K & 0.0136 & 0.015124 & 54812.33\pm0.51 & \rm H17 \\
\rm SN 2009ay & \rm II & r'-J,r'-H,r'-K & 0.0342 & 0.022182 & 54901.63\pm0.84 & \rm H17 \\
\rm SN 2009iz & \rm Ib & u'-B,r'-J,r'-H,r'-K & 0.0729 & 0.014199 & 55115.14\pm0.04 & \rm B14 \\
\rm SN 2009jf & \rm Ib & u'-B,r'-J,r'-H,r'-K & 0.0971 & 0.008148 & 55121.97\pm0.08 & \rm B14 \\
\rm SN 2010bq & \rm II & r'-J,r'-H,r'-K & 0.0191 & 0.030988 & 55295.10\pm0.34 & \rm H17 \\ 
\hline
\end{tabular}%
}
\label{Atab:cc_data}
$^a$From \citet{Bianco:2014} (B14 in table) and \citet{Hicken:2017} (H17 in table). The colors listed for each SN are used in the extrapolations, and the dates of peak brightness were measured in the course of this work.
\end{table*}
\begin{figure*}
\centering
\includegraphics[scale=1.2]{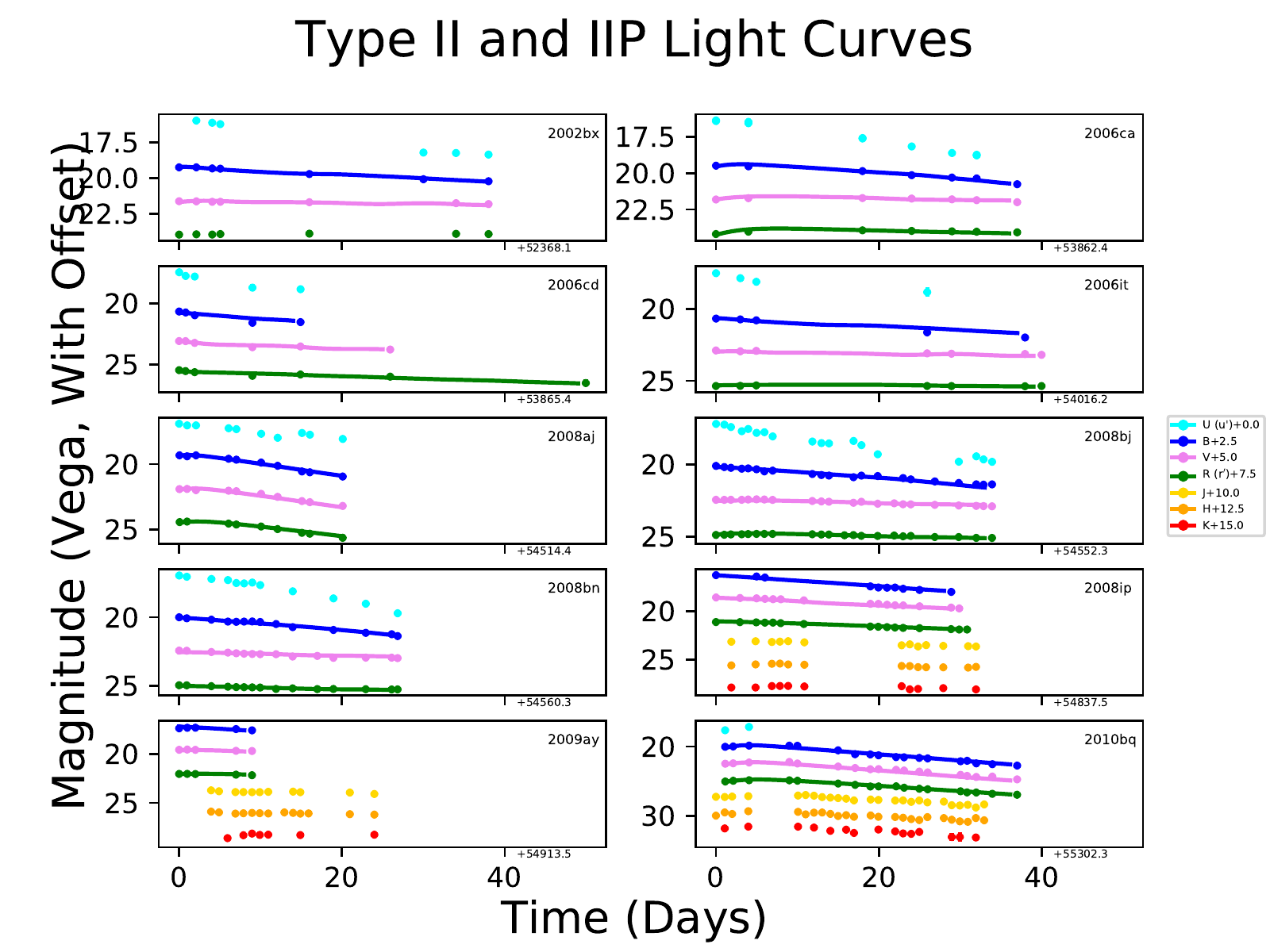}
\caption{\label{Afig:hickencurves} All of the SN~II and SN~IIP light curves from \citet{Hicken:2017} used in this work.}
\end{figure*}
\begin{figure*}
\centering
\includegraphics[scale=1.2]{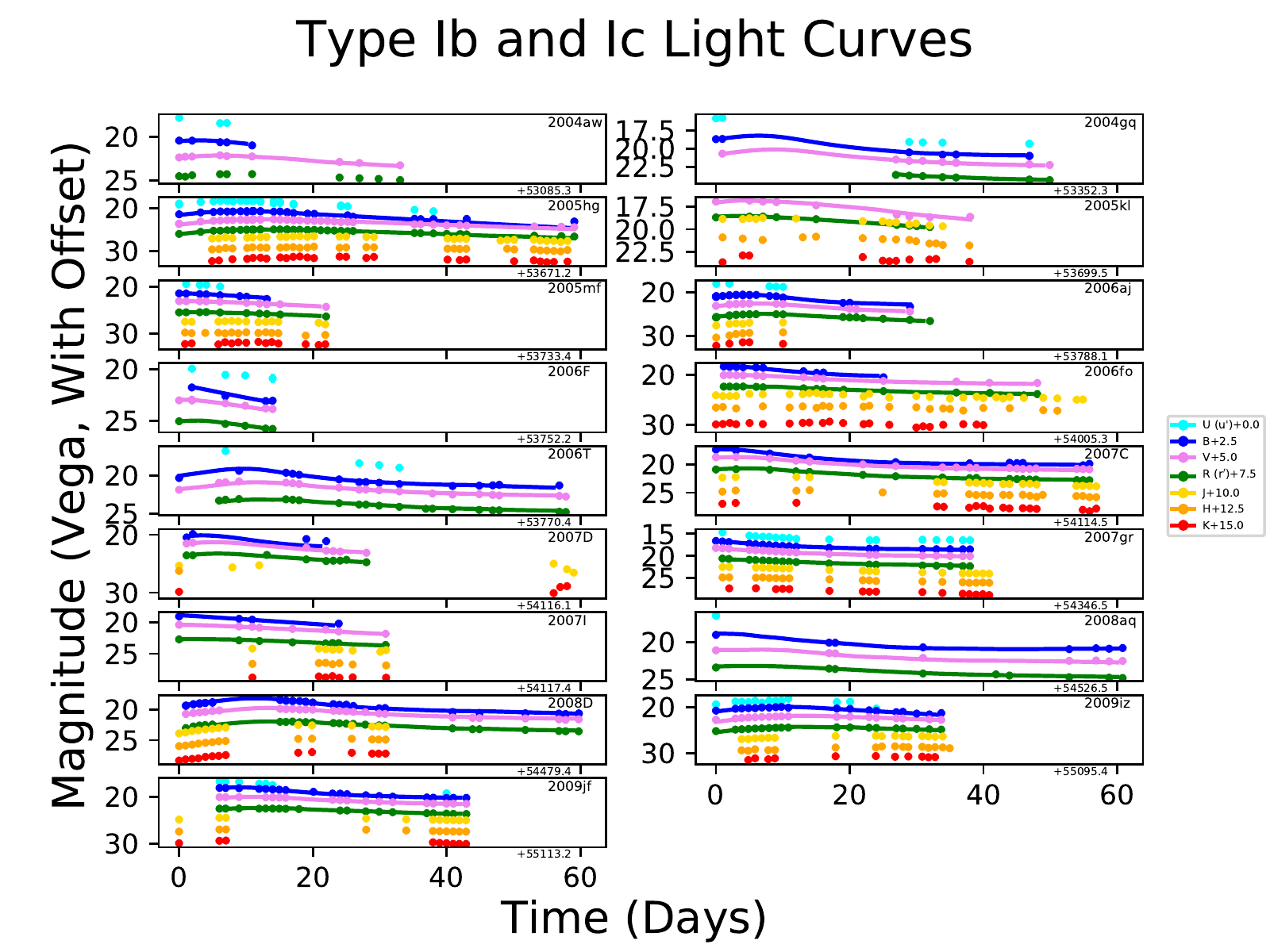}
\caption{\label{Afig:biancocurves} All of the SN~Ib and SN~Ic light curves from \citet{Bianco:2014} used in this work.}
\end{figure*}

\

\textbf{SNCosmo Fitting}

Using the classifications and redshifts provided by the sources of these data, the SNe are separated by subtype, and their light-curve data in magnitudes are converted into fluxes. The SNCosmo package is used to estimate a time of peak luminosity ($t_0$) for each SN by fitting models of matching classification and incorporating that SN’s measured redshift from Table \ref{Atab:cc_data}. For each SN, light-curve points more than 50 days from $t_0$ are removed, as those are beyond the temporal extent of our SEDs.

\begin{figure*}
\centering
\textbf{Type Ic Supernova Light-Curve Fitting}
\includegraphics[scale=.5,trim={0 15cm 0 0},clip]{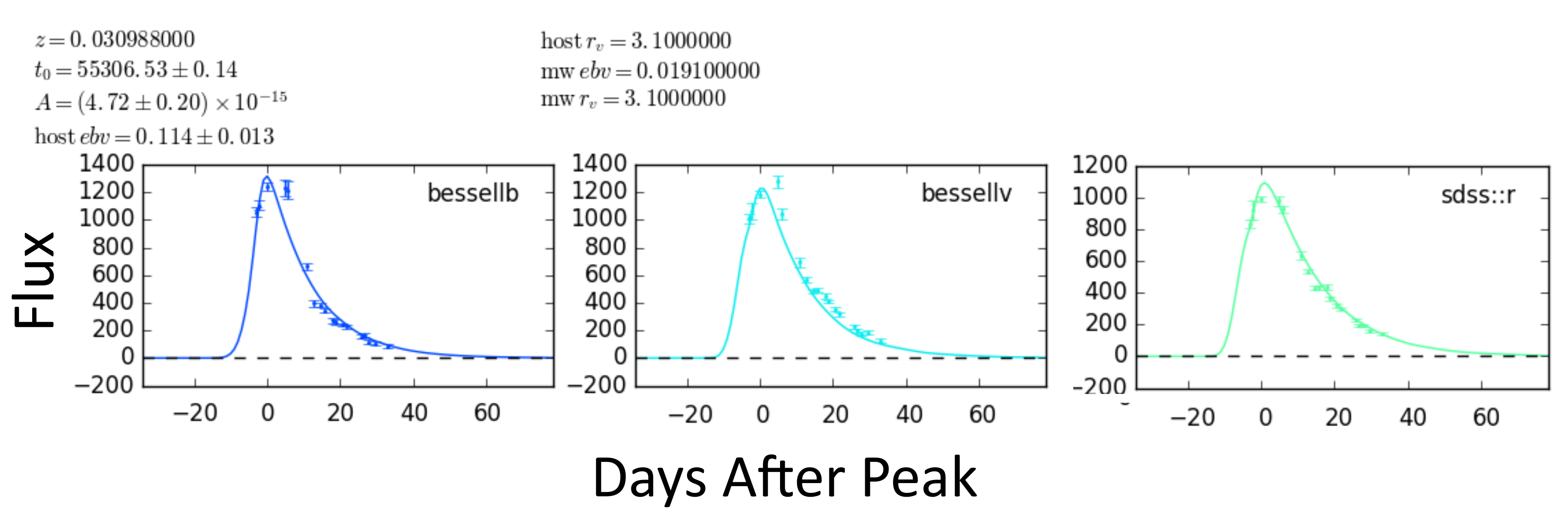}
\caption{\label{fig:SN_fitting} Example SNCosmo fitting results of optical colors ($B$, $V$, $\rprime$) for a SN~Ic.}
\end{figure*}

Once the initial estimate of $t_0$ is found, the light curves and initial parameters are provided to the \textit{SNSEDextend}\ package, which completes the fitting process. At this point the redshift, SN subtype, and Milky Way $E(B-V)$ (mag) parameters are all incorporated into the light-curve fitting based on the information provided in the literature for these SNe (Table \ref{Atab:cc_data}). The $R_V$ parameter for both the host and Milky Way galaxies are set to 3.1, and the host $E(B-V)$ is given reasonable bounds of $\pm1$ \citep{Cardelli:1989,odonnell:1994}. With all of these parameters in place, the \textit{SNSEDextend}\ package chooses the best-fit SNCosmo model from the original set matching the SN classification (Figure \ref{fig:SN_fitting}), and uses it to generate color tables for each SN subtype (see Section \ref{sub:curves}). 

\

\

\textbf{Color-Table Generation}

\begin{enumerate}
\item Use the fitted light-curve model from Section \ref{sub:fitting} to interpolate the color reference bands ($B$ or $r'$) to the epochs where observations exist in the extrapolation anchor bands ($U$, $\uprime$, $J$, $H$, $K$).  
\item Define a color (e.g., $r'-J$) from the difference of the interpolated magnitude in the reference band and the observed magnitude in the extrapolation anchor band. 
\item All SNe of like classification (e.g., all SNe~Ic) are collected together to make a merged table of dereddened colors, an example of which is reported in Table \ref{Atab:color_table}.
\item
First- and second-order polynomials are fit to each discrete, time-dependent color curve
(i.e., $U-B$, $r'-{JHK}$). The best-fit polynomial is determined by
minimizing the Bayesian Information Criterion (BIC), and the median model
is extracted from posterior predictive fitting for use in
extrapolation \citep{Salvatier:2015}. The curve flattens outside the edges of our data points, as we have no constraints in those regions and previous work suggests that the slope of a color curve far from peak diverges from the peak color-curve slope \citep{Krisciunas:2009}.
\end{enumerate}

\begin{table*}
\centering
\caption{Partial example of SN~Ic dereddened color table generated by \textit{SNSEDextend}.}
\resizebox{\textwidth}{!}{%
\begin{tabular}{@{}RCCCCCCCCC@{}}
\toprule
{\rm Days~After~Peak} & U-B & U-B~ {\rm Error} & r-J & r-J~{\rm Error} & r-H & r-H~{\rm Error} & r-K & r-K~{\rm Error} & {\rm SN} \\ \hline
-4.3736 & -0.30 & 0.11 & 0.81 & 0.04 & 0.61 & 0.17 & 1.27 & 0.14 \\
-3.5605 & -- & -- & 0.71 & 0.10 & 0.60 & 0.40 & -- & -- \\
-2.6903 & -0.34 & 0.09 & -- & -- & -- & -- & -- & -- \\
-2.4106 & -0.07 & 0.09 & 0.88 & 0.05 & 0.81 & 0.10 & 1.46 & 0.12 \\
-1.4136 & -- & -- & 0.91 & 0.20 & 1.03 & 0.08 & -- & -- \\
-0.4106 & -- & -- & 0.69 & 0.05 & 1.07 & 0.09 & 1.53 & 0.17 \\
0.4986 & -- & -- & 1.02 & 0.11 & 1.33 & 0.09 & 1.39 & 0.12 \\
0.5864 &  &  & 0.86 & 0.04 & 1.10 & 0.07 & 1.47 & 0.10 \\
0.6616 & 0.71 & 0.19 & -- & -- & 0.84 & 0.21 & -- & -- \\
1.4876 & -- & -- & 1.01 & 0.06 & 1.22 & 0.09 & 1.56 & 0.12 \\
2.6982 & 0.83 & 0.23 &  &  & ... &  &  & ... \\
3.4289 & 0.30 & 0.15 & -- & -- & -- & -- & -- & -- \\
3.5226 & -- & -- & -- & -- & 1.30 & 0.08 & -- & -- \\
3.5851 & 0.46 & 0.14 & -- & -- & -- & -- & -- & -- \\
3.6702 & 0.74 & 0.31 & -- & -- & -- & -- & -- & -- \\
4.4291 & 0.21 & 0.14 & -- & -- & -- & -- & -- & -- \\
4.5862 & 0.40 & 0.09 & -- & -- & -- & -- & -- & -- \\
5.4836 & -- & -- & 1.16 & 0.07 & 1.40 & 0.11 & 1.43 & 0.37 \\
5.6114 & 0.29 & 0.09 & 0.95 & 0.04 & 1.26 & 0.06 & 1.14 & 0.10 \\
6.4776 & -- & -- & 1.23 & 0.08 & 1.28 & 0.20 & 1.92 & 0.18 \\
 & ... &  &  &  & ... &  &  & ...
\end{tabular}%
}
\label{Atab:color_table}
$^a$A similar table is created for each SN subtype, which is then fit with a polynomial.
\end{table*}

\

\renewcommand{\arraystretch}{0.001}
\renewcommand{\baselinestretch}{0.8}
\begin{table*} 
\begin{center} 
\caption{Selection parameters and references used for SN~Ia sample analyzed in Section \ref{sec:typeIa}.}
\tiny
\begin{tabular}{l cccc cccc} 
\hline 
SN name & $\alpha$ (deg) & $\delta$ (deg) & $z^{a}_{\rm helio}$  & LC Data & $t_{B {\rm max}}$ & $\Delta m_{15}(B)^{c}$ & $E(B-V)^d_{\rm host}$  & $E(B-V)^e_{\rm MW}$ \\
&&&& source$^b$ & (MJD days) & (mag) & (mag) & (mag) \\[0.1cm]
\hline 
SN1998bu        &  161.69167  &  11.83528  &  0.0030 $\pm$ 0.000003 & CfA & 50953.11 $\pm$ 0.08 & 1.076 $\pm$ 0.012 &  0.351 $\pm$ 0.006 &  0.022 $\pm$ 0.0002 \\ 
SN1999ee        &  334.04167  &  -36.84444  &  0.0114 $\pm$ 0.000010 & CSP & 51469.61 $\pm$ 0.04 & 0.802 $\pm$ 0.007 &  0.384 $\pm$ 0.004 &  0.017 $\pm$ 0.0001 \\ 
SN1999ek        &  84.13167  &   16.63833   &  0.0176 $\pm$ 0.000007 & K04c & 51482.60 $\pm$ 0.19 & 1.113 $\pm$ 0.031 &  0.277 $\pm$ 0.014 &  0.479 $\pm$ 0.0187 \\ 
SN2000bh        &  185.31292  &  -21.99889  &  0.0229 $\pm$ 0.000027 & CSP & 51636.16 $\pm$ 0.25 & 1.055 $\pm$ 0.019 &  0.065 $\pm$ 0.012 &  0.047 $\pm$ 0.0064 \\ 
SN2000ca        &  203.84583  &  -34.16028  &  0.0236 $\pm$ 0.000200 & CSP & 51666.25 $\pm$ 0.18 & 0.917 $\pm$ 0.019 & -0.033 $\pm$ 0.010 &  0.057 $\pm$ 0.0025 \\ 
SN2000E         &  309.30750  &  66.09722   &  0.0047 $\pm$ 0.000003 & V03 & 51577.20 $\pm$ 0.13 & 1.041 $\pm$ 0.027 &  0.217 $\pm$ 0.011 &  0.319 $\pm$ 0.0086 \\ 
SN2001ba        &  174.50750  &  -32.33083  &  0.0296 $\pm$ 0.000033 & CSP & 52034.47 $\pm$ 0.17 & 0.997 $\pm$ 0.020 & -0.072 $\pm$ 0.009 &  0.054 $\pm$ 0.0017 \\ 
SN2001bt        &  288.44500  &  -59.28972  &  0.0146 $\pm$ 0.000033 & K04c & 52064.69 $\pm$ 0.07 & 1.199 $\pm$ 0.009 &  0.216 $\pm$ 0.008 &  0.056 $\pm$ 0.0007 \\ 
SN2001cn        &  281.57417  &  -65.76167  &  0.0152 $\pm$ 0.000127 & K04c & 52071.93 $\pm$ 0.19 & 1.044 $\pm$ 0.012 &  0.176 $\pm$ 0.008 &  0.051 $\pm$ 0.0008 \\ 
SN2001cz        &  191.87583  &  -39.58000  &  0.0155 $\pm$ 0.000027 & K04c & 52104.10 $\pm$ 0.10 & 0.956 $\pm$ 0.014 &  0.136 $\pm$ 0.008 &  0.079 $\pm$ 0.0005 \\ 
SN2001el        &  56.12750  &   -44.63972  &  0.0039 $\pm$ 0.000007 & K03 & 52182.38 $\pm$ 0.10 & 1.080 $\pm$ 0.019 &  0.277 $\pm$ 0.010 &  0.012 $\pm$ 0.0003 \\ 
SN2002dj        &  198.25125  &  -19.51917  &  0.0094 $\pm$ 0.000003 & P08 & 52451.04 $\pm$ 0.14 & 1.111 $\pm$ 0.019 &  0.093 $\pm$ 0.013 &  0.082 $\pm$ 0.0009 \\ 
SN2003du        &  218.64917  &  59.33444   &  0.0064 $\pm$ 0.000013 & St07 & 52766.01 $\pm$ 0.09 & 1.010 $\pm$ 0.015 & -0.033 $\pm$ 0.010 &  0.008 $\pm$ 0.0008 \\ 
SN2003hv        &  46.03875  &   -26.08556  &  0.0056 $\pm$ 0.000037 & L09 & 52891.49 $\pm$ 0.11 & 1.501 $\pm$ 0.006 & -0.092 $\pm$ 0.007 &  0.013 $\pm$ 0.0008 \\ 
SN2004ef        &  340.54175  &  19.99456   &  0.0310 $\pm$ 0.000017 & CSP & 53264.90 $\pm$ 0.05 & 1.422 $\pm$ 0.011 &  0.116 $\pm$ 0.006 &  0.046 $\pm$ 0.0013 \\ 
SN2004eo        &  308.22579  &  9.92853    &  0.0156 $\pm$ 0.000003 & CSP & 53278.90 $\pm$ 0.04 & 1.318 $\pm$ 0.006 &  0.077 $\pm$ 0.005 &  0.093 $\pm$ 0.0010 \\ 
SN2004ey        &  327.28254  &  0.44422    &  0.0158 $\pm$ 0.000003 & CSP & 53304.81 $\pm$ 0.04 & 1.025 $\pm$ 0.011 &  0.006 $\pm$ 0.004 &  0.120 $\pm$ 0.0139 \\ 
SN2004gs        &  129.59658  &  17.62772   &  0.0274 $\pm$ 0.000007 & CSP & 53356.75 $\pm$ 0.05 & 1.546 $\pm$ 0.006 &  0.189 $\pm$ 0.006 &  0.026 $\pm$ 0.0006 \\ 
SN2004S         &  101.43125  &  -31.23111  &  0.0093 $\pm$ 0.000003 & K07 & 53040.00 $\pm$ 0.29 & 1.052 $\pm$ 0.021 &  0.112 $\pm$ 0.014 &  0.086 $\pm$ 0.0014 \\ 
SN2005bo        &  192.42096  &  -11.09647  &  0.0139 $\pm$ 0.000027 & CfA & 53479.63 $\pm$ 0.15 & 1.310 $\pm$ 0.020 &  0.272 $\pm$ 0.007 &  0.044 $\pm$ 0.0006 \\ 
SN2005cf        &  230.38417  &  -7.41306   &  0.0064 $\pm$ 0.000017 & CfA & 53534.31 $\pm$ 0.06 & 1.072 $\pm$ 0.023 &  0.088 $\pm$ 0.010 &  0.084 $\pm$ 0.0013 \\ 
SN2005el        &  77.95300  &   5.19428 &  0.0149 $\pm$ 0.000017 & CSP & 53647.42 $\pm$ 0.04 & 1.370 $\pm$ 0.006 & -0.102 $\pm$ 0.005 &  0.098 $\pm$ 0.0004 \\ 
SN2005iq        &  359.63542  &  -18.70917  &  0.0340 $\pm$ 0.000123 & CSP & 53688.14 $\pm$ 0.06 & 1.280 $\pm$ 0.012 & -0.049 $\pm$ 0.006 &  0.018 $\pm$ 0.0007 \\ 
SN2005kc        &  338.53058  &  5.56842  &  0.0151 $\pm$ 0.000003 & CSP & 53698.31 $\pm$ 0.08 & 1.112 $\pm$ 0.023 &  0.350 $\pm$ 0.012 &  0.114 $\pm$ 0.0023 \\ 
SN2005ki        &  160.11758  &  9.20233  &  0.0195 $\pm$ 0.000010 & CSP & 53706.01 $\pm$ 0.04 & 1.365 $\pm$ 0.004 & -0.065 $\pm$ 0.004 &  0.027 $\pm$ 0.0009 \\ 
SN2005lu        &  39.01546  &   -17.26389 &  0.0320 $\pm$ 0.000037 & CSP & 53712.08 $\pm$ 0.23 & 0.834 $\pm$ 0.008 &  0.324 $\pm$ 0.011 &  0.022 $\pm$ 0.0009 \\ 
SN2005na        &  105.40258  &  14.13325  &  0.0263 $\pm$ 0.000083 & CfA & 53739.37 $\pm$ 0.30 & 1.027 $\pm$ 0.014 & -0.050 $\pm$ 0.012 &  0.068 $\pm$ 0.0025 \\ 
SN2006ac        &  190.43708  &  35.08528  &  0.0231 $\pm$ 0.000010 & CfA & 53781.55 $\pm$ 0.10 & 1.189 $\pm$ 0.008 &  0.066 $\pm$ 0.010 &  0.014 $\pm$ 0.0006 \\ 
SN2006ax        &  171.01442  &  -12.29144  &  0.0167 $\pm$ 0.000020 & CSP & 53827.78 $\pm$ 0.04 & 1.058 $\pm$ 0.012 & -0.009 $\pm$ 0.005 &  0.041 $\pm$ 0.0019 \\ 
SN2006bh        &  340.06708  &  -66.48508  &  0.0108 $\pm$ 0.000013 & CSP & 53834.14 $\pm$ 0.06 & 1.408 $\pm$ 0.007 & -0.043 $\pm$ 0.004 &  0.023 $\pm$ 0.0004 \\ 
SN2006bt        &  239.12721  &  20.04592  &  0.0321 $\pm$ 0.000007 & CSP & 53859.29 $\pm$ 0.26 & 1.093 $\pm$ 0.042 &  0.313 $\pm$ 0.023 &  0.042 $\pm$ 0.0013 \\ 
SN2006cp        &  184.81208  &  22.42722  &  0.0223 $\pm$ 0.000003 & CfA & 53897.45 $\pm$ 0.15 & 1.023 $\pm$ 0.046 &  0.134 $\pm$ 0.022 &  0.022 $\pm$ 0.0011 \\ 
SN2006D         &  193.14142  &  -9.77522  &  0.0085 $\pm$ 0.000017 & CfA & 53757.84 $\pm$ 0.08 & 1.460 $\pm$ 0.013 &  0.062 $\pm$ 0.009 &  0.039 $\pm$ 0.0004 \\ 
SN2006ej        &  9.74904  &  -9.01572  &  0.0204 $\pm$ 0.000007 & CSP & 53977.24 $\pm$ 0.25 & 1.394 $\pm$ 0.013 &  0.016 $\pm$ 0.011 &  0.030 $\pm$ 0.0008 \\ 
SN2006kf        &  55.46033  &   8.15694 &  0.0200 $\pm$ 0.000010 & CSP & 54041.86 $\pm$ 0.05 & 1.517 $\pm$ 0.008 &  0.007 $\pm$ 0.006 &  0.210 $\pm$ 0.0020 \\ 
SN2006lf        &  69.62292  &   44.03361 &  0.0132 $\pm$ 0.000017 & CfA & 54045.56 $\pm$ 0.06 & 1.406 $\pm$ 0.010 & -0.054 $\pm$ 0.010 &  0.814 $\pm$ 0.0503 \\ 
SN2006N         &  92.13000  &   64.72361 &  0.0143 $\pm$ 0.000083 & CfA & 53761.48 $\pm$ 0.15 & 1.457 $\pm$ 0.013 & -0.030 $\pm$ 0.007 &  0.083 $\pm$ 0.0010 \\ 
SN2007A         &  6.31942   & 12.88681  & 0.0176 $\pm$ 0.000087 & CSP & 54113.67 $\pm$ 0.13 & 1.037 $\pm$ 0.034 &  0.225 $\pm$ 0.014 &  0.063 $\pm$ 0.0016 \\ 
SN2007af        &  215.58763  &  -0.39378  &  0.0055 $\pm$ 0.000013 & CSP & 54174.97 $\pm$ 0.04 & 1.116 $\pm$ 0.010 &  0.183 $\pm$ 0.005 &  0.034 $\pm$ 0.0008 \\ 
SN2007ai        &  243.22392  &  -21.63019  &  0.0317 $\pm$ 0.000137 & CSP & 54174.03 $\pm$ 0.26 & 0.844 $\pm$ 0.021 &  0.339 $\pm$ 0.013 &  0.286 $\pm$ 0.0035 \\ 
SN2007as        &  141.90004  &  -80.17756  &  0.0176 $\pm$ 0.000460 & CSP & 54181.15 $\pm$ 0.23 & 1.120 $\pm$ 0.023 &  0.138 $\pm$ 0.010 &  0.123 $\pm$ 0.0007 \\ 
SN2007bc        &  169.81071  &  20.80903  &  0.0208 $\pm$ 0.000007 & CSP & 54200.82 $\pm$ 0.09 & 1.282 $\pm$ 0.012 &  0.039 $\pm$ 0.006 &  0.019 $\pm$ 0.0006 \\ 
SN2007bd        &  127.88867  &  -1.19944  &  0.0304 $\pm$ 0.000100 & CSP & 54207.43 $\pm$ 0.06 & 1.270 $\pm$ 0.012 & -0.018 $\pm$ 0.010 &  0.029 $\pm$ 0.0009 \\ 
SN2007ca        &  202.77421  &  -15.10183  &  0.0141 $\pm$ 0.000010 & CSP & 54228.20 $\pm$ 0.14 & 1.037 $\pm$ 0.024 &  0.376 $\pm$ 0.012 &  0.057 $\pm$ 0.0016 \\ 
SN2007co        &  275.76500  &  29.89722  &  0.0270 $\pm$ 0.000110 & CfA & 54264.91 $\pm$ 0.23 & 1.040 $\pm$ 0.040 &  0.208 $\pm$ 0.017 &  0.096 $\pm$ 0.0037 \\ 
SN2007cq        &  333.66833  &  5.08028  &  0.0260 $\pm$ 0.000080 & CfA & 54280.90 $\pm$ 0.10 & 1.062 $\pm$ 0.021 &  0.051 $\pm$ 0.011 &  0.092 $\pm$ 0.0020 \\ 
SN2007jg        &  52.46175  &   0.05683 &  0.0371 $\pm$ 0.000013 & CSP & 54366.64 $\pm$ 0.25 & 1.088 $\pm$ 0.034 &  0.150 $\pm$ 0.017 &  0.090 $\pm$ 0.0020 \\ 
SN2007le        &  354.70171  &  -6.52258  &  0.0067 $\pm$ 0.000003 & CSP & 54399.85 $\pm$ 0.07 & 1.027 $\pm$ 0.016 &  0.379 $\pm$ 0.008 &  0.029 $\pm$ 0.0003 \\ 
SN2007qe        &  358.55417  &  27.40917  &  0.0240 $\pm$ 0.000050 & CfA & 54429.59 $\pm$ 0.10 & 0.988 $\pm$ 0.023 &  0.069 $\pm$ 0.014 &  0.033 $\pm$ 0.0008 \\ 
SN2007sr        &  180.47000  &  -18.97269  &  0.0055 $\pm$ 0.000030 & CSP & 54449.73 $\pm$ 0.19 & 1.084 $\pm$ 0.015 &  0.173 $\pm$ 0.009 &  0.040 $\pm$ 0.0010 \\ 
SN2007st        &  27.17696  &   -48.64939 &  0.0212 $\pm$ 0.000030 & CSP & 54455.09 $\pm$ 0.32 & 1.486 $\pm$ 0.019 &  0.101 $\pm$ 0.018 &  0.014 $\pm$ 0.0004 \\ 
SN2008af        &  224.86875  &  16.65333  &  0.0334 $\pm$ 0.000007 & CfA & 54499.69 $\pm$ 0.43 & 1.178 $\pm$ 0.010 & -0.028 $\pm$ 0.023 &  0.029 $\pm$ 0.0012 \\ 
SN2008ar        &  186.15800  &  10.83817  &  0.0262 $\pm$ 0.000010 & CSP & 54535.22 $\pm$ 0.07 & 1.032 $\pm$ 0.014 &  0.081 $\pm$ 0.008 &  0.031 $\pm$ 0.0011 \\ 
SN2008bc        &  144.63012  &  -63.97378  &  0.0151 $\pm$ 0.000120 & CSP & 54550.41 $\pm$ 0.08 & 1.015 $\pm$ 0.019 &  0.003 $\pm$ 0.008 &  0.225 $\pm$ 0.0042 \\ 
SN2008bf        &  181.01208  &  20.24517  &  0.0235 $\pm$ 0.000167 & CSP & 54555.31 $\pm$ 0.06 & 0.967 $\pm$ 0.012 & -0.013 $\pm$ 0.006 &  0.030 $\pm$ 0.0027 \\ 
SN2008C         &  104.29804  &  20.43714  &  0.0166 $\pm$ 0.000013 & CSP & 54466.60 $\pm$ 0.23 & 1.075 $\pm$ 0.019 &  0.239 $\pm$ 0.010 &  0.072 $\pm$ 0.0023 \\ 
SN2008fl        &  294.18683  &  -37.55125  &  0.0199 $\pm$ 0.000103 & CSP & 54721.85 $\pm$ 0.13 & 1.328 $\pm$ 0.006 &  0.080 $\pm$ 0.005 &  0.157 $\pm$ 0.0058 \\ 
SN2008fr        &  17.95475  &   14.64083 &  0.0390 $\pm$ 0.002001 & CSP & 54733.93 $\pm$ 0.26 & 0.920 $\pm$ 0.014 & -0.002 $\pm$ 0.011 &  0.040 $\pm$ 0.0012 \\ 
SN2008fw        &  157.23321  &  -44.66544  &  0.0085 $\pm$ 0.000017 & CSP & 54732.29 $\pm$ 0.15 & 0.844 $\pm$ 0.009 &  0.112 $\pm$ 0.008 &  0.112 $\pm$ 0.0030 \\ 
SN2008gb        &  44.48792  &   46.86583 &  0.0370 $\pm$ 0.000167 & CfA & 54748.22 $\pm$ 0.34 & 1.183 $\pm$ 0.014 &  0.080 $\pm$ 0.018 &  0.171 $\pm$ 0.0035 \\ 
SN2008gg        &  21.34600  &   -18.17244 &  0.0320 $\pm$ 0.000023 & CSP & 54750.93 $\pm$ 0.34 & 1.036 $\pm$ 0.028 &  0.155 $\pm$ 0.013 &  0.019 $\pm$ 0.0010 \\ 
SN2008gl        &  20.22842  &   4.80531 &  0.0340 $\pm$ 0.000117 & CSP & 54768.70 $\pm$ 0.09 & 1.319 $\pm$ 0.010 &  0.030 $\pm$ 0.006 &  0.024 $\pm$ 0.0008 \\ 
SN2008gp        &  50.75304  &   1.36189 &  0.0330 $\pm$ 0.000070 & CSP & 54779.62 $\pm$ 0.04 & 1.017 $\pm$ 0.008 & -0.018 $\pm$ 0.004 &  0.104 $\pm$ 0.0051 \\ 
SN2008hj        &  1.00796   & -11.16875   &  0.0379 $\pm$ 0.000130 & CSP & 54802.26 $\pm$ 0.12 & 1.055 $\pm$ 0.027 &  0.038 $\pm$ 0.012 &  0.030 $\pm$ 0.0008 \\ 
SN2008hm        &  51.79542  &   46.94444 &  0.0197 $\pm$ 0.000077 & CfA & 54804.74 $\pm$ 0.21 & 0.993 $\pm$ 0.025 &  0.182 $\pm$ 0.014 &  0.380 $\pm$ 0.0085 \\ 
SN2008hs        &  36.37333  &   41.84306 &  0.0174 $\pm$ 0.000070 & CfA & 54812.94 $\pm$ 0.14 & 1.531 $\pm$ 0.015 &  0.122 $\pm$ 0.024 &  0.050 $\pm$ 0.0003 \\ 
SN2008hv        &  136.89192  &  3.39225  &  0.0126 $\pm$ 0.000007 & CSP & 54817.65 $\pm$ 0.04 & 1.328 $\pm$ 0.006 & -0.065 $\pm$ 0.006 &  0.028 $\pm$ 0.0008 \\ 
SN2008ia        &  132.64646  &  -61.27794  &  0.0219 $\pm$ 0.000097 & CSP & 54813.67 $\pm$ 0.09 & 1.340 $\pm$ 0.009 &  0.003 $\pm$ 0.007 &  0.195 $\pm$ 0.0050 \\ 
SN2009aa        &  170.92617  &  -22.27069  &  0.0273 $\pm$ 0.000047 & CSP & 54878.81 $\pm$ 0.04 & 1.172 $\pm$ 0.008 &  0.020 $\pm$ 0.005 &  0.029 $\pm$ 0.0009 \\ 
SN2009ab        &  64.15162  &   2.76417 &  0.0112 $\pm$ 0.000020 & CSP & 54883.89 $\pm$ 0.08 & 1.288 $\pm$ 0.016 &  0.050 $\pm$ 0.010 &  0.184 $\pm$ 0.0028 \\ 
SN2009ad        &  75.88908  &   6.65992 &  0.0284 $\pm$ 0.000003 & CSP & 54886.91 $\pm$ 0.07 & 0.949 $\pm$ 0.013 &  0.020 $\pm$ 0.007 &  0.095 $\pm$ 0.0011 \\ 
SN2009ag        &  107.92004  &  -26.68508  &  0.0086 $\pm$ 0.000007 & CSP & 54890.23 $\pm$ 0.16 & 1.088 $\pm$ 0.019 &  0.343 $\pm$ 0.009 &  0.218 $\pm$ 0.0012 \\ 
SN2009al        &  162.84196  &  8.57853  &  0.0221 $\pm$ 0.000080 & CfA & 54897.20 $\pm$ 0.18 & 1.079 $\pm$ 0.033 &  0.236 $\pm$ 0.020 &  0.021 $\pm$ 0.0004 \\ 
SN2009an        &  185.69750  &  65.85111  &  0.0092 $\pm$ 0.000007 & CfA & 54898.56 $\pm$ 0.09 & 1.327 $\pm$ 0.010 &  0.063 $\pm$ 0.010 &  0.016 $\pm$ 0.0003 \\ 
SN2009bv        &  196.83542  &  35.78444  &  0.0366 $\pm$ 0.000017 & CfA & 54927.07 $\pm$ 0.20 & 0.948 $\pm$ 0.033 & -0.026 $\pm$ 0.019 &  0.008 $\pm$ 0.0008 \\ 
SN2009cz        &  138.75008  &  29.73531  &  0.0212 $\pm$ 0.000010 & CSP & 54943.50 $\pm$ 0.09 & 0.899 $\pm$ 0.014 &  0.102 $\pm$ 0.007 &  0.022 $\pm$ 0.0003 \\ 
SN2009D         &  58.59512  &   -19.18172 &  0.0250 $\pm$ 0.000033 & CSP & 54841.65 $\pm$ 0.11 & 1.025 $\pm$ 0.024 &  0.054 $\pm$ 0.009 &  0.044 $\pm$ 0.0012 \\ 
SN2009kk        &  57.43458  &   -3.26444 &  0.0129 $\pm$ 0.000150 & CfA & 55126.37 $\pm$ 0.20 & 1.189 $\pm$ 0.006 & -0.055 $\pm$ 0.011 &  0.116 $\pm$ 0.0025 \\ 
SN2009kq        &  129.06292  &  28.06722  &  0.0117 $\pm$ 0.000020 & CfA & 55154.81 $\pm$ 0.17 & 0.992 $\pm$ 0.025 &  0.089 $\pm$ 0.010 &  0.035 $\pm$ 0.0005 \\ 
SN2009Y         &  220.59938  &  -17.24678  &  0.0093 $\pm$ 0.000027 & CSP & 54877.10 $\pm$ 0.10 & 1.063 $\pm$ 0.023 &  0.169 $\pm$ 0.010 &  0.087 $\pm$ 0.0005 \\ 
SN2010ai        &  194.85000  &  27.99639  &  0.0184 $\pm$ 0.000123 & CfA & 55277.50 $\pm$ 0.08 & 1.421 $\pm$ 0.016 & -0.075 $\pm$ 0.016 &  0.008 $\pm$ 0.0010 \\ 
SN2010dw        &  230.66792  &  -5.92111  &  0.0381 $\pm$ 0.000150 & CfA & 55358.25 $\pm$ 0.35 & 0.844 $\pm$ 0.058 &  0.177 $\pm$ 0.028 &  0.080 $\pm$ 0.0009 \\ 
SN2010iw        &  131.31250  &  27.82278  &  0.0215 $\pm$ 0.000007 & CfA & 55497.14 $\pm$ 0.26 & 0.876 $\pm$ 0.019 &  0.084 $\pm$ 0.012 &  0.047 $\pm$ 0.0006 \\ 
SN2010kg        &  70.03500  &   7.35000 &  0.0166 $\pm$ 0.000007 & CfA & 55543.96 $\pm$ 0.10 & 1.194 $\pm$ 0.011 &  0.183 $\pm$ 0.014 &  0.131 $\pm$ 0.0022 \\ 
SN2011ao        &  178.46250  &  33.36278  &  0.0107 $\pm$ 0.000003 & CfA & 55639.61 $\pm$ 0.11 & 1.012 $\pm$ 0.018 &  0.035 $\pm$ 0.019 &  0.017 $\pm$ 0.0001 \\ 
SN2011B         &  133.95208  &  78.21750  &  0.0047 $\pm$ 0.000003 & CfA & 55583.38 $\pm$ 0.06 & 1.174 $\pm$ 0.005 &  0.112 $\pm$ 0.008 &  0.026 $\pm$ 0.0011 \\ 
SN2011by        &  178.94000  &  55.32611  &  0.0028 $\pm$ 0.000003 & CfA & 55690.95 $\pm$ 0.05 & 1.053 $\pm$ 0.008 &  0.067 $\pm$ 0.005 &  0.012 $\pm$ 0.0002 \\ 
SN2011df        &  291.89000  &  54.38639  &  0.0145 $\pm$ 0.000020 & CfA & 55716.40 $\pm$ 0.11 & 0.923 $\pm$ 0.015 &  0.072 $\pm$ 0.010 &  0.112 $\pm$ 0.0034 \\ 
SNf20080514-002 &  202.30625  &  11.26889  &  0.0219 $\pm$ 0.000010 & CfA & 54612.80 $\pm$ 0.00 & 1.360 $\pm$ 0.000 & -0.143 $\pm$ 0.000 &  0.027 $\pm$ 0.0014 \\[0.1cm]
\hline \\[0.1cm]
\multicolumn{9}{l}{\small $^a$ Heliocentric Redshift from NED or the literature.} \\
\multicolumn{9}{l}{\small $^b$ Light-curve (LC) data source. CfA:  \citealt{Woodvasey:2008}; \citealt{Hicken:2009b}; \citealt{Hicken:2012};} \\
\multicolumn{9}{l}{\small \citealt{Friedman:2015};} \\
\multicolumn{9}{l}{\small \citealt{Marion:2016}, CSP: \citealt{Contreras:2010}; \citealt{Stritzinger:2010};}\\ \multicolumn{9}{l}{\small \citealt{Stritzinger:2011}; \citealt{Krisciunas:2017}, } \\
\multicolumn{9}{l}{\small Others: K04c: \citealt{Krisciunas:2004}; V03: \citealt{Valentini:2003}; K03: \citealt{Krisciunas:2003};} \\
\multicolumn{9}{l}{\small P08: \citealt{Pignata:2008}; St07: \citealt{Stanishev:2007}; L09: \citealt{Leloudas:2009}; } \\
\multicolumn{9}{l}{\small K07: \citealt{Krisciunas:2007}. Also see Table 3 of \citet{Friedman:2015} for references. } \\
\multicolumn{9}{l}{\small $^c$ LC-shape parameter: apparent-magnitude decline between $B$-band peak luminosity and 15 days after peak.} \\
\multicolumn{9}{l}{\small $^d$ Host-galaxy color excess, as measured by SNooPy fits to the optical and NIR LCs. } \\
\multicolumn{9}{l}{\small $^e$ Milky-Way color excess, from the \citet{Schlafly:2011} Milky Way dust maps.} \\
\end{tabular} 
\label{Atab:ia_selections} 

\end{center} 
\end{table*}
\renewcommand{\arraystretch}{1}
\renewcommand{\baselinestretch}{1}

\

\begin{figure*}
\centering
\includegraphics{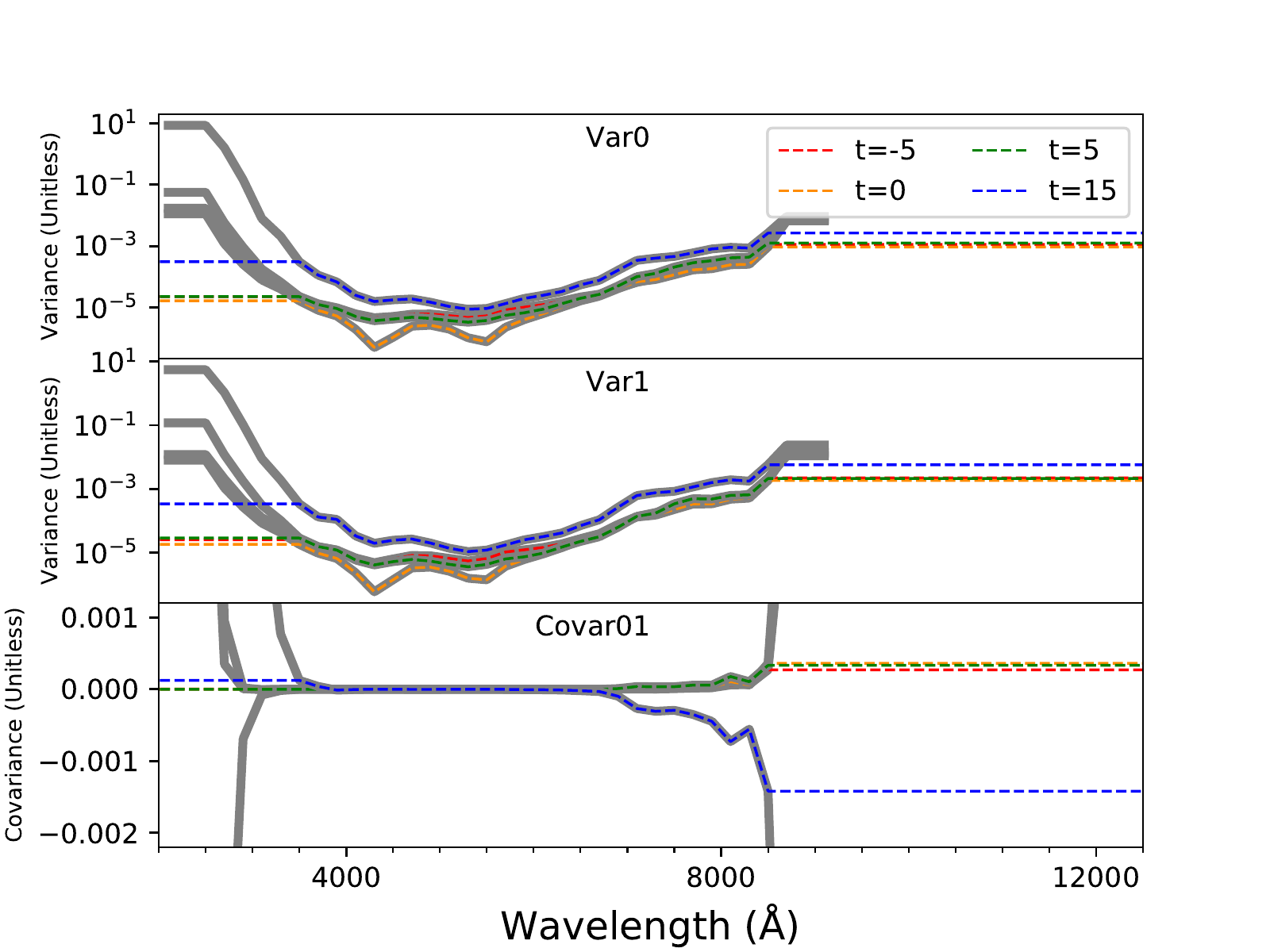}
\caption{\label{Afig:SALT2_variance} Extrapolation of the SALT2 variance and covariance tables to UV and IR wavelengths, at four phases.
The top and middle panels show the variance of model components $M_0$ and $M_1$, respectively, and the bottom panel plots the ($M_0$, $M_1$) covariance.  
In each panel the pre-existing variance and covariance components are plotted as thick gray lines, and the newly extrapolated versions are shown as thin dashed lines, with color-coding indicating the phase being plotted.  
In  each case the extrapolation is done as a ``flat-line'' extrapolation holding a constant value into the UV and NIR wavelengths.}
\end{figure*}

\begin{figure*}
\centering
\includegraphics{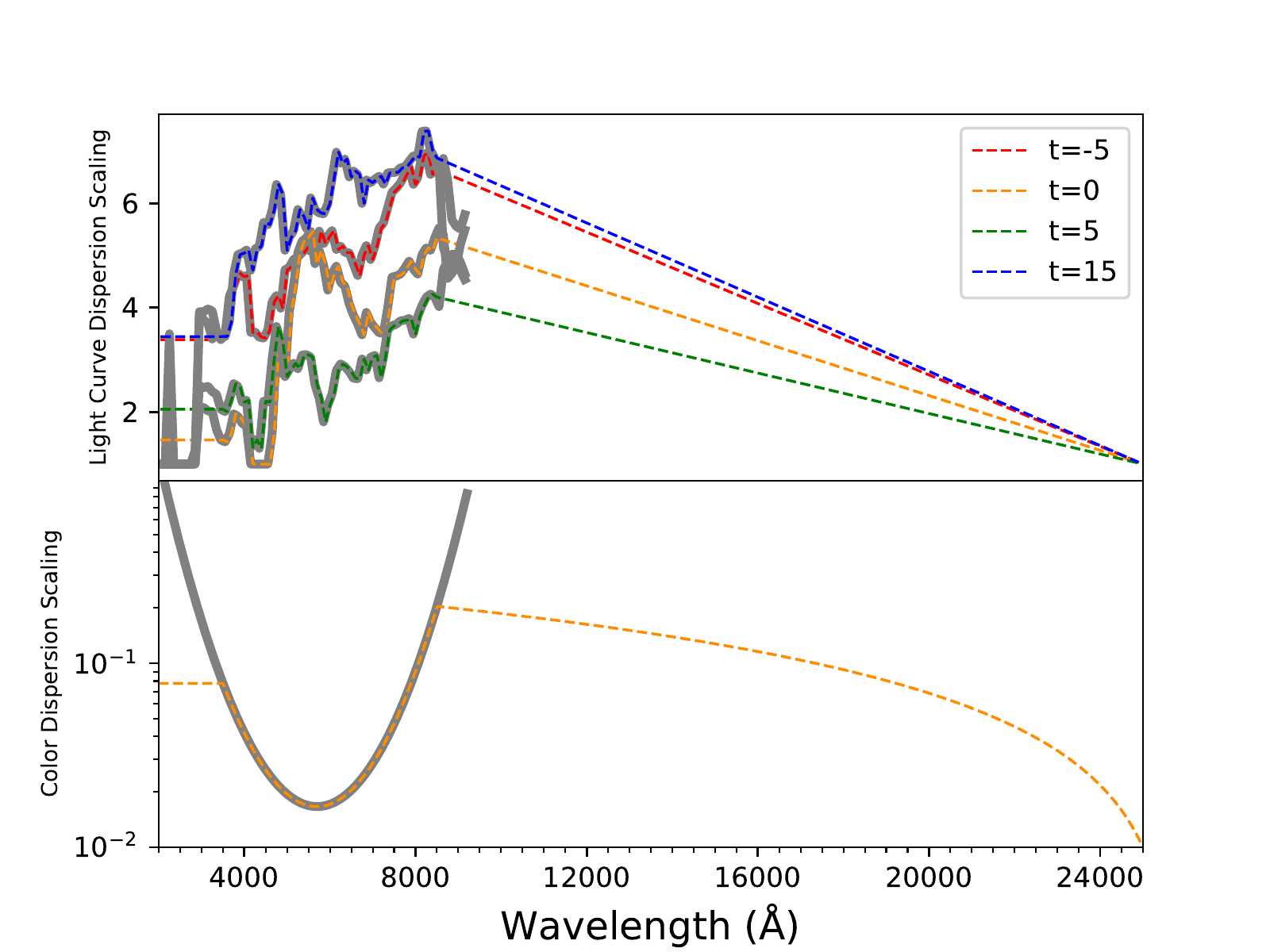}
\caption{\label{Afig:SALT2_dispersion} Extrapolation of the SALT2 dispersion tables to UV and IR wavelengths. 
The top panel shows the light-curve dispersion scaling and the bottom panel shows the color dispersion scaling array.  
In both panels the pre-existing dispersion arrays are plotted as thick gray lines, and the newly extrapolated versions are shown as thin dashed lines. For the top panel, color-coding indicates the phase being plotted.  
In both cases, on the NIR side the extrapolation begins at 8500~\mbox{\normalfont\AA}\ and extends with a downward linear extrapolation to 2.5~$\mu$m.   On the UV side we apply a ``flat-line'' extrapolation, holding the SALT2 value at 3500~\mbox{\normalfont\AA}\ constant down to 1700~\mbox{\normalfont\AA}.}
\end{figure*}

\end{document}